\documentclass[
aps,
prx,
showpacs,
preprintnumbers,
twocolumn,
superscriptaddress,
]{revtex4-2}

\usepackage[T1]{fontenc}
\usepackage{newtxtext}
\usepackage{newtxmath}
\usepackage[normalem]{ulem}

\usepackage{enumitem}

\usepackage{makecell}

\usepackage{physics}
\usepackage{amsmath}
\usepackage{mathtools}
\usepackage{dsfont}

\usepackage{comment}

\usepackage{appendix}

\usepackage{placeins}

\usepackage{color}
\usepackage[dvipsnames]{xcolor}
\usepackage{colortbl}
\usepackage{hyperref}
\hypersetup{colorlinks,citecolor=blue,linkcolor=blue,urlcolor=blue}

\definecolor{darkgreen}{rgb}{0.55, 0.71, 0.00}
\definecolor{Gray}{gray}{0.9}

\usepackage{xpatch}

\makeatletter
\patchcmd{\@ssect@ltx}
    {\addcontentsline{toc}{#1}{\protect\numberline{}#8}}
    {}
    {}
    {}
\makeatother

\usepackage{xspace}
\usepackage{layouts}
\usepackage{tcolorbox}

\usepackage{graphicx}
\usepackage[all]{hypcap}
\usepackage{tikz} % https://www.bu.edu/math/files/2013/08/tikzpgfmanual.pdf

\graphicspath{{FiguresSI/}{FiguresMain/}}

\newcommand{\Hdd}{H_\mathrm{dd}}

\newcommand{\id}{\mathds{1}}

\newcommand{\C}{$^{13}$C}

\newcommand{\xd}{\delta}
\newcommand{\xe}{\epsilon}

\newcommand{\tm}{{\text -}}

\newcommand{\tacq}{t_{\R{acq}}}
\newcommand{\xg}{\gamma}

\newcommand{\xph}{\phi}

\newcommand{\app}{\approx}

\newcommand{\Bpol}{\textbf{B}_{\R{pol}}}

\newcommand{\Cs}{{}^{13}\R{C}}

\newcommand{\fac}{f_{\R{AC}}}
\newcommand{\fres}{f_{\R{res}}}

\newcommand{\Bac}{B_{\R{AC}}}
\newcommand{\Pac}{\Phi_{\R{AC}}}

\newcommand{\mHdd}[0]{\mH_{\R{dd}}}

\newcommand{\xy}[0]{\xhat\tm\yhat}

\newcommand{\xD}{\Delta}

\newcommand{\xPh}{\Phi}

\newcommand{\mH}[0]{\mathcal{H}}

\newcommand{\beq}{\begin{equation}}
\newcommand{\eeq}{\end{equation}}
                  
\newcommand{\benum}{\begin{enumerate}}
\newcommand{\eenum}{\end{enumerate}}
                    
\newcommand{\bit}{\begin{itemize}}
\newcommand{\eit}{\end{itemize}}
\newcommand{\xhat}{\hat{\T{x}}}
\newcommand{\yhat}{\hat{\T{y}}}
\newcommand{\zhat}{\hat{\T{z}}}

\newcommand{\bea}{\begin{eqnarray}}
\newcommand{\eea}{\end{eqnarray}}

\newcommand{\bev}{\begin{verbatim}}
\newcommand{\eev}{\end{verbatim}}

\newcommand{\zt}{\times}

\newcommand{\qt}{\tau}

\newcommand{\T}[1]{\textbf{#1}}
\newcommand{\I}[1]{\textit{#1}}
\newcommand{\R}[1]{\textrm{#1}}
\newcommand{\zfl}[1]{\protect\label{fig:#1}}
\newcommand{\zfr}[1]{\figurename\,\ref{fig:#1}}

\newcommand{\zsl}[1]{\label{sec:#1}}
\newcommand{\zsr}[1]{Sec.\ref{sec:#1}}

\newcommand{\expec}[1]{\left\langle #1\right\rangle}

\newcommand{\ba}{\left\{ \begin{array}{lr}}
\newcommand{\ea}{\end{array}\right.}

\newcommand{\blist}[1]{
 \begin{list}{#1}%$\ast\circ\bullet\Right
 \begin{align}
	 arrow
 \end{align}
 $\checkmark\star
  { \setlength{\itemsep}{3pt}
     \setlength{\parsep}{2pt}
     \setlength{\topsep}{3pt}
     \setlength{\partopsep}{0pt}
     \setlength{\leftmargin}{1em}
     \setlength{\labelwidth}{1em}
     \setlength{\labelsep}{0.5em} } }
\newcommand{\elist}{
  \end{list}  }

\newcommand{\bef}
{
\begin{figure}[htbp]
\centering
}

\newcommand{\eef}{\end{figure}}

\renewcommand{\figurename}{Fig.}

\newcommand{\beginmethods}{%
        \setcounter{table}{0}
        \renewcommand{\thetable}{M\arabic{table}}%
        \setcounter{figure}{0}
        \renewcommand{\thefigure}{M\arabic{figure}} %
        \renewcommand{\theHfigure}{M\arabic{figure}} %fixes linking to figures
		\renewcommand{\figurename}{Fig.} 
        \setcounter{equation}{0}
        \renewcommand{\theequation}{M\,\arabic{equation}}
     }

\newcommand{\lemma}[2]{
\setcounter{equation}{0}
\renewcommand{\theequation}{L\,\arabic{equation}}

\twocolumngrid
\onecolumngrid
\begin{tcolorbox}[colback=white!95!gray, colframe=black, boxrule=0.5pt, arc=2mm, width=\textwidth]
\hypertarget{proof:main}{}
\begin{flushleft}
#1
\end{flushleft}
\label{#2}
\end{tcolorbox}
\twocolumngrid
}
\newcommand{\beginsupplement}{%
        \setcounter{table}{0}
        \renewcommand{\tablename}{Supplementary Table}
        \renewcommand{\thetable}{\arabic{table}}%
        \setcounter{figure}{0}
        \renewcommand{\thefigure}{S\arabic{figure}} %
        \renewcommand{\theHfigure}{S\arabic{figure}} %fixes linking to figures
		\setcounter{page}{1}
		\renewcommand{\figurename}{Fig.} 
		\renewcommand{\thesection}{\:S\arabic{section}}
		\setcounter{section}{0}
        \setcounter{equation}{0}
        \renewcommand{\theequation}{S\,\arabic{equation}}
     }

\newcommand{\affA}{Department of Chemistry, University of California, Berkeley, Berkeley, CA 94720, USA.}
\newcommand{\affB}{Max Planck Institute for the Physics of Complex Systems, N\"othnitzer Str.~38, 01187 Dresden, Germany.}
\newcommand{\affC}{Chemical Sciences Division,  Lawrence Berkeley National Laboratory,  Berkeley, CA 94720, USA.}

\newcommand{\affF}{CIFAR Azrieli Global Scholars Program, 661 University Ave, Toronto, ON M5G 1M1, Canada.}

\begin{document}

%%%%%%%%%%%%%%%%%%%%%%%%%%%%%%%%%%%%%%%%%%%%%%%%%%%%%%%%%%%%%%%%%%%%%%%%%%%
%                        Main Document
%%%%%%%%%%%%%%%%%%%%%%%%%%%%%%%%%%%%%%%%%%%%%%%%%%%%%%%%%%%%%%%%%%%%%%%%%%%
%%%%%%%%%%%%%%%%%%%%%%%%%%%%%%%%%%%%%%%%%%%%%%%%%%%%%%%%%%%%%%%%%%%%%%%%%%%

%\title{Discrete Time Crystal Quantum Sensing}
\title{Sensing with discrete time crystals}

\author{Leo~Joon~Il~Moon}
\thanks{equal contribution}
\affiliation{\affA}
\affiliation{\affC}

\author{Paul~M.~Schindler}
\thanks{equal contribution}
\affiliation{\affB}

\author{Ryan~J.~Smith}
\thanks{equal contribution}
\affiliation{\affC}

\author{Emanuel~Druga}
\affiliation{\affA}

\author{Zhuo-Rui Zhang}
\affiliation{\affA}

\author{Marin~Bukov}
\email{mgbukov@pks.mpg.de}
\affiliation{\affB}

\author{Ashok~Ajoy}
\email{ashokaj@berkeley.edu}
\affiliation{\affA}
\affiliation{\affC}
\affiliation{\affF}

\begin{abstract}
Prethermal discrete time crystals~(PDTCs) are a nonequilibrium state of matter characterized by long-range spatiotemporal order, and exhibiting a subharmonic response stabilized by many-body interactions under periodic driving. The inherent robustness of time crystalline order to perturbations in the drive protocol makes DTCs promising for applications in quantum technologies. 
We exploit the susceptibility of PDTC order to deviations in its order parameter to devise highly frequency-selective quantum sensors for time-varying (AC) magnetic fields in a system of strongly-driven, dipolar-coupled $\Cs$ nuclear spins in diamond. 
Integrating a time-varying AC field into the PDTC allows us to exponentially increase its lifetime, with improvements of up to three orders of magnitude (44,204 cycles), and results in a strong resonant response in the time crystalline order parameter. The linewidth of our sensor is limited by the PDTC lifetime alone, as strong interspin interactions help stabilize DTC order. 
The sensor operates in the $0.5{-}50$~kHz range -- a challenging frequency regime for sensors based on atomic vapor or electronic spins -- and attains a competitive sensitivity.  
PDTC sensors are resilient to errors in the drive protocol and sample inhomogeneities, and are agnostic to the macroscopic details of the physical platform: the underlying physical principle applies equally to superconducting qubits, neutral atoms, and trapped ions.
\end{abstract}

\maketitle

\section{Introduction}

Non-equilibrium matter has emerged as a frontier in modern many-body physics, displaying novel phenomena beyond restrictions imposed by thermal equilibrium.
A milestone is the demonstration~\cite{zhang2017observation,choi2017observation,kyprianidis2021observation,rovny2018observation,pal2018temporal,autti2018observation,randall2021many,frey2022realization,Zhang2022Observation,DTC_Beatrez2022,mi2022time,stasiuk2023_U1enhanced,he2024experimental} of discrete time crystals (DTCs)~\cite{sacha2015_dtc,khemani2016phase,Yao17_DTC,sacha2017time,pizzi2019period,Else2020,mcginley2022absolutely,xiang2024long,Zaletel2023_DTC_Review}, a new form of non-equilibrium matter that breaks time-translation symmetry, akin to ordinary crystals breaking spatial symmetry. A hallmark of DTCs is their robust period-doubling response, stabilized by many-body interactions of mean strength $J$, making them resilient to errors in the protocol creating them. Most observed time-crystalline states rely on Floquet prethermalization~\cite{Abanin2015_Heating,Mori2016_Heating,Weidinger2017,Abanin2017_prethermalization,Abanin2017_Rigorous,peng2021floquet}, where periodically driven quantum states are preserved for durations parametrically controlled by the drive frequency, resulting in lifetimes $T_2^\prime$ far exceeding the system’s interaction-dominated intrinsic decay time $T_2^{\ast}$ (${\propto}J^{-1}$).

The robustness and long lifetimes of DTCs in the presence of interactions make them promising for quantum technologies, such as simulating complex systems~\cite{Estarellas2020_DTCSimulating}, topologically protected quantum computation~\cite{Bomantara2018_DTC_QC}, and robust generation of entangled states~\cite{bao2024_schrodingercatsgrowing60}. Experimental work has demonstrated the use of continuous time crystals as a DC-field sensor~\cite{greilich2024robust}. Separately, theoretical proposals have suggested using discrete time crystals for enhanced quantum sensing~\cite{choi2017_quantummetrologybasedstrongly,Lyu2020_DTCbeyondHeisenberg,Iemini2024_DTC_ACsensor,yousefjani2024_DTCphaseSensing,gribben2024quantum}. However, experimental realizations using DTCs as quantum sensors have been challenging due to the necessity of using strongly correlated states~\cite{choi2017_quantummetrologybasedstrongly} or fine-tuned systems~\cite{Lyu2020_DTCbeyondHeisenberg,Iemini2024_DTC_ACsensor,yousefjani2024_DTCphaseSensing}.

In this work, we develop a new approach for using DTCs to construct highly frequency-selective quantum sensors for time-varying (AC) magnetic fields, and demonstrate it experimentally in an ensemble of randomly positioned, hyperpolarized, $\Cs$ nuclear spins in diamond. The sensor operates in the 0.5-50 kHz range -- typically a challenging frequency regime for sensors based on atomic vapor~\cite{Budker07} or electronic spins~\cite{Fabricant23}, while achieving competitive sensitivity. The scheme leverages the robustness of DTC order, requires no preparation of strongly correlated states, and is broadly applicable in platforms exhibiting prethermal $U(1)$ DTC order~\cite{choi2017observation,rovny2018observation,pal2018temporal,randall2021many,DTC_Beatrez2022,stasiuk2023_U1enhanced}.

%Our approach is based on the observation that, under certain conditions, DTC order can be significantly additionally stabilized by the presence of an AC field, with the stabilization being highly frequency-selective, enabling it to be exploited for sensing.
Our approach is based on the observation that, under certain conditions, the stability of DTC order can be significantly enhanced by the presence of an AC field; this stabilization is highly frequency-selective, enabling the DTC to be exploited for sensing. Specifically, the AC field $\Bac(t)$ couples to the DTC order parameter only when its frequency $\fac{=}\fres$ matches the DTC oscillations, protecting the DTC order from symmetry-breaking perturbations and exponentially enhancing its lifetime $T_2'$~\cite{PrethermalWithoutTemperature_LuitzEtAl2020}. We demonstrate that the lifetime extension can be as much as three orders of magnitude, yielding a record for coupling-normalized DTC lifetimes \(JT_2’ {\approx} 14051\); and importantly that it is a strongly \I{resonant} effect, producing a narrow AC frequency response, with linewidths under $70\,$mHz around $\fres$, determined solely by $(T_2^\prime)^{-1}$.

We leverage this, using a DTC excited via two-tone Floquet drive~\cite{DTC_Beatrez2022}, to construct a noise-rejected, continuously interrogated, AC sensor that operates for extended periods without re-initialization. Unlike traditional quantum sensing~\cite{Degen17}, which avoids interactions between sensor spins~\cite{Bauch20}, the DTC sensor here intimately relies on these interactions, and thermalization, to establish a $T_2^\prime$-limited sensor linewidth, while remaining robust against drive errors and on-site disorder (see SI Sec.\ref{supmat:sec:thy_properties}).
We additionally demonstrate that the AC-field mediated lifetime extension applies equally to DTCs excited along both transverse and longitudinal axes, suggesting wide applicability across diverse platforms including spin systems~\cite{choi2017observation,rovny2018observation,pal2018temporal,randall2021many,DTC_Beatrez2022,stasiuk2023_U1enhanced}, superconducting qubits~\cite{mi2022time,bao2024_schrodingercatsgrowing60,shinjo2024unveiling}, and cold atoms~\cite{zhang2017observation,kyprianidis2021observation}.

\begin{figure*}[t]
    \centering
    \includegraphics[width=\textwidth]{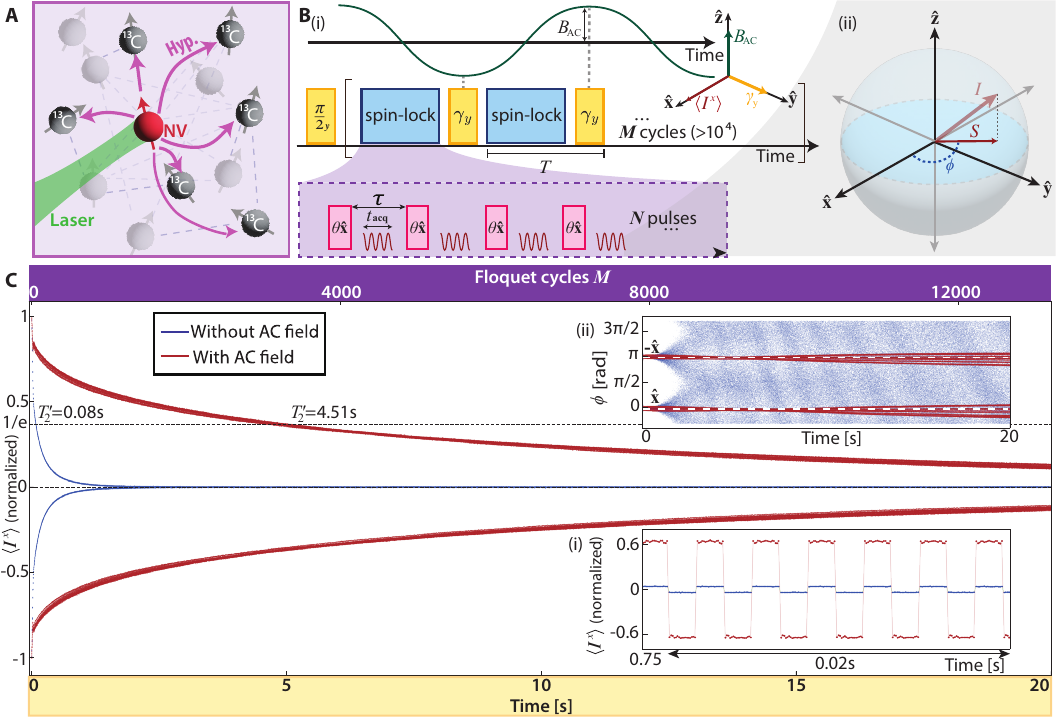}
    \caption{\T{System and Principle.}
(A) \T{\I{System}} consists of dipolar interacting $\Cs$ nuclear spins, hyperpolarized by NV centers using optical and chirped microwave excitation.
(B) \T{\I{Protocol}} (i) evolution under native dipole-dipole interactions, $\mHdd$~(see main text) is interrupted by a concatenated two-tone drive with $N$ \textit{spin-locking} (pink) $\theta (\xhat)$-pulses separated by $\qt$, interspersed with \textit{spin-flip} $\gamma (\yhat)$-pulses (yellow).
This time block, of total period $T$, is repeated $M$ times.
The protocol causes switching between $\xhat{\leftrightarrow}-\xhat$ every $t{=}N\tau$, while remaining robust against deviations $\gamma_y=\pi+\xe$, forming a prethermal DTC.
Additionally, a $\zhat$-oriented AC field (green) with amplitude $\Bac$ and frequency $\fac$ is applied; shown is the resonant case $\fac{=}\fres=1/(2T)$.
(ii) Net spin magnetization $I$ is monitored during acquisition time $\tacq {\sim}13.6\,\mu$s between the $\theta_x$-pulses.
Projection $S$ onto the $\Cs$ nuclear spin's rotating-frame $\xy$ plane and its phase $\phi$ are measured.
(C) \T{\I{Main result.}} Magnetization $\expval{I^x}$ for the PDTC protocol without (blue) and with (red) applied resonant AC field. Here, $N{=}16$; pulse separation $\tau$ ($\approx36\,\mu$s);$(\pi/2)_{y}$ and $\theta_{x}$ pulses ($\approx 50.25\,\mu$s); $\gamma_y$ pulse ($\approx 98.5\,\mu$s); $\Bac{=}82.4\,\mu$T with $\fres{=}330\,$Hz. Upper axis indicates number of flips $M$. Dashed line indicates the $1/e$-intercept, yielding lifetimes of $T_2^\prime{=}80\,$ms without AC field and $T_2^\prime{=}4.51\,$s with it. Data shows $M{>}13000$ $\expval{I^x}$ flips sustained over $t{=}20$s.
(i) \T{\I{Zoomed view}} of data in a small $20\,$ms window at $t{=}0.76\,$s, displaying magnetization switching from $-\xhat$ to $\xhat$. Lifetime extension under AC field is evident from increased amplitude of red data.
(ii) \T{\I{Tracked phase $\phi$}} for data from main panel, displaying coherent signal far beyond $1{/}e$ decay time. $\xhat (-\xhat)$ rails correspond to phases $0 (\pi)$ respectively. Decoherence in non-AC case leads to $\phi$ spread uniformly in $[-\pi,\pi]$ (blue points). AC field leads significant lifetime increase (red).
}
%Comments:  
    \zfl{fig1}
\end{figure*}

\begin{figure}[t]
    \centering  \includegraphics[width=0.495\textwidth]{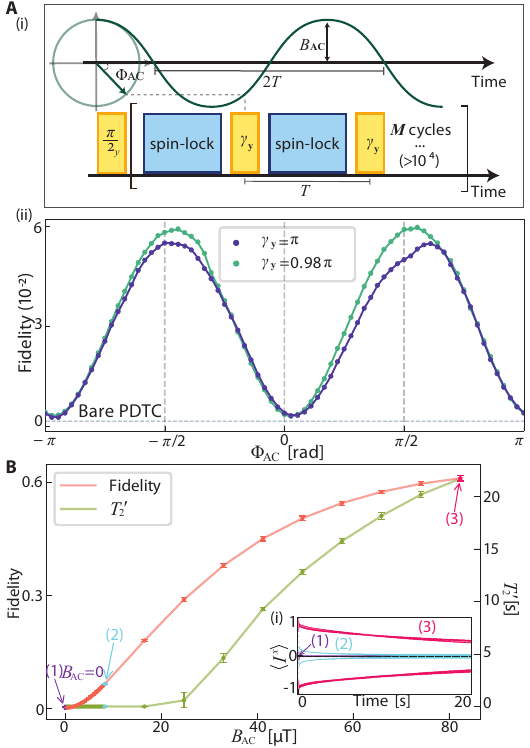}
   \caption{\T{PDTC lifetime extension under resonant AC fields.}
(A) \T{\I{Effect of AC field phase}} \(\xPh_\text{AC}\). 
(i) \T{\I{Schematic}:} AC field phase \(\xPh_\text{AC}\) is measured relative to the application of the \(\xg_y\) pulses, with \(\xPh_\text{AC}{=}\pm\pi/2\) indicating AC field troughs and peaks align with center of \(\xg_y\) kicks.  
(ii) \T{\I{Lifetime extension fidelity}} $F$ (Eq.~\eqref{eq:fidelity_metric}) as a function of AC field phase \(\xPh_\text{AC}\) at fixed amplitude \(\Bac{=}8.24\,\mu\)T on resonance. Signal increase is strongest at \(\xPh_\text{AC}{=}\pi/2\); for \(\xPh_\text{AC}{=}0\) there is minimal lifetime increase over bare PDTC (dashed line). Blue (green) data points show cases for $\xg_y$ pulses on ($\xe{=}2\%$ away from) the $\xg_y{=}\pi$ PDTC stable point. Qualitatively the same behavior is observed, indicating PDTC robustness. $(\pi/2)_{y}$ and $\theta_{x}$ pulses ($\approx 50.25\,\mu$s), $\pi$ pulse ($\approx 100.5\,\mu$s), $\tau$ ($\approx 36\,\mu$s) and $0.98\pi$ pulse ($\approx 98.5\,\mu$s).
(B) \T{\I{Effect of AC field amplitude}} \(\Bac\). Data points show fidelity $F$ as a function of resonant amplitude \(\Bac\) with \(\xPh_\text{AC}{=}\pi/2\). Solid line is spline fit guide to the eye. Normalized lifetime extension is shown on the right axis, demonstrating a ${>}3500$-fold $T_2'$ increase for \(82.4\,\mu\)T fields.  
\T{\I{Inset:}} Time-domain profiles of representative points in (B): (1) no AC field, (2) intermediate field \(\Bac {=} 8.24\,\mu\)T, and (3) \(\Bac {=} 82.4\,\mu\)T.
}
%Comments:  

    \label{fig:fig2}
\end{figure}

\begin{figure*}[t]
    \centering
    \includegraphics[width=\textwidth]{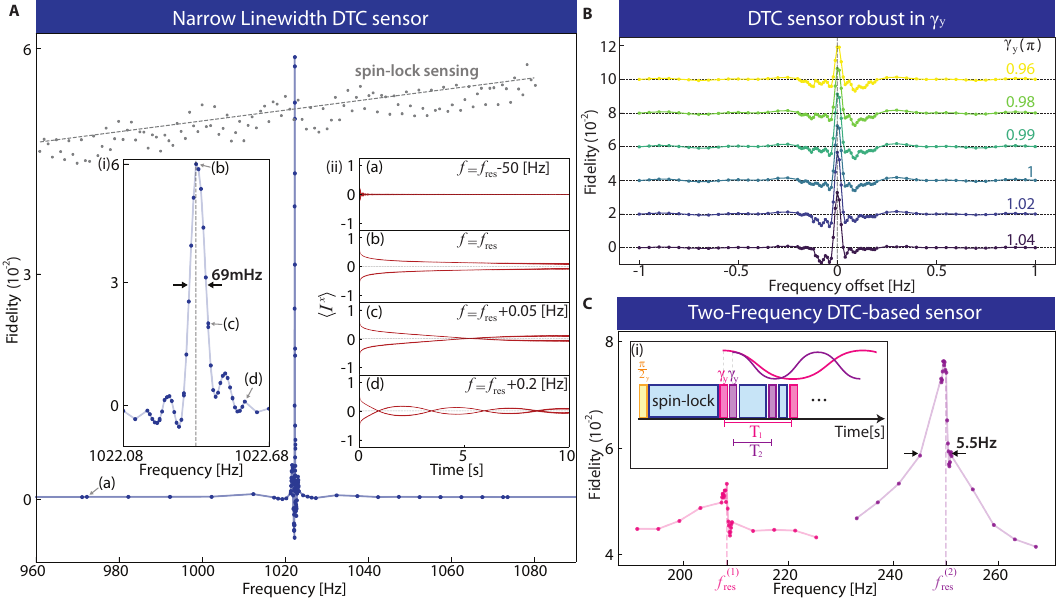}
     \caption{\T{PDTC based AC magnetic field sensing.}
(A) \T{\I{Narrow linewidth AC sensing}:} The fidelity $F$ (blue points) is measured by sweeping the frequency $\fac$ with $\Bac = 8.24\,\mu\mathrm{T}$ and $N=4$, while keeping other parameters consistent with \zfr{fig1}C. A sharp increase in PDTC lifetime, and hence fidelity, occurs at the resonance condition $\fac = \fres$. In contrast, the spin-lock sensing scheme introduced in \cite{Sahin22} (grey points) lacks frequency selectivity.
(i) \T{\I{Zoomed view}} into the resonance feature, showing a narrow linewidth, $\xD\ell\app70\,$mHz, determined by $(T_2^\prime)^{-1}$. Points \I{(b)}-\I{(d)} are marked on the spectral wing.
(ii) \T{\I{Time-domain PDTC profiles}} of $\expec{I^x}$ at points \I{(a)}-\I{(d)} in (A) and (i) at various offset frequencies from resonance. \I{(a)} Far off-resonance: fast signal decay, similar to bare PDTC case. \I{(b)} On resonance: significantly extended PDTC lifetime. \I{(c, d)} Slightly off-resonance, showing long lifetimes with beat pattern at frequency $\delta f{=}\fac{-}\fres$, resulting in $T_2^\prime$-limited AC sensing. See SI~\ref{supmat:sec:noise_reject} for exploiting this for noise-rejected sensing. $(\pi/2)_{y}$ and $\theta_{x}$ pulses ($\approx 51.5\,\mu$s); $\gamma_{y}$ pulse ($\approx103\,\mu$s); $\tau$ ($ \approx 36\,\mu$s).
(B) \T{\I{Robustness of the resonance feature}} to deviations in $\xg_y$ kick angle, $\xg_y=\pi-\xe$ (colorbar). The fidelity baselines for different $\xg_y$ kick angles were offset by $2\times10^{-2}$ to prevent overlap. Data shows that spectral width $\xD\ell$ remains independent of $\xe$. Similar experiment mapping PDTC phase diagram with respect to $\xg_y$ is shown in SI. $(\pi/2)_{y}$ and $\theta_{x}$ pulses ($\approx 52.25\,\mu$s); $\tau (\approx 36\,\mu$s); $\gamma_{y}$ pulse length scales linearly with its angle, with $\pi$-pulse ($\approx 104.5 \,\mu$s).
(C) \T{\I{Tunable sensor profile}} for two-frequency sensing. \I{Inset (i):} Sequence with two interspersed $\xg_y$-pulse blocks, leading to two resonance conditions, $\fres^{(1)}$ and $\fres^{(2)}$. $\Bac{=}32.96\,\mu\mathrm{T}$. \I{Main panel:} Measured frequency response, similar to (A), showing a two-tone response centered at $208\,$Hz and $250\,$Hz, with a narrow linewidth $\xD\ell{\app}5.5\,$Hz. $(\pi/2)_{y}$ and $\theta_{x}$ pulses ($\approx 50.25\,\mu$s); $\gamma_{y}$ pulse ($\approx 100.5 \,\mu$s). $t_{acq} \approx 13.6\,\mu$s for (C).
}
%Comments:
    \label{fig:fig3}
\end{figure*}

\begin{figure}[t]
    \centering
    \includegraphics[width=0.5\textwidth]{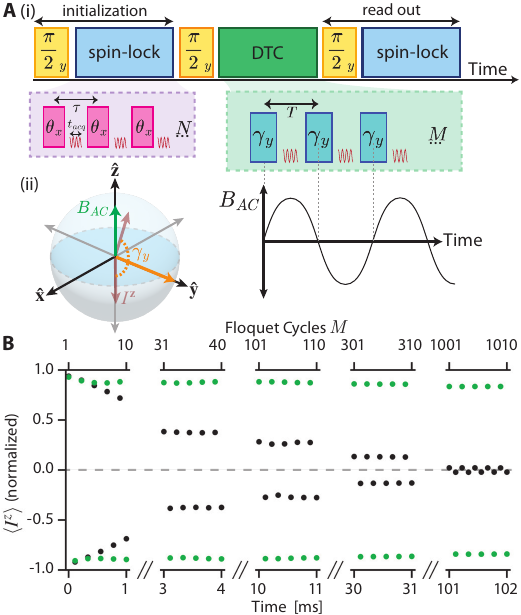}
     \caption{\T{Single-axis PDTC extension under AC field:}
(A) (i) \T{\I{Protocol}:} PDTC sequence (green) consists of a train of $M$ 100.5~$\mu$s $\gamma_y{=}\pi$-pulses along $\yhat$, spaced by 300~$\mu$s, flipping the spins between $+\zhat$ and $-\zhat$, robust to deviations from $\gamma_y{=}\pi$~\cite{DTC_Beatrez2022}. Spin-locking trains before and after the DTC sequence read initial and final spin population $\expec{I_z}$, enhancing signal-to-noise ratio.
(ii) \textit{Spin motion on the Bloch sphere}. PDTC oscillates along the same axis as the applied AC field ($\zhat$), unlike the two-tone case where the axes are orthogonal.
(B) \T{\I{PDTC lifetime extension}.} \I{Black points:} Normalized single-axis PDTC signal $\expec{I_z}$ without AC field, decaying with $T_2' {=} 0.01\,$s, or $M{=}35$ flips.  
\I{Green points:} DTC under a resonant AC field with $\Bac {=} 40\mu T$, showing a significantly extended lifetime of $T_2' {=} 20\,$s, or $M{=}70,000$ spin flips — a $2000$-fold increase. Each data point requires restarting the experiment.}
\label{fig:fig4}
\end{figure}

\section{Results}

\subsection{Principle: PDTC lifetime extension by AC fields}

The system consists of a diamond crystal with $\Cs$ nuclear spins hyperpolarized by optically pumped nitrogen-vacancy (NV) centers (\zfr{fig1}A). The $\Cs$ nuclei, at natural abundance, are randomly distributed and influenced by fluctuating fields from lattice NV and P1 centers~\cite{Ajoy19,Beatrez2023electron}. Spins interact via magnetic dipole interactions, $\mHdd = \sum_{k<l}J_{kl} (3I_{k}^z I_{l}^z - \T{I}_k\cdot \T{I}_{l})$, with spin-spin coupling strengths $J_{kl}$, and median coupling strength $J{\app}0.6\,$kHz~\cite{Beatrez21_90s} determined via free induction decay; $I_k^\alpha$ are spin-$1/2$ operators for $\Cs$ nuclear spin $k$, total polarization is $I^\alpha=\sum_kI^\alpha_k$, $\alpha=x,y,z$.

A $U(1)$ prethermal DTC (PDTC) is created using the two-tone drive protocol from Ref.~\cite{DTC_Beatrez2022} (\zfr{fig1}B(i)). Hyperpolarization initializes the $\Cs$ nuclear spins in the $\xhat$-polarized state $\rho_0{\propto} I^x$, after which the two-tone drive is activated. The first spin-locking drive, consisting of $\xhat$-oriented $\theta$ pulses separated by period $\tau$, realizes an effective Hamiltonian $H_\mathrm{eff}$ with emergent $U(1)$ symmetry; $\comm{H_\mathrm{eff}}{I^x}{=}0$. 

The second drive (period $T$) establishes the PDTC order via superimposed $\yhat$ pulses of angle $\gamma_y$~(${\approx} \pi$), applied after every $N$th spin-lock pulse. The $\Cs$ nuclear spins are inductively interrogated between pulses via an RF cavity; a down-sampling technique~\cite{moon24_rondeau}~(see Methods) enables quasi-continuous monitoring of their projection onto the $\xy$ plane directly in the rotating frame (\zfr{fig1}B(ii)). Net projection is denoted as $S$, while phase in the $\xy$-plane is $\phi$ (\zfr{fig1}B(ii)). Continuous interrogation with the two-tone drive allows full time-trace readout in a single shot, key for quantum sensing, and distinguishes it from single-tone drives commonly employed in other systems~\cite{rovny2018observation,stasiuk2023_U1enhanced,frey2022realization} (see Sec.~\eqref{sec:single_tone}).

PDTC order, arising from emergent $U(1)$-symmetry, is characterized by robust period doubling, seen in the long-lived oscillation of polarization $\expval{I^x}$ with period $2T$~\cite{DTC_Beatrez2022}. This is shown by the blue data points in \zfr{fig1}C, with the number of cycles $M$ of the DTC $\gamma_y$-drive on the upper axes. The decay of $\expval{I^x}$ has a characteristic $1/e$ time, $T_2' {=} 79\,$ms, that is much longer than $T_2^* {=} 1.5$ms \cite{Beatrez21_90s}. Purple points in the inset (\zfr{fig1}C(i)) provide a zoomed view at $t {=} 0.75$s.

The PDTC decay can be understood by noting that the initial state~($\rho_0{\sim} \id + \mu I^x$) corresponds to a zero-energy state with respect to the effective Hamiltonian, $\expval{H_\mathrm{eff}}_{\rho_0}{=}0$. The eigenstate thermalization hypothesis~(ETH)~\cite{Srednicki1995,rigol2008thermalization,Deutsch2018,dalessio2016quantum} implies that, without conservation laws, the system prethermalizes to a featureless infinite temperature~($\mathcal{T}{=}\infty$) state, $\rho_{\mathcal{T}{=}\infty} \sim \id$. For $U(1)$ quasi-conservation, prethermalization is restricted to states with the same polarization. However, small symmetry-breaking perturbations restore prethermalization to infinite temperature. In particular, higher-order corrections to $H_\mathrm{eff}$ in the two-tone drive break $U(1)$ conservation, leading to an inverse decay time (heating rate) $\Gamma_e = 1/T_e \propto (JT)^2$ ~\cite{Beatrez21_90s,DTC_Beatrez2022}~(SI Sec.\ref{supmat:sec:engineering}).

We now consider the effect of a resonant AC magnetic field, $\Bac(t) I^z$, with frequency $\fres$, aligned along $\zhat$ and locked to the DTC $\gamma_y$-kicks (green line in \zfr{fig1}B(i)). We show (SI Sec.\ref{supmat:sec:engineering}) that it \textit{exponentially} extends the lifetime of $U(1)$ PDTC order by Floquet-engineering a finite energy density~\cite{PrethermalWithoutTemperature_LuitzEtAl2020}, forming the basis for the sensor operation.
 
To understand this, note that the AC field induces an effective coupling to the PDTC order parameter $(-1)^\ell I^x$, i.e., $H_\mathrm{eff}\to H_\mathrm{eff, \ell, AC} {=} H_\mathrm{eff} + (-1)^\ell B I^x$, which, like the DTC response, alternates in sign each $T$-period $\ell$, where the coupling $B{\propto} \Bac$ is proportional to the AC field strength. Considering the effective Hamiltonian every even period, $H_\mathrm{eff, AC}{=}H_\mathrm{eff} + B I^x$, the DTC-ordered initial state, $\rho_0{\sim} \id+\mu I^x$, acquires a finite energy density, $\expval{H_\mathrm{eff,AC}}_{\rho_0}{=}\expval{H_\mathrm{eff}}_{\rho_0}{+}\expval{I^x}_{\rho_0}{\propto} \mu B$, controlled by the AC field (see SI Sec.\ref{supmat:sec:engineering}).

Thus, the AC field creates a finite energy density from the PDTC order, leading the PDTC to prethermalize to a finite temperature state, $\rho_\mathcal{T} {\propto} \exp(-H_\mathrm{eff, AC}/\mathcal{T})$, even with symmetry-breaking perturbations [we set Boltzmann's constant to unity]. This enhances its robustness, energetically protecting the PDTC state from prethermalization to infinite temperature, and results in a Floquet heating rate that is now exponentially suppressed in the driving period $T$, $\Gamma_e^\text{AC} {\propto} \exp(-1/JT)$. Note that experimentally observing this exponential extension of the lifetime is challenging due to technical limitations (see SI Sec.\ref{supmat:sec:summary_engineering}).  

Red data in \zfr{fig1}C shows the PDTC under an AC-field with $\Bac{=}82.4\,\mu$T and $\fac{=}\fres{=}330.023$\,Hz. The $1/e$ lifetime is extended ${>}50$-fold to $T_2^\prime{=}4.51\,$s, and corresponding to over $M{=}2900$ $\xhat{\leftrightarrow}-\xhat$ spin-flips. This manifests also in the larger signal in the zoomed view in \zfr{fig1}C(i). Notably, lifetime extension is not limited to the specific case $\gamma_y{=}\pi$ but applies throughout the entire stability regime of DTC order~(see SI~\zsr{dome} and \zfr{dome-fig}).

Spin evolution remains observable far beyond the value naively suggested by the $1/e$ time, as shown by the phase signal $\phi$ remaining coherent for several seconds (see Fig.~\ref{fig:fig1}C(ii)). The $\xhat$ and $-\xhat$ rails correspond to phase values \(\phi {=} 0,\pi\), with each point tracking $\phi$ after every $\theta$-pulse  (over 500k in total). Heating of the conventional PDTC (blue points) towards the infinite-temperature state, $\rho_{\mathcal{T}{=}\infty}{=}\id$, is evident as $\phi$ disperses across the \([- \pi, \pi]\) phase space within ${\app}2\,$s.
In contrast, under the AC field (red data in Fig.~\ref{fig:fig1}C(ii)), the PDTC signal remains stable for over $20\,$s and $544,000$ pulses (spin-lock plus $\gamma$-kick). We also note that with the AC field applied, micromotion among the interpulse spacings within a single period of the DTC sequence causes an apparent splitting of the signal into multiple strands. As described in Ref.~\cite{sahin22_trajectory}, this is due to the distinct prethermal plateaus corresponding to each stroboscopic frame within the period of the Floquet cycle. Similar micromotion is also evident in the signal shown in Fig.~\ref{fig:phase-DTC}.

% ————————————————————————————————————————————
\subsection{Robust, high-resolution, AC magnetic field sensing}

The lifetime enhancement as in \zfr{fig1}C(i), also yields a change in measured signal at every fixed time $t$ compared to the case without additional AC field, and hence enables a means to \I{sense} the AC field.
We now consider how this extension applies to the AC field characteristics (\zfr{fig2}A(i)), $\Bac(t) = \Bac \sin(2\pi \fac t + \Phi_\mathrm{AC})$, i.e., its (i) phase $\Phi_\mathrm{AC}$ (ii) amplitude $\Bac$, and (iii) frequency $\fac$.

To quantify the signal enhancement, we devise a fidelity metric that remains accurate even when the signal approaches the noise floor,
\begin{equation}
\label{eq:fidelity_metric}
    F = \frac{1}{N'}\sum_{i = 1}^{N'} \expec{I^x(t_i)}  P(t_i)\,,
\end{equation}
where \(P(t)\) represents the ideal DTC toggling response, alternating between ${\pm}1$ as spins flip between the $\pm\xhat$ axes, and the normalized sum is carried out over all $N'$ time points $t_{i}\in[t_{1}, t_{N'}]$; $F$ yields the largest value when the DTC oscillations are strongest and most stable.
Formally, $F$ corresponds to a weighted summation over the Fourier harmonics $\ell \fres$ ($\ell \in \mathbb{N}_{>0}$); when approaching the noise floor, it is more robust than the standard approach of estimating the PDTC response from only the period-doubling (i.e., $\fres$) component. Lacking an analytical model for the DTC decay, the fidelity metric provides a simple, profile-agnostic measure of stability by integrating the observed magnetization. A more detailed understanding
of the form of the DTC decay and its response to external magnetic field could provide a more optimized metric to improve the sensitivity of our approach.

Using this metric, \zfr{fig2}A(ii) examines the impact of the AC field phase $\xPh_\text{AC}$ on resonance $\fac{=}\fres$ and $\gamma_y{=}\pi$ (blue points). Maximum lifetime extension occurs at \(\xPh_\text{AC}{=}\pi/2\), where the AC field peaks align with the center of the $\gamma_y$ pulses, as predicted theoretically~(SI Sec.\ref{supmat:sec:thy_properties}). When the AC field nodes coincide with the $\gamma_y$ pulses, there is a minimal effect on the PDTC lifetime. Optimal sensing therefore occurs when \(\xPh_\text{AC}{=}\pi/2\). We observe a slight phase shift in the response from the expected maximum at $\Phi_{\text{AC}} = \pi/2$, due to unaccounted pulse transients when setting the AC field phase based on the pulse length in \zfr{fig2}A(i). This apparent shift arises because the actual pulse applied to the probe is slightly longer than the one generated by the Arbitrary Waveform Transceiver (AWT) \cite{mehring1972phase}.

Additionally, the blue points in \zfr{fig2}A(ii) show the response to slight deviations from the small point, here $\gamma_y{=}0.98\pi$. The data confirms the robustness of the PDTC order. In SI Fig.~\ref{fig:dome-fig}, we display the entire experimentally mapped PDTC phase diagram for all $\gamma_y$ values, demonstrating a large stable region around $\gamma_y{=}\pi$, independent of $t$.

\zfr{fig2}B studies the effect of the AC field strength $\Bac$, set at resonance with \(\xPh_\text{AC} {=} \pi/2\).
The fidelity profile shows a gradual increase, followed by a linear rise, and eventually plateaus at higher field strengths when \(\Bac\) becomes comparable to the Rabi field of the \(\theta_x\) Floquet pulses.
The right vertical axis shows the corresponding $T_2'$ lifetimes; the maximum extension, corresponding to $T_2'{=}21.3$s is ${>}3000$-fold.
Data here is for $N{=}4$; SI \zsr{extension_N} discusses the extensions obtained as a function of $N$.
\zfr{fig2}B illustrates time profiles of $\expec{I^x}$ for three cases: (i) no field \(\Bac{=}0\), (ii) $\Bac{=}8.24\mu$T, and (iii) $\Bac{=}82.4\mu$T. \zfr{fig2}B indicates this can be used for sensing at appropriately chosen bias points.
The sensitivity is determined by comparing the response of the fidelity $F$ to perturbation in AC field amplitude $B_\text{AC}$ (as shown in Fig.~\ref{fig:fig2}B) with the fluctuations in $F$ obtained for multiple initializations at fixed $B_\text{AC}$ (see Methods and SI Sec.~\ref{supmat:sec:sensitivity}). We obtain a sensitivity of $880\,\mathrm{pT}/\sqrt{\mathrm{Hz}}$ with an optimum bias field of $B_\text{AC}$=415~nT.

A distinguishing feature of the AC-field mediated lifetime extension is its strongly resonant nature. \zfr{fig3}A examines the fidelity $F$ across a range of AC field frequencies $\fac$ under identical conditions. Off-resonant frequencies have negligible impact on the DTC lifetime, matching the bare PDTC ($\Bac{=}0$, $F{\app }0$). In contrast, a significant lifetime increase is observed on resonance $\fres$, as shown in \zfr{fig3}A. A zoomed-in view in \zfr{fig3}A(i) reveals a narrow linewidth of $\xD f{\app}70\,$mHz set by the maximum integration time ($t_{N'}$ in Eq.(\ref{eq:fidelity_metric}))), positively correlated with the inverse of the PDTC lifetime $(T_2')^{-1}$. We also note a weak additional response at sub-harmonics, especially $\fres/2$, at large $\Bac$. This is discussed in SI \zsr{harmonics} (see SI \zfr{harmonics}).

The sharp $\fres$ response is further clarified using the representative points marked in \zfr{fig3}A(i). The corresponding time domain PDTC profiles are shown in \zfr{fig3}A(ii).
Far off resonance~\I{(a)}, the dynamics remain unaffected by the AC field. Exactly on resonance~\I{(b)}, the significant lifetime increase is observed. Slightly off-resonance~\I{(c-d)}, a distinctive beating in the fidelity $F$ appears, reflecting the frequency offset \(\delta f {=} \fac {-} \fres\). The integration of this beating pattern over time leads to the $(t_N')^{-1}$ linewidth. In sensing applications, this can enable precise reconstruction of unknown signals within the narrow resonance band $\xD f$ via a Fourier transform of the DTC temporal dynamics. The sensor bandwidth itself is determined by the shortest possible pulse lengths, and could span the 0.5-50kHz range (SI \zfr{sensor-comparison}). 
 
The dynamics of the tracked phase $\xph$ corresponding to \zfr{fig3}A(ii) is presented in SI \zsr{phase}. The data (\zfr{phase-DTC}) reveals intricate micromotion dynamics and demonstrates the ability to measure it for periods well beyond the $1/e$ lifetimes, exceeding 60,000 $\xg$-kicks, with high clarity.

We emphasize that the lifetime-limited linewidth is a feature of DTC-based sensing, distinguishing it from other methods such as magnetometry using spin-locked prethermal states introduced in~\cite{Sahin22}. The grey data points in \zfr{fig3}A illustrate the response of the spin-lock sensing scheme over the same frequency range where we measured the DTC's response. Unlike DTC-based sensing, the spin-lock sensing scheme can detect multiple frequencies without modifying pulse sequence parameters, but as indicated by the data, it exhibits monotonically increasing response with frequency, lacking frequency selectivity. Even on resonance, the linewidth of the spin-lock sensing scheme can extend to several hundred Hz -- at least four orders of magnitude broader -- primarily dominated by interspin couplings, and largely independent of $T_{2}'$. A detailed analysis of the spin-lock sensing data is provided in \zsr{spin_lock_sensing}. 
In contrast, the narrow linewidth of the DTC sensing scheme enables \I{tuning} into specific fields that meet the resonance condition, effectively rejecting non-resonant fields (see SI \zfr{noise-rej}). More broadly, when compared to conventional quantum sensors based on electronic spins~\cite{Wolf15}, the two-tone cavity-interrogated nuclear PDTC allows single-shot, quasi-continuous sensing for ${>}5T_2' ({>}100$s) (see \zfr{fig1}C(ii) and \zfr{phase-DTC}) without sensor re-initialization, with the resonant lifetime extension enhancing the sensor precision.

Another consequence of PDTC order is that the narrow sensing linewidth remains highly robust to pulse errors $\epsilon$ in the $\xg_y$-pulses away from $\gamma_y {=} \pi$. This is shown in \zfr{fig3}B with $\gamma_y$ denoted by the colorbar. The sensor linewidth (zoomed in \zfr{fig3}B(i)) remains unaffected by these errors. 
Additionally, the system exhibits a remarkable tolerance to on-site disorder~(SI Sec.~\ref{supmat:sec:noise}), and fluctuations in the spin-lock $\theta_x$-drive, evidenced in the capacity of reliably applying ${>}10^6$ $\theta_x$-pulses even with realistic imperfections (due to Rabi frequency hetereogenity) in these experiments. 

The two-tone PDTC discussed so far (\zfr{fig1}B(i)) hosts a single resonance frequency $\fres$, tunable via the sequence parameter $T$. However, it is possible to expand the number of resonance frequencies and adjust the DTC sensing spectrum by modifying the PDTC sequence. For instance, \zfr{fig3}C(i) introduces a three-tone PDTC, establishing two resonance conditions at $\fres^{(1)}$ and $\fres^{(2)}$, achieved through two different interleaved periods for the $\xg_y$-pulses, interspersed with spin-locking $\theta_x$ pulses. The experimental response in \zfr{fig3}C shows two distinct frequencies separated by ${\sim}42\,$Hz. We observe asymmetric spectra with a stronger x-component when the period $T$ is decreased, leading to a more frequent overlap between the AC field anti-nodes and the $\gamma_y$ pulses. This increased overlap enhances the x-component response (see SI~\ref{supmat:sec:two_tone}). The two-frequency linewidths, around $5\,$Hz, remain significantly narrower than the sensor's linewidth without DTC order ($\sim$223Hz)\cite{Sahin22}, although single-frequency linewidths~(\zfr{fig3}A) are narrower due to longer $T_2^\prime$ lifetimes in the two-tone case.

% ————————————————————————————————————————————
\subsection{AC field mediated lifetime extension for single-tone PDTC}
\label{sec:single_tone}

The lifetime enhancement from AC field-mediated finite energy density applies broadly to all $U(1)$-PDTCs, not just the two-tone PDTC. To demonstrate this, we consider a conventional single-tone DTC that alternates between the $+\zhat$ and $-\zhat$ states on the Bloch sphere. This approach is widely used across platforms, including superconducting qubits~\cite{mi2022time,frey2022realization}, cold atoms~\cite{Giergiel20}, and NMR~\cite{rovny2018observation, pal2018temporal,stasiuk2023_U1enhanced}. Unlike the two-tone DTC, which enables non-destructive inductive readout in the $\xhat$-$\yhat$ plane to monitor decay dynamics in a single shot, the single-tone DTC requires restarting the experiment for each data point.

The sequence is shown in \zfr{fig4}A(i), and consists of \(M\) spin-flip $\gamma_y$-pulses along $\yhat$ (schematically descibed in \zfr{fig4}A(ii)). We utilize the exact $U(1)$ symmetry of the dipole-dipole Hamiltonian $\Hdd$, which conserves $\zhat$-magnetization, $\left[\Hdd, I^z \right]{=}0$. For efficient readout, spins are tipped onto the $\xy$ plane and spin-locked using a train of \(\theta_x\)-pulses along $\xhat$. Unlike the two-tone case in \zfr{fig1}, the data here is collected point-by-point for different values of \(M\).

Results are shown in \zfr{fig4}B. Without an AC field (black points), we observe robust period-doubling dynamics with a $1/e$ decay time \(T_1' {\approx} 0.01\,\)s and $M{=}40$ $\zhat{\leftrightarrow}-\zhat$ flips. With a resonant AC field of $\Bac{=}40\,\mu$T (green points), the lifetime is markedly prolonged, extended by more than three orders of magnitude to $M{=}70,000$ and \(T_1' {=} 20\,\)s. The phase response to $\Pac$ is opposite to that in \zfr{fig2}A(ii) as the AC field direction aligns with the PDTC oscillation axis; here being maximal near $\Pac{=}0$ (see Fig.~\ref{supmat:fig:single_frequency_phase}). While the quasi-continuous measurement from the two-tone drive is more suitable for sensing applications, the data in \zfr{fig4}B demonstrates that this lifetime extension mechanism applies broadly to $U(1)$-PDTCs.

\section{Discussion}

This work introduces several novel features. The key conceptual result is the \I{resonant} extension of PDTC lifetime via an AC field, exponentially suppressing the heating rate relative to the driving period $T$ (SI Sec.\ref{supmat:sec:engineering}). As shown in \zfr{fig2}B, we extend the PDTC $1/e$ lifetime to $T_2^\prime{>}20\,$s and $M{=}40,000$ spin-flipping Floquet cycles. Compared to previous work~\cite{choi2017observation,zhang2017observation,randall2021many,mi2022time,kyprianidis2021observation,stasiuk2023_U1enhanced,DTC_Beatrez2022}, this sets a new record for both parameters, representing over two orders of magnitude improvement in the total number of DTC spin-flips. SI Table~\ref{tab:lifetime_comparison} provides a detailed comparison, highlighting the significantly increased DTC lifetime in our case, despite similar interaction strengths and pulsing rates to previous works.

The methodology introduced here is not limited to nuclear spins and can be applied to a wide range of quantum and classical systems. We anticipate immediate applications to cavity-interrogated NV centers~\cite{Eisenach21} for quantum sensing in the 1MHz-1GHz range~\cite{Carmiggelt23}.  The underlying principles themselves are also applicable to superconducting qubits~\cite{mi2022time,frey2022realization}, cold atoms~\cite{Giergiel20}, and ions~\cite{kyprianidis2021observation}.

Crucially, sensing based on many-body DTC order naturally tolerates strong interspin couplings, leading to a lifetime-limited linewidth ${\sim }(T_2^\prime)$ rather than being dominated by couplings ${\sim}J^{-1}$. This enables sensing at high sensor densities, extending beyond the conventional regime of dilute, non-interacting sensors~\cite{Bauch20}; all else being identical, this can yield significant sensitivity improvements~\cite{Barry20}. Moreover, sensing here inherits the robustness of prethermal DTC order against pulse sequence errors and sample inhomogeneities; we demonstrate this in SI Sec.~\ref{supmat:sec:noise}, showing the resilience of AC sensing to strong on-site disorder. Finally, this approach does not require the preparation of strongly entangled states or fine-tuning to a critical point.  All these factors indicate broad applicability to a wide range of systems. 
 
\paragraph*{Acknowledgements.}
This work was funded by DNN NNSA (FY24-LB-PD3Ta-P38), ONR (N00014-20-1-2806), CIFAR Azrieli Foundation (GS23-013), and European Union (ERC, QuSimCtrl, 101113633). RJS acknowledges the Hoffman fellowship and NNSA NextGen program. Numerical simulations were performed on the MPIPKS HPC cluster.
Views and opinions expressed are those of the authors only and do not necessarily reflect those of the European Union or the European Research Council Executive Agency. Neither the European Union nor the granting authority can be held responsible for them.

\paragraph*{Contributions.}
LJIM and PMS conceived the idea of using two-frequency DTC as a magnetic field sensor. 
RJS conceived and realized the idea of applying an AC field to a single-axis PDTC. 
LJIM conceived and realized the idea of using a three-tone drive for two-frequency magnetic field sensing.
LJIM, RJS, and ED implemented the experiment. 
LJIM and RJS discovered the first experimental response of the two-frequency DTC. 
RJS and LJIM devised the fidelity metric. PMS and LJIM worked out the theory details. 
RJS and LJIM collected and analyzed the experimental data. PMS implemented and ran the simulations. 
MB and AA supervised the theory and experimental work.
LJIM, RJS, PMS, ZRZ, and AA made the figures in the manuscript.
All authors contributed to the manuscript.

%\paragraph*{Competing Interest.}

%\paragraph*{Data availability.}

%\paragraph*{Code availability.}

%%%%%%%%%%%%%%%%%%%%%%%%%%%%%%%%%%%%%%%%%%%%%%%%%%%%%%%%%%%%%%%%%%%%%%%%%%%
%                        Methods
%%%%%%%%%%%%%%%%%%%%%%%%%%%%%%%%%%%%%%%%%%%%%%%%%%%%%%%%%%%%%%%%%%%%%%%%%%%
%%%%%%%%%%%%%%%%%%%%%%%%%%%%%%%%%%%%%%%%%%%%%%%%%%%%%%%%%%%%%%%%%%%%%%%%%%%

\beginmethods

\section{Methods}
\label{sec:methods}
\T{Materials -- } The sample used in this work is a single-crystal diamond measuring $3.4{\zt} 3.2 {\zt} 2.1$ mm, containing a natural abundance of $\Cs$ nuclei and NV centers at ${\sim}$1 ppm concentration. This same sample has been characterized in prior studies~\cite{Beatrez21_90s,DTC_Beatrez2022}, allowing for direct comparisons to the lifetime extensions observed. The sample is oriented parallel to the $\Bpol$ magnetic field, ensuring simultaneous hyperpolarization of the four NV center axes, and experiments are carried out underwater to provide more uniform illumination and aid in thermal management~\cite{Sarkar22}. 

\T{Experimental setup -- }
Instrumentation for hyperpolarization and $\Cs$ readout follows previous works, with detailed descriptions available in those studies~\cite{Beatrez21_90s,DTC_Beatrez2022,Sahin22,Beatrez2023electron,moon24_rondeau}. Data here is taken at room temperature and $B_0{=}7$T. Polarization is carried out at a low field center (\(B_{\text{pol}} {\approx} 38\) mT) located below the magnet, driven by optically excited NV centers and chirped microwave excitation. The polarization mechanism involves successive Landau-Zener anti-crossings in the rotating frame~\cite{Zangara19,Pillai23}. Sample shuttling to the $B_0$ high-field occurs in under 1s, with an NMR saddle coil used for inductive readout of the $\Cs$ precession signal.

$\Cs$ interrogation is performed using a home-built NMR spectrometer based on a high-speed arbitrary waveform transceiver (Proteus P9484M)~\cite{moon24_rondeau}. The AWT device generates NMR pulses and digitizes the nuclear precession directly at the Larmor frequency, eliminating insertion losses typically encountered with intermediate frequencies. The device's high memory capacity (16GB) and large sampling rate (up to 2.7GS/s) enable continuous interrogation of $\Cs$ spin precession in windows between pulses. In typical experiments, we apply 0.2-1M pulses with a readout window of ${\app}13.6\mu$s. The entire Larmor precession can be sampled every 0.74 ns and mixed with an on-board numerically controlled oscillator (NCO) at the Larmor frequency, allowing us to track both the amplitude and phase of the spins directly in the rotating frame~\cite{sahin22_trajectory,moon24_rondeau}.

The receive chain amplifies the signal using low-noise preamplifiers (Advanced Receiver Research P75VDG and Pasternack PE15A1011), while NMR pulse generation is via a Herley TWT amplifier through a Techmag transcoupler. For AC magnetometry, the spins are exposed to a weak magnetic field applied via a secondary coil, with the field parallel to $\zhat$ and positioned within the NMR probe. The field is applied via a Tektronix source, amplified by a Techron 7224 amplifier, and the field strength is calibrated through the voltage drop across a high-power 250W 4 Ohm resistor.

\T{Classification of temporal order -- }
To place our work in the broader context of temporal order in closed systems, let us recall the different ways temporal order can be realized.
In closed periodically driven systems, as of present date, three mechanisms are known to realize temporal order, namely (i) many-body localized discrete time crystals, (ii) prethermal discrete time crystals, and (iii) $U(1)$ prethermal DTCs. 
The many-body localized DTCs require stable many-body localization~\cite{Nandkishore15_MBLreview,Abanin19_MBLreview} which may only exist in $1$D short-range quantum systems in the presence of strong disorder~\cite{Luitz2017_BathMBL}; under these conditions, the emerging spatiotemporal eigenstate order~\cite{khemani2016phase,Else2016_FTC} is entirely robust out to infinite times~\cite{Keyserlingk2016_absolutestableDTC}.
In contrast, the prethermal discrete time crystals require the existence of (pre)-thermal order, i.e., a low-temperature symmetry breaking state, that is by the Mermin-Wagner theorem only possible in two and higher dimensions. The resulting prethermal spatio-temporal order has a finite lifetime determined by Floquet-heating, and thus, exponentially suppressed in the driving period for short-range interacting systems~\cite{Abanin2015_Heating,Mori2016_Heating}.
Finally, $U(1)$ prethermal DTCs only require a (quasi-)conserved $U(1)$ symmetry and an initial state that breaks this symmetry, irrespective of the effective temperature of this state. 
However, melting of the prethermal temporal order is dominated by symmetry-breaking perturbations, which in case of an emergent symmetry only lead to power-law suppression of heating in the driving frequency.

Our work focuses on the last class, the $U(1)$ prethermal DTCs, demonstrating that the lifetime of temporal order can be exponentially enhanced by coupling the system to the order parameter.
Thus, the AC-enriched $U(1)$ PDTCs effectively mimic the behaviour of prethermal DTCs in terms of lifetime. We emphasize that adding the AC field to the $U(1)$ prethermal DTC is not sufficient to realize a thermally ordered state required for the PDTC.

\T{Sensitivity calculation --}
The sensitivity to AC magnetic fields based on the described effect of DTC extension is calculated by 
\begin{equation}
\label{eq:sensitivity}
    \R{sensitivity} = \frac{\sigma_F}{\partial F /\partial B_{\text{AC}}}\sqrt{t_{int}}\,,
\end{equation}
i.e., the response of the signal metric (here the fidelity $F$) to perturbation in the field strength $B_{\text{AC}}$ is compared to the noise $\sigma_F$ in the metric at fixed field strength, normalized for the noise reduction achievable by integrating the signal over the measurement time $t_{int}$. Sensitivity is determined via a scan of the amplitude $B_{\text{AC}}$. For each chosen amplitude, the DTC experiment is repeated several times. The fluctuation $\sigma_F$ in the fidelity metric is determined by the standard deviation amongst trials at a fixed amplitude $B_{\text{AC}}$. The mean fidelity for trials at each amplitude is used in calculating the centered difference between adjacent amplitude points $B_{\text{AC}}$ to estimate the response $\partial F /\partial B_{\text{AC}}$. As shown in Fig.~\ref{fig:fig2}B, the amplitude dependence of the DTC coherence extension is nonlinear; the response $\partial F /\partial B_{\text{AC}}$ increases with $B_{\text{AC}}$ up to $\approx$8~$\mu$T. This could suggest that the sensor should be operated with a bias AC field near 8~$\mu$T for best performance. However, the fluctuations in fidelity are also found to grow with $B_{\text{AC}}$, even more steeply than the response $\partial F /\partial B_{\text{AC}}$. Thus, the best sensitivity is actually obtained for a small bias of 415~nT. Fig.~\ref{fig:sensitivity} shows the calculated sensitivity as a function of bias. The cause of the sharp rise in $\sigma_F$ with applied field is unclear; this could be due to an increased instability of the AC source itself, or due to increased susceptibility of the DTC to drifts in the instrument as a result of the bias field. The sensitivity and opportunities to improve it are further discussed in Sec.~\ref{supmat:sec:sensitivity}.

\T{Floquet engineering finite energy density --}
As we sketch below, in both single-tone and two-tone DTC the AC-induced signal enhancement is the result of coupling the system to the DTC order parameter via the added AC field.
The complete derivation can be found in the SI.

\I{Single-tone DTC.} 
For the single-tone DTC, the order parameter corresponds to the $\zhat$-magnetization~($\expval{I^z}$) flipping sign every period $T$, i.e., $\mathcal{O}_\mathrm{1DTC}(\ell T)=\expval{(-1)^\ell I^z}$, therefore oscillating with a period $2T$. 
Thus, naturally an AC-field in the $\zhat$-directions couples to the order parameter, such that, the system every full DTC period~($2\ell T$) is effectively described by $H_\mathrm{AC, eff}= H_\mathrm{dd}+ B_\mathrm{eff} I^z$, with the effective field $B_\mathrm{eff}$ proportional to the AC amplitude, $B_\mathrm{eff}\propto \Bac$.

\I{Two-tone DTC.}
For the two-tone DTC a key difference is that the order parameter oscillates in between $\xhat$ and $-\xhat$, $\mathcal{O}_\mathrm{2DTC}(\ell T)=\expval{(-1)^\ell I^x}$, orthogonal to the AC field.
However, crucially the $\yhat$ pulses implementing the DTC sequence are of finite time duration $\tau_y$, i.e., a finite AC field is present also during the application of the $\yhat$-pulses.
Indeed, one can show [see SI~\ref{supmat:sec:two_tone}] that for the $\gamma_y=\pi$ pulses, applying the $I^y$ field and $I^z$ simultaneously corresponds to applying an $I^x$ and an $I^y$ field separately, i.e., 
\begin{equation*}
    e^{-i \left(\gamma_y I^y + (-1)^n B_{n,z} \tau_y I^z \right)} \approx e^{-i\gamma_y I^y} e^{-i \frac{(-1)^n B_{n,z} \tau_y}{\gamma_y}I^x}\,,
\end{equation*}
up to an error scaling as $O\left((B_z \tau_y/\gamma_y)^2\right)$, where $n$ labels $\yhat$-pulse and $B_{z,n} = \abs{\int^{\tau_y} \Bac(t) \mathrm{d}t/\tau_y}$ is the amplitude of the AC field accumulated during the $n$'th $\yhat$ pulse.
So indeed, the $\zhat$ AC field effectively induces an $\xhat$ AC field.
Notably, this induced AC field is strongest if the minima and maxima of the AC field align with the $\yhat$-pulses, in agreement with experimental results in Fig~\ref{fig:fig2}; for a derivation of the other effects in Figs.~\ref{fig:fig2} and Fig.~\ref{fig:fig3}, see SI~\ref{supmat:sec:thy_properties}.
Thus, similar to the single-tone case, the system in presence of AC field is effectively described by $H_\mathrm{AC, eff}= H_\mathrm{eff}+ B_\mathrm{eff} I^x$ after every full DTC-period~($2\ell T$), for the effective interacting Hamiltonian $H_\mathrm{eff}\approx \sum_{k<l}J_{kl}\pqty{I^x_k I^x_l - \mathbf{I}_k \cdot \mathbf{I}_l}$ and effective field $B_\mathrm{eff}\propto \Bac$; for the derivation see SI Sec.~\ref{supmat:sec:two_tone}.

\I{Lifetime enhancement via eigenstate thermalization hypothesis.}
In both cases the system over a full DTC cycle~($2T$) is effectively described by an interaction part and an emergent static field with strength proportional to the AC field strength, $H_\mathrm{AC, eff}=H_\mathrm{int} + B_\mathrm{eff} I^\alpha$ where $\alpha=z,x$ for single- and two-tone drive, respectively.
Note that, for small imperfections in the drive $\gamma_y\neq\pi$ the `interaction' terms $H_\mathrm{int}$ will break the $U(1)$-conservation law, $\comm{H_\mathrm{int}}{I^\alpha}{\neq}0$.
Therefore, the only conserved quantity is the quasi-conserved energy, $\expval{H_\mathrm{AC, eff}}$, in the prethermal plateau. 
Thus, the eigenstate thermalization hypothesis~\cite{Srednicki1995,rigol2008thermalization,Deutsch2018,dalessio2016quantum} predicts that a generic system starting in an initial state, $\ket{\psi_0}$, will relax at `late' times to a state, $\ket{\psi_\infty}$ that is indistinguishable from a thermal state, $\rho_\mathcal{T}\propto e^{-H_\mathrm{AC, eff}/\mathcal{T}}$, with temperature $\mathcal{T}$ determined by energy conservation. 
This means, for spatially local observables $A$, like the energy or magnetization, that the expectation value in the late time state is equal to the thermal expectation value, $\expval{A}_{\psi_\infty}=\expval{A}_{\rho_\mathcal{T}}$. 
The same holds for the initial mixed state, $\rho_0$.

Importantly, in the absence of an AC field the energy of the initial state, $\rho_0\sim \id+\mu I^\alpha$, for both single- and two-tone DTC vanishes, $\expval{H_\mathrm{int}}_{\rho_0}=0$, thus leading to (pre)thermalization to a featureless infinite temperature state, $\rho_{\mathcal{T}{=}\infty}\propto \id$. 
By contrast, in the presence of an AC field the initial energy becomes finite, $\expval{H_\mathrm{eff, AC}}_{\rho_0}\propto \mu \Bac$, thus the system prethermalizes to a finite temperature state $\rho_{\mathcal{T}\neq\infty}\propto e^{-H_\mathrm{AC, eff}/\mathcal{T}}$, with finite magnetization $\expval{I^\alpha}_{\rho_{\mathcal{T}\neq\infty}}\neq0$.
This prethermal value persists until the ultimate melting of the prethermal plateau which is exponentially suppressed in the period $T$, $\Gamma_e^\mathrm{AC}\propto \exp(-c/JT)$ for some constant $c$.
Thus, the lifetime can be exponentially enhanced by introducing the additional AC field~(see SI Sec.\ref{supmat:sec:engineering} for details).

\T{Overview of requirements for AC-induced signal enhancement --}
While the above analysis focused on dipole-coupled nuclear spins in diamond, the mechanism applies more generally.
In a nutshell, the key ingredients are:
(i) (emergent) symmetry-protected period doubling response;
(ii) the system should (pre-)thermalize in agreement with ETH,
(iii) high-temperature initial state such that lifetime is limited by symmetry-breaking terms;
and 
(iv) ability to (effectively) couple the system to the DTC order parameter, e.g., via Floquet engineering.
Note that, since the DTC order parameter itself oscillates in time with period $2T$, inducing a coupling to the order parameter necessarily requires adding an additional time-varying (AC) field.
Then, by adding this time-varying field one can exponentially extend the lifetime of the PDTC order using the procedure presented above.

%%%%%%%%%%%%%%%%%%%%%%%%%%%%%%%%%%%%%%%%%%%%%%%%%%%%%%%%%%%%%%%%%%%%%%%%%%%
%                    Bibliography
%%%%%%%%%%%%%%%%%%%%%%%%%%%%%%%%%%%%%%%%%%%%%%%%%%%%%%%%%%%%%%%%%%%%%%%%%%%
%%%%%%%%%%%%%%%%%%%%%%%%%%%%%%%%%%%%%%%%%%%%%%%%%%%%%%%%%%%%%%%%%%%%%%%%%%%

%\clearpage
\bibliography{bilbio_dtcsensing}
\vspace{-1mm}

%%%%%%%%%%%%%%%%%%%%%%%%%%%%%%%%%%%%%%%%%%%%%%%%%%%%%%%%%%%%%%%%%%%%%%%%%%%
%                        SI
%%%%%%%%%%%%%%%%%%%%%%%%%%%%%%%%%%%%%%%%%%%%%%%%%%%%%%%%%%%%%%%%%%%%%%%%%%%
%%%%%%%%%%%%%%%%%%%%%%%%%%%%%%%%%%%%%%%%%%%%%%%%%%%%%%%%%%%%%%%%%%%%%%%%%%%

\clearpage
\onecolumngrid
%\begin{widetext}
\begin{center}
\textbf{\large{\textit{Supplemental Information:} \\ \smallskip Sensing with discrete time crystals}} \\\smallskip
\end{center}

\twocolumngrid

\beginsupplement
\tableofcontents

%%%%%%%%%%%%%%%%%%%%%%%%%%%%%%%
%            Text
%%%%%%%%%%%%%%%%%%%%%%%%%%%%%%%
%%%%%%%%%%%%%%%%%%%%%%%%%%%%%%%

\begin{table*}
    \centering
\begin{tabular}{|c||c|c|c|c|c|c|}
    \hline
    & Platform & DTC type & Mean interactions & Period & Spin lifetime & Floquet cycles 
    \\
    \hline
    \hline
    Mi et.al. 2022 \cite{mi2022time} & Superconducting qubits & MBL & $-$ & $-$ & $\approx 6.4\,\mu$s & $\approx50{-}100$ 
    \\
    \hline
    Zhang et.al. 2017 \cite{zhang2017observation}   & Trapped Ions  & prethermal    &   $\approx0.04{-}0.25\,$kHz  &   $\approx 74\,\mu$s   &  $\approx7\,$ms   &   $\approx100$
    \\
    \hline
    Kyprianidis et.al. 2021 \cite{kyprianidis2021observation} & Trapped Ions & prethermal & $\approx0.33\,$kHz & $\approx280{-}500\,\mu$s & $\approx12{-}19\,$ms & $\approx50{-}100$ 
    \\
    \hline
    Choi et.al. 2017 \cite{choi2017observation} & NV centers & $U(1)$ & $\approx105\,$kHz & $\approx92{-}998\,$ns & $\approx60\,\mu$s & $\approx50$ 
    \\
    \hline
    Rovny et.al. 2018 \cite{rovny2018observation}   &  ADP ${}^{31}$P    &   $U(1)$  & $\approx 508\,$Hz  &   $10\,\mu\mathrm{s}{-}1\,\mathrm{s}$ & $-$  & $\approx 50$
    \\
    \hline
    Pal et.al. 2018 \cite{pal2018temporal}  & Acetonitrile, TMP, TTSS ${}^{1}$H   &   $U(1)$  & $\approx2.5{-}136\,$Hz    & $-$  & $-$ &   $\approx 20{-}60$
    \\
    \hline 
    Stasiuk et.al. 2023 \cite{stasiuk2023_U1enhanced} & Fluorapatite ${}^{19}$F & $U(1)$ & $\approx0-5.2\,$kHz & $\approx120\,\mu$s & $\approx9.3\,$ms & $\approx 80$ 
    \\
    \hline
    Randall et.al. 2021 \cite{randall2021many} & Diamond \C & prethermal/MBL & $\approx6.7\,$Hz & $\approx 5\,$ms & $\approx2.5\,$s & $\approx500$ 
    \\
    \hline
    Beatrez et.al. 2022 \cite{DTC_Beatrez2022}  & Diamond \C    &   $U(1)$  &   $\approx 0.66\,$kHz & $\approx 5{-}50\,$ms  &   $\approx 14\,$s &   $\approx 500$
    \\
    \hline
    \cellcolor{blue!10} This work & \cellcolor{blue!10} Diamond \C & \cellcolor{blue!10} $U(1)$ & \cellcolor{blue!10} $\approx0.66\,$kHz & \cellcolor{blue!10} $\approx 1.5\,$ms & \cellcolor{blue!10} $\approx21.29\,$s & \cellcolor{blue!10} $\approx44,200$ 
    \\
    \hline
\end{tabular}
    \caption{
        \textbf{Comparison of DTC lifetimes in other experiments and platforms.}
        We compare the lifetimes of other DTC experiments in the literature with the lifetimes achieved in this work.
    }
    \label{tab:lifetime_comparison}
\end{table*}

\begin{figure*}[t]
    \centering
    \includegraphics[width=0.99\textwidth]{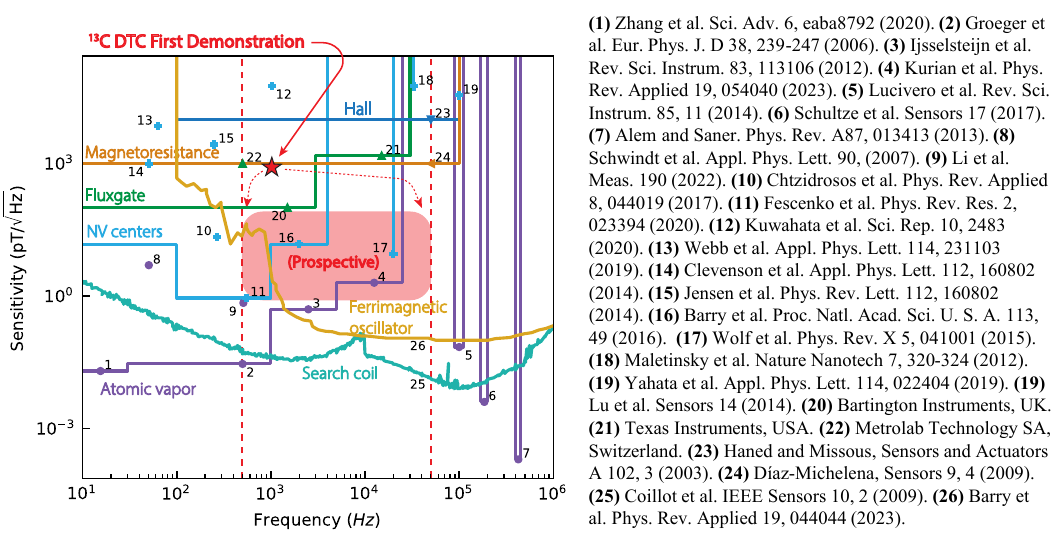}
    \caption{
    \T{Comparison of magnetometer sensitivity} as a function of frequency. Red dashed lines indicate the frequency regime in which the sensor presented here is expected to operate. The red shaded area indicates prospected sensitivity with improvements described in SI Sec.~\ref{supmat:sec:sensor_comparison}. Contours indicate the best sensitivity the authors have identified at a given frequency for a given type of magnetometer. For the NV magnetometers labeled 17-19, no bandwidth was identified, hence only vertical contours are ascribed to these. The sensor concept demonstrated in this paper is best suited for sensing of $\sim$kHz frequencies, a range at which other existing sensor technologies lose substantial sensitivity. Comparison to the most sensitive magnetometers operating at similar frequencies is made in SI Sec.~\ref{supmat:sec:sensor_comparison}.
    Note that superconducting quantum interference devices (SQUIDs) and certain atomic magnetometers are not included here. While these can offer excellent sensitivity, operational constraints limit their applications. Namely, some atomic magnetometers can only function in a shielded, low-field environment, and SQUIDs require cryogenic cooling.}
\label{fig:sensor-comparison}
\end{figure*}

\begin{figure*}[t]
    \centering
    \includegraphics[width=0.99\textwidth]{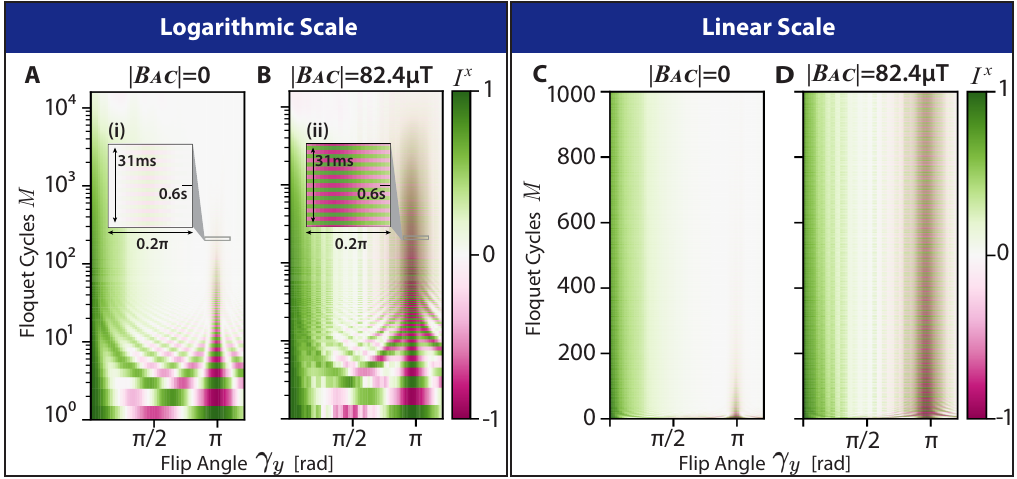}
    \caption{\T{Phase diagram mapping of PDTC revealing origin of lifetime extension}. Panels show mapped PDTC phase diagram for 100 \(\gamma_y\)-flip angles (vertical slices) in range $[0, 1.3\pi]$. In this experiment, pulse angle is assumed to be proportional to pulse length, with $\gamma_y=\pi$ corresponding to a pulse length of 103.5~$\mu$s. The y-axis represents number of Floquet cycles (\(\gamma_y\) kicks), while right axis shows total absolute time. Color bars represent $\expec{I^x}$ signal intensity.
    (A) Data on logarithmic scale without AC field. There is a characteristic stability dome around PDTC stable points at $\xg_y{=}\{0,\pi\}$, as described in Ref.~\cite{DTC_Beatrez2022}. Without AC field, the DTC phase dissipates rapidly (shown for \(N {=} 16\)). (B) With \(\Bac {=} 82 \mu T\), the DTC phase lasts significantly longer, extending beyond 12,000 cycles, an extension factor over 200. Data reveals that with \(\Bac\), the characteristic sharpening of the dome with increasing Floquet cycles is absent. Instead, there is a stability region around \(\gamma_y\) that is independent of \(\gamma_y\) and depends only on the AC field strength, matching theoretical expectations.
    (C-D) Right panels show same data on linear scale. Extension in DTC lifetime is evident.
    Brown color [not present on colorbar] is an optical illusion due to fast alternating green and magenta cycles on a log scale.
    }
\label{fig:dome-fig}
\end{figure*}

\begin{figure}[t]
    \centering
    \includegraphics[width=0.5\textwidth]{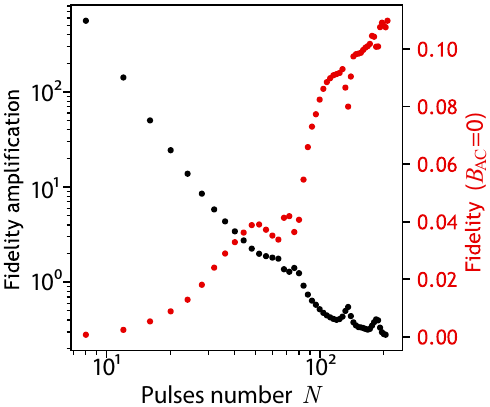}
    \caption{
    \T{Amplification of fidelity} 
    by application of $82.4\,\mu$T AC field relative to the zero AC case as a function of number of spin-lock pulses $N$ interspersed between $\gamma_y$ kicks, shown by black data points corresponding to left axis. For all $N$, spacings of 36~$\mu$s between each pulse are used, and the pulse length for $\gamma_y=\pi$ is 102.5~$\mu$s. The AC field increases fidelity by a larger factor for smaller $N$, while the baseline fidelity without AC (red, right axis) increases with $N$. For $N>80$, AC is observed to decrease fidelity, opposite to the coherence extension focused on in this paper.}
\label{fig:N-dependence}
\end{figure}

\begin{figure*}[t]
    \centering
    \includegraphics[width=0.95\textwidth]{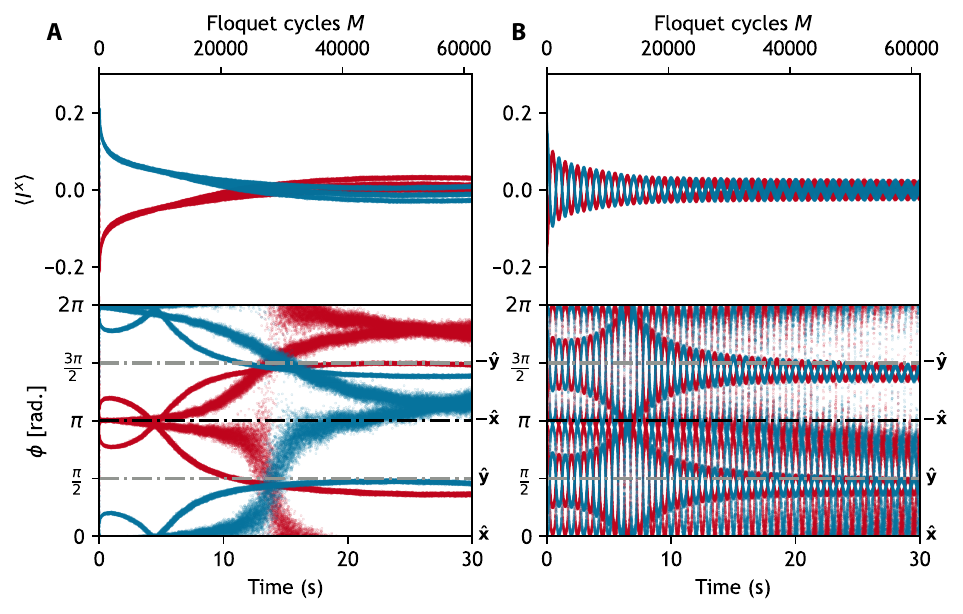}
    \caption{\T{Dynamics of DTC with off-resonant AC mapped in phase} $\phi$ (lower) as well as polarization $\langle I^x \rangle$ (upper), shown for AC frequencies with small offsets of $\xd f{=}0.02$ Hz (A) and $\xd f{=}1$ Hz (B) from resonance. Solid boundaries at $\phi{=}0,2\pi$ correspond to the $+\xhat$ axis along which spins point upon initiating the DTC, while the dash-dotted black line at $\phi{=}\pi$ indicates the $-\xhat$ axis. Dash-dotted gray lines at $\phi{=}\pi/2,3\pi/2$ indicate the $+\yhat$ and $-\yhat$ axes respectively. Blue points represent measurements after an even number of $\gamma$ pulses, while red represent measurements after an odd number of flips. Each plot has over 200000 total points, allowing us to rapidly and quasi-continuously track spin motion along the Bloch sphere equator, revealing intricate micromotion dynamics. Data is shown up to $t{=}30$s and 60,000 \(\gamma_y\) pulses (upper axis). Beating at the offset frequency from resonance is observed - after each half beat, the direction of the $I^x$ component is observed to invert. We also observed an increasing magnitude of the $I^y$ component during the sequence. Beyond the separation between even and odd number of $\gamma$-kicks, we observe further subdivision into multiple strands associated with micromotion between spacings between each of the $N=4$ spin lock pulses.}
\label{fig:phase-DTC}
\end{figure*}

\begin{figure}[t]
    \centering
    \includegraphics[width=0.5\textwidth]{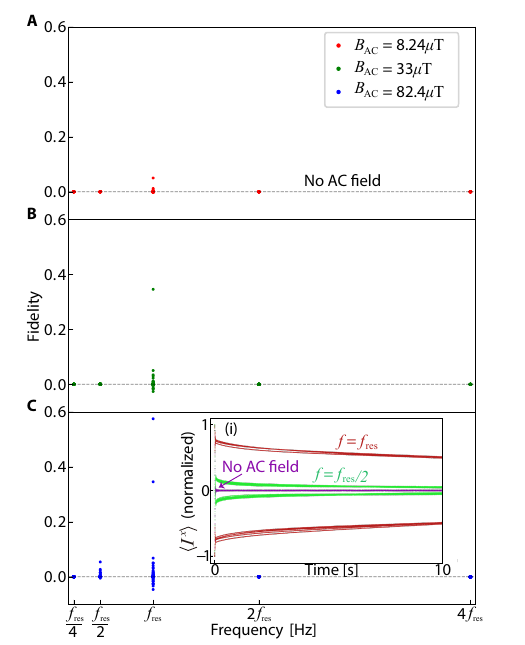}
    \caption{\T{Subharmonic response} in the AC field for different AC field amplitudes, following the methodology of \zfr{fig3}A in the main paper, with \(N{=}4\). (A-B) For smaller $B_{\text{AC}}$ values, as indicated in \zfr{fig3}A, the response is dominated by the first harmonic at $\fres$, corresponding to fitting a single AC field period within one Floquet cycle. (C) For larger AC field amplitudes, subharmonic responses appear at $\fres/2$, respectively. This corresponds to fitting a single AC field period within two resonance Floquet cycles. (i) \I{Inset} shows the oscillatory response of $\langle I^{\text{x}} \rangle$ over time for three cases: No Ac field (purple), subharmonic (green), and resonant (red) AC field with a magnitude of $B_{\text{AC}} = 82.4\mu\text{T}$ is applied. $(\pi/2)_{y}$ and $\theta_{x}$ pulses ($\approx 50.73\,\mu$s); $\tau (\approx 36.03\,\mu$s); $\gamma_{y}$ pulse ($\approx 101.55 \,\mu$s).}
\label{fig:harmonics}
\end{figure}

\begin{figure}[t]
    \centering
    \includegraphics[width=0.5\textwidth]{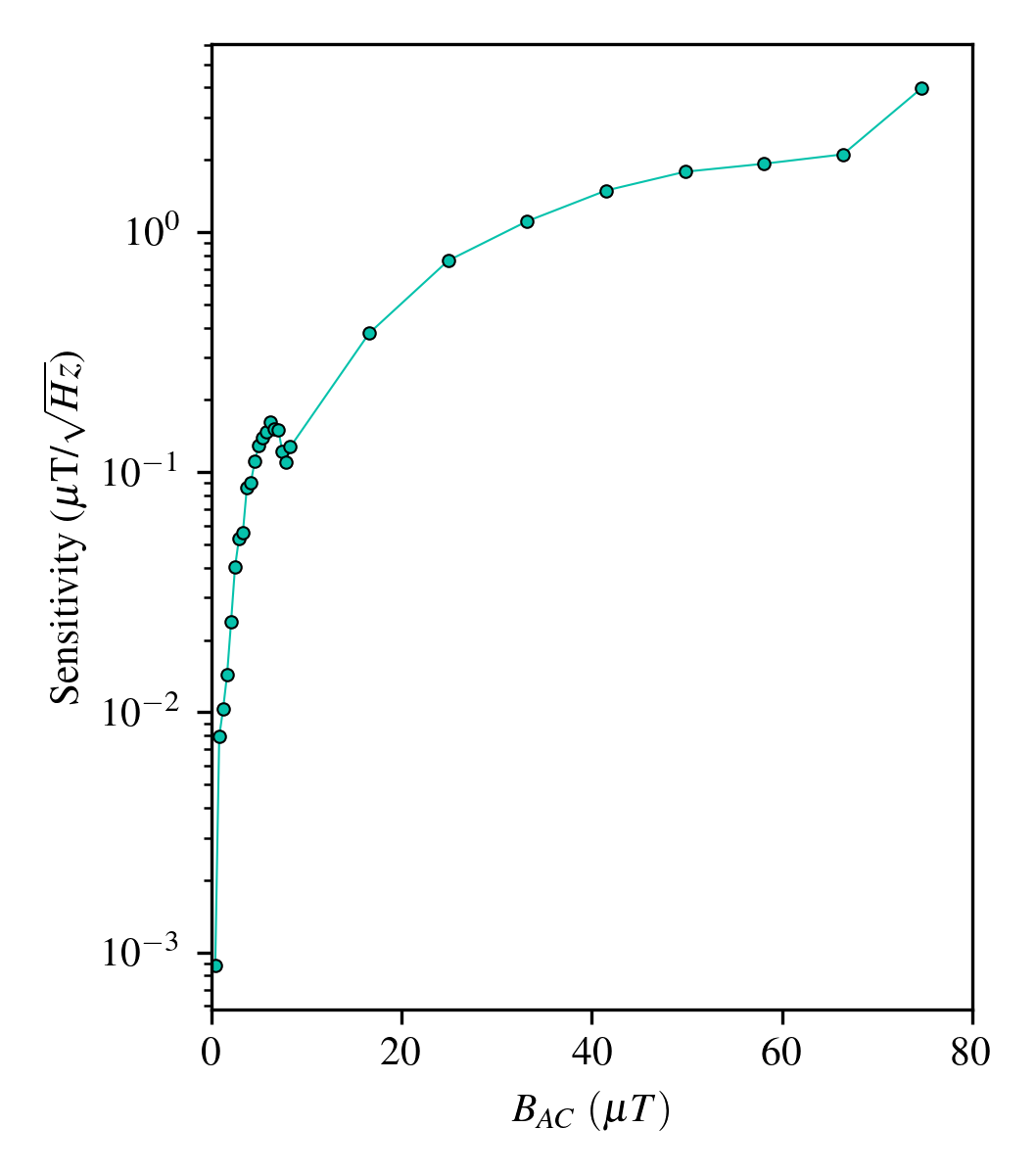}
    \caption{\T{Effective AC field sensitivity} measured from fidelity as a function of AC amplitude, following \zfr{fig2}B. Optimum sensitivity is obtained for small  AC amplitudes.}
\label{fig:sensitivity}
\end{figure}

\begin{figure}[t]
    \centering
    \includegraphics[width=0.5\textwidth]{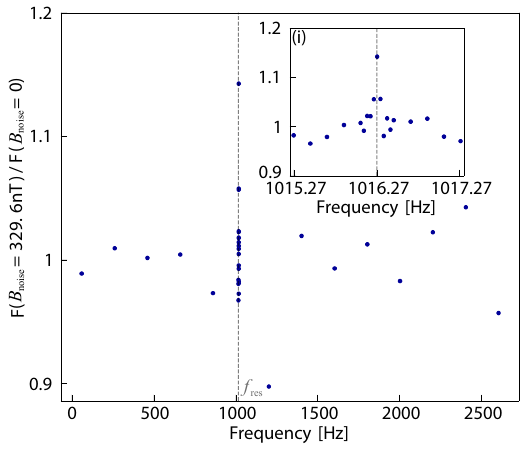}
    \caption{\T{Noise rejection of the DTC sensor} The sensor is tuned to a resonant frequency with an applied on-resonant bias AC field of $B_{\text{bias}} = 3.29\,\mu\text{T}$, while an order of magnitude smaller noise fields, $B_{\text{noise}} = 329.6\,\text{nT}$, of varying frequencies are introduced. The plot shows the relative increase in fidelity compared to the baseline fidelity $F(B_{\text{noise}} = 0)$. Off-resonant noise does not significantly affect the baseline, whereas a resonant signal of the same magnitude leads to a noticeable increase in the fidelity. The inset highlights a zoomed-in window of $\pm$1 Hz around the resonant frequency $f_{\text{res}} = 1016.27$ Hz. $(\pi/2)_{y}$ and $\theta_{x}$ pulses ($\approx 51.96\,\mu$s); $\tau (\approx 36\,\mu$s); $\gamma_{y}$ pulse ($\approx 104.01 \,\mu$s)}
\label{fig:noise-rej}
\zfl{fig:noise-rej}
\end{figure}

\section{Lifetime comparison}
\zsl{table}
In Table~\ref{tab:lifetime_comparison} we compare the DTC lifetimes achieved in this work with other lifetimes in the literature, including other platforms and DTC orders. 
Note that, in principle the MBL DTC realized using superconducting qubits~\cite{mi2022time} has an infinite lifetime, however, the DTC lifetime is limited by the coherence of the device.
Our AC-enhanced $U(1)$ PDTC exceeds previous experimental DTC realizations by at least one order of magnitude considering the total number of Floquet pulses applied, here $40,000$.

\section{Sensor comparison}
\label{supmat:sec:sensor_comparison}

To compare with other magnetometers, we illustrate the demonstrated and prospective sensitivity of the $^{13}$C DTC sensor along with various others, as a function of the magnetic field frequencies they probe, in Fig.~\ref{fig:sensor-comparison}. Sensor numbers 1-26 correspond to Refs.~\cite{sensor-1,sensor-2,sensor-3,sensor-4,sensor-5,sensor-6,sensor-7,sensor-8,sensor-9,sensor-10,sensor-11,sensor-12,sensor-13,sensor-14,sensor-15,sensor-16,sensor-17,sensor-18,sensor-19,sensor-20,sensor-21,sensor-22,sensor-23,sensor-24,sensor-25,sensor-26}. Although sensors 3, 4, 20 and 25 do achieve better sensitivity than we have yet demonstrated at frequencies between 1~kHz and 10~kHz, these are significantly larger sensors at 100~mm$^3$, 39000~mm$^3$, 5300~mm$^3$, and 1257~mm$^3$ respectively compared to the sensing volume 22.7~mm$^3$ of our diamond sample. Larger sensing volume necessitates greater standoffs from local field sources, which offsets sensitivity advantages. Larger sensors are also more vulnerable to performance degradation due to ambient field gradients, and are less apt for mapping inhomogeneous fields at high resolution.

Sensors 16 and 26 are the only ones identified which can sense fields between 1~kHz and 20~kHz  with higher sensitivity in a smaller volume than ours. For practical applications, we note that certain additional properties might advantage one sensor over another among this small set. For example, our sensor is demonstrated at a high bias field of 7~T, compared to the bias fields of 0.178~T and 0.7~mT for 26 and 16 respectively. This is advantageous for instance in NMR applications, since the higher bias strengthens the field source from nuclear spins (offsetting discrepancy in the magnetometer sensitivity). It also increases chemical shift, enabling better resolution of NMR spectra. We also note that robustness to environmental conditions may differ among these sensors. For instance, vibrational noise can pollute the magnetic field signal in the case of the ferrimagnetic oscillator \cite{sensor-26}, while we do not anticipate a direct imprint on DTC stability from vibrations.

\section{Phase diagram mapping: action of AC field}
\zsl{dome}
To demonstrate the efficacy of the applied AC field in extending the DTC phase, we systematically mapped the DTC phase diagram using the technique from Ref.~\cite{DTC_Beatrez2022}. Unlike other DTC experiments, which record the trace point by point after each Floquet pulse, our approach captures the entire trace in a single shot. This enables mapping the phase diagram by measuring multiple slices as a function of the $\gamma_y$-pulse angle.

\zfr{dome-fig}A shows the DTC phase diagram without an AC field, similar to Beatrez et al.~\cite{DTC_Beatrez2022}. In this experiment, we use $N{=}16$, meaning $\gamma_y$ pulses are applied in rapid succession, separated by only 16 spin-lock pulses. This smaller $N$ results in a faster DTC decay compared to Ref.~\cite{DTC_Beatrez2022}.

Data in \zfr{dome-fig}A is presented in both linear and log scales. The y-axis represents the total number of Floquet cycles, indicated by the number of $\gamma_y$-pulses applied, while the x-axis represents the $\gamma_y$ flip-angle. Colors indicate the obtained signal, with alternating green and magenta stripes representing the period-doubling response as the spins flip between $+\xhat$ and $-\xhat$. Two stable points are evident, centered at $\gamma_y{=}0$ and $\gamma_y{=}\pi$. White regions indicate where the spins undergo complete decoherence.

In the conventional DTC case, signal decay occurs in $\sim$100 Floquet cycles for $N{=}16$, similar to Ref.~\cite{DTC_Beatrez2022} where $N{=}32$. \zfr{dome-fig}B contrasts this with the effect of a \(\Bac{=}82 \mu\)T AC magnetic field, showing a significantly extended stability region near \(\gamma_y{=}\pi\).

Data is presented with a split axis for clarity, as the lifetime extension is over 200 times larger compared to without the AC field, sustaining over 10,000 $\gamma$-kicks. We observe a very stable period-doubling spin-flipping response. Importantly, the stability \I{"dome"} around \(\gamma_y{=}\pi\) is retained, reflecting the characteristic DTC behavior.

\section{Extended data}
\subsection{Lifetime extension as a function of $N$}
\zsl{extension_N}
In the two-tone driving scheme, the number of pulses \(N\) effectively sets the timescale for prethermalization. Consequently, one expects that the extension of the PDTC lifetime under the AC field depends on the total number of spin-lock pulses employed. We note that in previous work demonstrating a long-lived PDTC ($T_2^\prime {=} 4.3$s) and mapping its phase diagram, \(N\) was chosen to be a large number ($N{=}300$).

\zfr{N-dependence} examines the effect of lifetime extension as a function of \(N\). The data show that the lifetime extension is most dramatic for small \(N\) and appears to saturate for \(N {>} 100\). We rationalize this as follows: as \(N\) gets larger, the period $T$ lengthens, shifting away from the high-frequency regime. This inhibits the spins to prethermalize into finite energy density states under the combined AC field and PDTC two-tone driving, ultimately preventing lifetime extension.

The small \(N\) regime, where PDTC extension is greatest (\zfr{N-dependence}), is also particularly important for sensing of higher-frequency RF AC fields with improved sensitivity.

%\subsection{Lifetime extension as function of $\Bac$}
%\zsl{extension_Bac}
%As a complement to \zfr{fig3}C of the main paper, we investigate the extent to which the lifetime extension depends on the amplitude of the $\Bac$ field applied. For this, we study the lifetime extension for a fixed value of \(N {=} 16\) under different $\Bac$ values. These experiments are complicated by non-mono-exponential decays, and the variable time taken to prethermalize, making simple $1/e$ crossings unreliable for measuring lifetime.

%To address this, we measure the lifetime after the spins have had sufficient time to prethermalize, focusing on decays at times beyond $t {=} 10$s. Data in Figure 2B shows the resulting lifetime extension values, indicating almost no dependence on the AC field amplitude, as expected from theory.

\subsection{Phase measurement for off-resonance fields}
\zsl{phase}

In \zfr{fig3}A(ii) of the main paper, we show the polarization component $\expec{I^x}$ as a function of slight offsets from the exact resonance condition. In the extended data in \zfr{phase-DTC}, we present similar data but now include the phase information for clarity. The phase information, representing the instantaneous phase \(\phi\) of the spins on the Bloch sphere equator, is particularly revealing and shows the full extent of the intricate micromotion dynamics. This also illustrates the power of our measurement methodology, which can quasi-continuously track phase and amplitude information on the Bloch sphere.

We present two examples: one at a slight off-resonance of \(\delta f = 0.02\) Hz (\zfr{phase-DTC}A) and another at a more significant offset of \(\delta f = 1\) Hz (\zfr{phase-DTC}B). Upper panels show \(\langle I^x \rangle\), while lower panels  show $\phi$, with the rails (0, \(\pi\)) referring to the $\xhat, -\xhat$ axes, respectively. The beating observed is exactly at the resonance offset frequency as expected. The phase data particularly reveals multiple strands, corresponding to the micromotion between multiple stroboscopic prethermal plateaus corresponding to each of the interpulse spacings within a single Floquet cycle. The AC field generates a component to the prethermal axis that is transverse to $\hat{x}$ and is rotated by the spin-lock pulses about $\hat{x}$, causing nontrivial micromotion. A more detailed discussion of similar micromotion can be found in Ref.~\cite{sahin22_trajectory}.

\subsection{Sub-harmonic response at large $\Bac$}
\zsl{harmonics}
While the dominant effect of the AC field, shown in \zfr{fig3}A, is the first harmonic response (i.e., a strong extension of lifetime at the resonance frequency $\fres$), we also observe subharmonic responses (dominated by $\fres/2$). This occurs when two or more effective \(\gamma_y\)-kick cycles matches one AC field period.

This phenomenon becomes visually apparent at larger $\Bac$ values. \zfr{harmonics} studies this response, tracking the frequency for different $\Bac$ values. For small $\Bac$ values \zfr{harmonics}A, the response is dominated by the first harmonic. However, at larger $\Bac$ values \zfr{harmonics}B-C, subharmonic responses, especially at $\fres/2$, become more evident.

% \subsection{Strength of the $I^x$ and $I^y$ Components}
% To gain more insight into the micromotion dynamics for different $\Bac$ values, we measure the $\expec{I^x}$ and $\expec{I^y}$ components separately. Data in \zfr{signal-scaling} is presented for (i) a wide range of large $\Bac$ values (\zfr{signal-scaling}A) and (ii) a zoomed-in view of the small $\Bac$ regime (\zfr{signal-scaling}B).

% The \(I^x\) component, as shown in \zfr{fig2}C of the main paper, increases with $\Bac$, reflecting the extension of the PDTC lifetime. The \(I^y\) component undergoes more complex dynamics, initially increasing, reaching a peak, and then decreasing. This identifies three distinct regimes of operation. For low $\Bac$ values (\zfr{signal-scaling}B), \(I^x\) component increases quadratically, while \(I^y\) increases almost linearly. In \zfr{signal-scaling}B, the solid lines represent polynomial fits, with quadratic and linear fits for short times, respectively. At intermediate $\Bac$ values, the spins have a significant prethermalization component at \(I^y\), while for large $\Bac$ the micromotion is dominated by \(I^x\) (\zfr{signal-scaling}A).  Error bars are determined by standard deviations from 10 consecutive shots of the experiment.

\subsection{Sensitivity}
\label{supmat:sec:sensitivity}
As described in Sec~\ref{sec:methods}, the sensor achieves a sensitivity of 880~pT$/\sqrt{Hz}$ with an applied bias of 415~nT. The sensitivity is worsened with increasing bias as shown in Fig.~\ref{fig:sensitivity}. This is because we find that the shot-to-shot variance in fidelity for DTCs prepared with nominally identical conditions grows with AC amplitude more quickly than the fidelity response to AC perturbations. One possible source of shot-to-shot variance is fluctuations in the initial level of hyperpolarization; it was observed in Ref.~\cite{Beatrez2023electron} that hyperpolarization not only scales the magnetization signal but also affects coherence time. Minimizing such sources of shot-to-shot variance, so that uncertainty in $F$ is instead dominated by the readout noise within in a single shot, could disproportionately improve sensitivity. Not only would this decrease noise $\sigma_F$ in the fidelity metric at any fixed AC bias, but it would allow operation in a higher bias regime where the response $\partial F /\partial B_{\text{AC}}$ to AC perturbation is greater. Mitigation of shot-to-shot variance could be achieved by corrections in offline analysis as well as by improvements to the apparatus (e.g. laser stability). Reducing shot-to-shot variance so that fluctuations $\sigma_F$ in the fidelity metric at fixed $B_{\text{AC}}$ are dominated by the readout of the magnetization would allow us to exploit the stronger response $\partial F/\partial B_{\text{AC}}$ at biases around 8~$\mu$T. We estimate sensitivity could be improved by a factor of $\approx$10 in this case.
Beyond minimizing shot-to-shot variance, sensitivity could be substantially improved by reducing uncertainty on the fidelity $F$ estimated for a single shot. A simple way to achieve this is increasing sample volume, as the present sample only occupies a small portion of the 8~mm dimensions of the inductive readout coil. Increasing the fill factor of the diamond in the coil could improve sensitivity by a factor of $\approx$25. The signal-to-noise ratio within the magnetization readout could also be improved by increasing hyperpolarization of the $^{13}$C nuclei or enriching their concentration above the 1\% natural abundance in the current sample. Noise could also be reduced by upgrades to the readout electronics. With these improvements made in combination, an order-of-magnitude enhancement in sensitivity is expected. The prospective sensitivity is indicated in Fig.~\ref{fig:sensor-comparison}.

In addition to these technical considerations, we finally note that $\gamma_{y}$ pulses play a dual role in determining the sensitivity, as they set both the intrinsic DTC lifetime and the strength of the coupling to the AC field. Because these two effects compete, optimizing the $\gamma_{y}$-pulse duration is nontrivial and remains an important direction for future work.

\subsection{Noise rejection of the DTC sensor}
\label{supmat:sec:noise_reject}

We demonstrate in \zfr{fig3}A that the DTC sensor exhibits a frequency linewidth of 69~mHz. This narrow linewidth allows the sensor to reject off-resonant noise, as shown in \zfr{fig:noise-rej}. In this experiment, we first apply a bias field of strength $B_{\text{bias}} = 3.29\,\mu\text{T}$, which is resonant with the $\gamma_y$ pulses of the DTC sequence. We then introduce AC fields an order of magnitude smaller than the bias field $B_{\text{noise}} = 329.6\,\text{nT}$ with varying frequency to illustrate that only the resonant signal affects the sensor significantly, while off-resonant noise has minimal effect. 

In \zfr{fig:noise-rej}, we measure the increase in fidelity when AC fields, with amplitudes an order of magnitude smaller than the bias field and with varying frequencies, are applied along with the bias field. This is compared to the baseline case where only the bias field is present. The plot shows that a marked increase in fidelity only occurs when the AC field is resonant, indicated by the gray dotted line at $f_{\text{res}}$. This result highlights the DTC sensor's capability to detect signals at a specific frequency in a noisy environment containing different frequencies.

\subsection{Estimating magnetic field from the $\zhat$-coil}
\begin{figure}[t]
    \centering
    \includegraphics[width=0.5\textwidth]{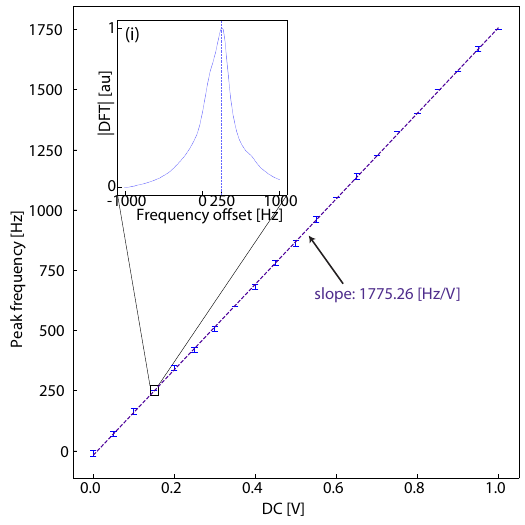}
    \caption{ \T{Estimating B-field from an AC coil} Free Induction Decay (FID) signal from the coil is measured for different DC voltages from the Tektronix source. The FID signal is mixed down to zero frequency with an on-chip Numerically Controlled Oscillator (NCO). The inset (i) shows the absolute magnitude of the Discrete Fourier Transformed (DFT) FID data with an applied bias B-field, where the voltage of the Tektronix source is set to $0.15,$V. A linear fit (purple dotted line) illustrates how the peak frequency shifts linearly with increasing DC voltage from the Tektronix source, with the slope indicating the rate of frequency shift.}
\label{fig:FID_DC}
\end{figure}

\zsl{calibration}
To generate an AC field aligned with the $B_{0} = $7T, a secondary coil is positioned inside the NMR probe to generate B-field parallel to $\mathbf{\hat{z}}$. To estimate the B-field applied to the diamond sample, we measure how the peak of the Fourier-transformed free induction decay (FID) signal shifts as the voltage of the Tektronix device is varied. Fig.\ref{fig:FID_DC} shows how the shift in peak frequency ($f_{\text{peak}}$), as the B-field from the secondary coil changes with increasing DC voltage from the Tektronix source. The change in frequency, $\Delta f_{\text{peak}}$, is given by $\Delta f_{\text{peak}} = \gamma_n (1 - \sigma) \Delta B = \gamma_n (1 - \sigma) (\Delta B / \Delta V) \Delta V$, where $\gamma_n$ is the gyromagnetic ratio of the $\Cs$ nuclear spin, $\sigma$ is the chemical shift, $\Delta V$ is the change in voltage from the voltage source, and $\Delta B$ is the corresponding change in the magnetic field. The slope shown in Fig.\ref{fig:FID_DC} corresponds to $\gamma_n (1 - \sigma) (\Delta B / \Delta V)$, indicating that a 1V change from the Tektronix source corresponds to approximately 164.85 $\mu$T. This is based on the value of $\gamma_n (1 - \sigma) = 7\text{T}/75.38\text{MHz} = 9.29\times10^{-8}\text{T}/\text{Hz}$.

\subsection{Three-tone drive}
\begin{figure}[t]
    \centering
    \includegraphics[width=0.5\textwidth]{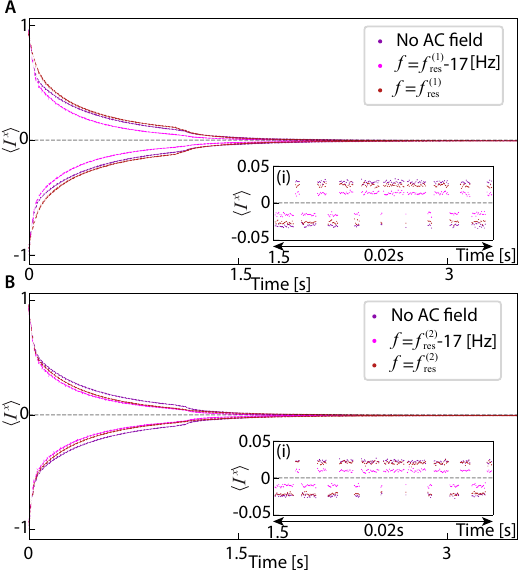}
    \caption{
    \T{Three-tone drive} depicted in the inset of \zfr{fig3}C functions as a two-frequency sensor, detecting two distinct resonant frequencies: $\fres^{(1)}=$208Hz and $\fres^{(2)}=$250Hz. (A) Time vs normalized $\xhat$-magnetization, $\expval{I^x}$, for the three tone drive under three conditions: no AC field (dark purple), an AC field with frequency $f = \fres^{(1)} - 17\text{Hz}$ (magenta), and an AC field at frequency $f = \fres^{(1)}$ (dark red). (B) Time vs $\expval{I^x}$ under similar conditions for $\fres^{(2)}$: no AC field(dark purple), an AC field with frequency $f = \fres^{(2)} - 17\text{Hz}$ (magenta), and AC field at frequency $f = \fres^{(2)}$ (dark red). Insets (A)(i) and (B)(i) provide a zoomed-in view of $\expval{I^x}$ for the time interval between 1.5 and 1.52 seconds. Both insets show that $\expval{I^{x}}$ is the most stable at later times ($>1.5$s) with no-AC field, while off-resonant AC fields cause the most decoherence in the $\xhat$ magnetization.Pulse sequence dimensions listed in \zfr{fig3}C.
   }

\label{fig:three-tone}
\end{figure}

Fig.\ref{fig:three-tone} shows how the normalized $\xhat$-magnetization $\expval{I^x}$ changes in time under a three tone drive. The pulse sequence, shown in the inset of Fig.\ref{fig:fig3}C, begins with a $(\frac{\pi}{2})_{y}$ pulse to tip the $\Cs$ nuclear spins' magnetization to the $\xhat$-axis of their rotating frame. Then, $\gamma_{y}$ pulses are applied at two distinct frequencies $2f_{\text{res}}^{(1)}$ and $2f_{\text{res}}^{(2)}$, interleaved with the spin-lock sequence. The signal from $\Cs$ nuclear spins is readout between the pulses, while AC fields with a magnitude of $B_{\text{AC}} = 329.6\text{nT}$ and varying frequencies are applied. 

We observe resonant responses in two cases: when the AC field frequency is $f_{\text{res}}^{(1)}$ and when the anti-nodes of the AC field align with the $\gamma_{y}$ pulses at a frequency $2f_{\text{res}}^{(1)}$. A similar resonant condition is observed for the AC field at $f_{\text{res}}^{(2)}$.

The three-tone drive exhibits a different resonant response compared to the two-tone drive. Notably, regardless of the two resonant conditions, $\expval{I^x}$ decreases when an AC field is applied, particularly at later times ($>1.5\text{s}$). This decline is clearly depicted in the insets of Fig.\ref{fig:three-tone}A and Fig.\ref{fig:three-tone}B, where more $\langle I^x\rangle$ is preserved in the absence of an AC field compared to the both off-resonant and resonant cases for $f_{\text{res}}^{(1)}$ and $f_{\text{res}}^{(2)}$. This behaviour occurs because, regardless of whether the AC field is applied at $f_{\text{res}}^{(1)}$ or $f_{\text{res}}^{(2)}$, off-resonant $\gamma_{y}$ pulses are always present, preventing the spins from prethermalizing into an effective Hamiltonian. 

However, Fig.\ref{fig:fig3}C shows that, experimentally, the signal decreases the least in the two on-resonant cases, compared to when off-resonant AC-fields of the same magnitude are applied. These two distinct resonant responses can be leveraged to realize a two-frequency sensor.

\subsection{Spin-lock sensing}
\zsl{spin_lock_sensing}

\begin{figure}[!ht]
    \centering
    \includegraphics[width=0.5\textwidth]{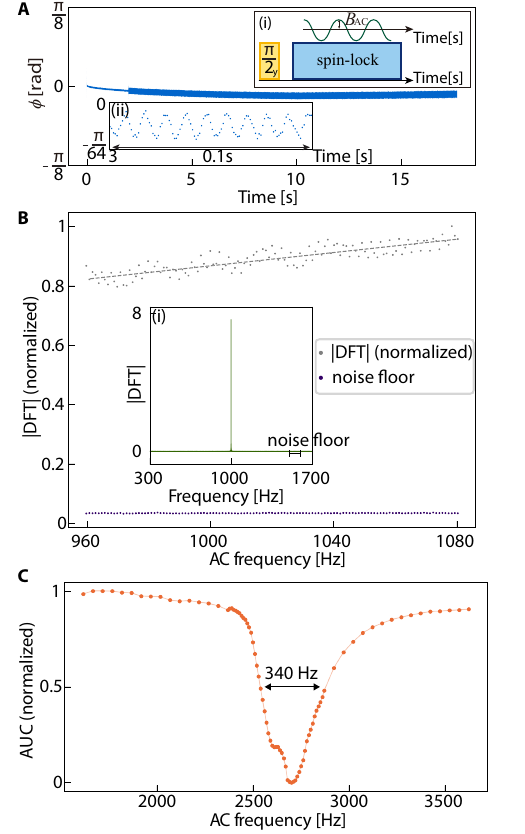}
    \caption{
    \T{Spin-lock sensing} (A) (i) The spin-lock sequence is applied for approximately 17.7 seconds, during which an AC field with $B_{\text{AC}} = 8.24\mu\text{T}$ is introduced in the $\pm\zhat$ direction after 2 seconds. The oscillations in the $\Cs$ nuclear spins, induced by the AC field, are imprinted in the phase ($\phi$) of their rotating frame. (ii) A zoomed-in view of the data over a $0.1\,$second window at $3.05\,$seconds is presented. $(\frac{\pi}{2})_{y}$ and $\theta_{x}$ pulses ($\approx 52.24 \mu s$). $\tau$ ($\approx 36.03\mu s$). (B) Each grey point represents the mean of the Discrete Fourier Transform (DFT) spectrum of the phase signal from $\Cs$ nuclear spins in the vicinity of its peak for different applied AC field frequencies, with the values normalized relative to the maximum across all frequencies. (B)(i) shows the full DFT spectrum for an applied of $1000$Hz, plotted in green. The average signal between 1600 and 1700Hz, with the frequency range denoted by a black error bar in B(i), is used to measure the noise floor. Dark purple points in (B) indicate the noise floor (normalized using the same maximum value as the grey points) across different AC field frequencies. (C) The FWHM (Full Width at Half-Maximum) corresponds to approximately $340\text{Hz}$. Pulse length of the spin lock is $\approx 55.4\mu\text{s}$, and pulse spacing is $\approx 40.0 \mu\text{s}$.}

\label{fig:spin-lock}
\end{figure}

Fig.\ref{fig:spin-lock} illustrates how the spin-lock sensing data presented in Fig.\ref{fig:fig3}A is obtained. Fig.\ref{fig:spin-lock}A(i) depicts the sensing protocol introduced in \cite{Sahin22}: $(\frac{\pi}{2})_{y}$ pulse is first applied to tip the $\Cs$ nuclear spins to the x-axis of their rotating frame. A train of $\theta_{x}$ pulse (spin-lock sequence) is applied, and the $\Cs$ nuclear spin signal is read out in between the pulses. Fig.\ref{fig:spin-lock}A shows how $\Cs$ nuclear spins imprint the applied AC field with a frequency of $f = 1000\text{Hz}$ and a magnitude of $B_{\text{AC}} = 8.24\mu\text{T}$ in the phase ($\phi$) of their rotating frame. A zoomed-in window in Fig.\ref{fig:spin-lock}A(ii) clearly displays the imprinted oscillations.

We vary the frequency of the applied AC field and measure the absolute value of the Discrete Fourier Transform (DFT) spectrum, shown in Fig.\ref{fig:spin-lock}B(i) for $f = 1000\text{Hz}$. For each frequency, we calculate the mean of the absolute value of the DFT spectrum near its peak, represented as grey points in Fig.\ref{fig:spin-lock}B. The linear fit is further shown in Fig.\ref{fig:spin-lock}B as grey dotted line to serve as a visual guide.

The linear profile of the mean of the absolute value of the DFT spectrum near its peak demonstrates that while the spin-lock sensing scheme can detect multiple frequencies with a single pulse parameter, it lacks the frequency selectivity offered by the DTC sensing scheme. 

To illustrate the spin-lock sensing scheme's linewidth on resonance as reported in \cite{Sahin22}, we vary the frequency of the AC field and measure the integrated signal. As shown in Fig.~\ref{fig:spin-lock}C, the linewidth is approximately 340Hz, which is at least four orders of magnitude larger than the linewidth of the DTC sensing scheme. This indicates that the spin-lock sensing scheme is not limited by the lifetime of the \C. Note that the reported linewidth in \cite{Sahin22} is 224Hz, acheived using a different set of pulse parameters and an alternative console to generate the $\pi/2$ pulses. The lack of narrow linewidth in the spin-lock sensing highlights the novelty of the DTC sensing scheme, which can be precisely tuned to a desired frequency.

\subsection{Phase-dependence of single-tone DTC}
\label{supmat:sec:phase_dependence_singletone}
\begin{figure}[t]
    \centering
    \includegraphics[width=\linewidth]{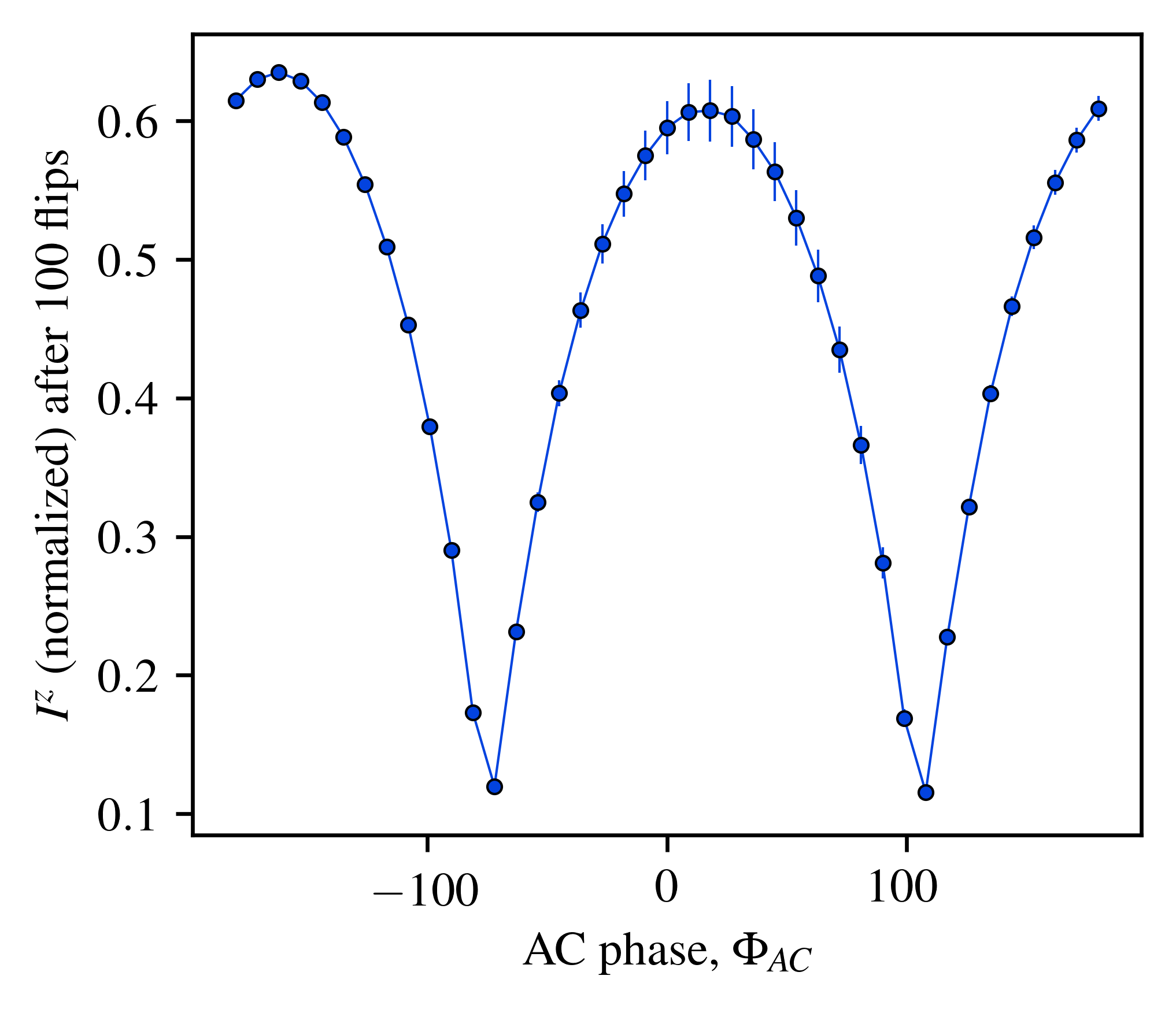}
    \caption{
        \textbf{Phase dependence of single-tone DTC extension.} In contrast to the two-tone DTC, extension of the single-tone DTC is maximized when AC nodes (rather than extrema) occur during $\gamma$ pulses. This is explained by the treatment in Secs.~\ref{supmat:sec:one_tone} and \ref{supmat:sec:two_tone}: the effect in the single-tone case arises due to the AC field between pulses, while in the two-tone case an effective field arises due to the field applied during the pulses. 
    }
    \label{supmat:fig:single_frequency_phase}
\end{figure}

In Fig.~\ref{supmat:fig:single_frequency_phase}, we report on the phase dependence of the AC response of the single-tone DTC; see Fig.~\ref{fig:fig4}A for details of the sequence.
In contrast to the two-tone DTC, see Fig.~\ref{fig:fig2}A(ii), for the single-tone DTC, the optimal response is achieved when the AC field and pulse sequence are in-phase, $\Phi_\mathrm{AC}=0$.
As we detail in Sec.~\ref{supmat:sec:one_tone}, this is a direct consequence of the AC field oscillating in the same axis~($\zhat$) as the signal response of the single-tone DTC; therefore, the largest signal is accumulated when no sign change between two $\yhat$-pulses occurs.
%========================================%
%
%     Theory & Simulation content
%
%========================================%

\section{Numerical Algorithm}
\label{supmat:sec:algorithm}

We perform (closed-system) quantum simulations of the experimental sequence, detailed in Fig.~\ref{fig:fig1}, on a small number $L$ of spins~($L=15$) using the QuSpin python library\cite{quspin2017,quspin2019} and a slightly modified version of the algorithm used in Ref.~\cite{DTC_Beatrez2022}.

\paragraph*{System.}
The experimental setup consists of a macroscopic ensemble of NV-centers each surrounded by a cluster of $1,000$-$10,000$ nuclear spins that are randomly distributed on the vertices of the diamond lattice. The nuclear spins interact via the dipole-dipole coupling
\begin{equation}
\label{supmat:eq:Hdd}
    \Hdd = \sum_{k<\ell}^L J_{k\ell} \left( 3 I_k^z I_\ell^z - \boldsymbol{I}_k \cdot \boldsymbol{I}_\ell\right)
\end{equation}
with $J_{k\ell}=c_\mathrm{exp}(3\cos^2\theta_{k\ell} - 1)/r^3_{k\ell}$, where $r_{k\ell}$ is the distance between two spins on sites $k$ and $\ell$ and $\theta_{k\ell}$ is the angle between the vector connecting the two spins and the direction of the magnetic field~($\zhat$); $c_\mathrm{exp}$ is a sample-dependent constant.

While the clusters of nuclear spins can be thought of as isolated during the experimental time scale, the effective coherent system size in the experiment exceeds thousands of spins. 
In stark contrast, our exact numerical simulations are limited to few~($L=15$) spins due to the exponentially increasing Hilbert space dimension.
To mimic the experiment with this small system size we use a specifically tailored random graph instead of placing the spins randomly on a diamond lattice. 
Concretely, to make the best use of the small system size, we want to avoid (i) spins that are too weakly coupled to the rest of the system and (ii) spins that are too strongly coupled, since both would result in spins being effectively decoupled from the rest of the system.
Therefore, we use the procedure from Ref.~\cite{DTC_Beatrez2022}: Spin positions are drawn randomly one by one in a $3$D cube such that (a) each spin has a maximal distance $r_\mathrm{max}$ to at least one other spin and (b) each spin has at least a distance $r_\mathrm{min}$ to all other spins; thus, avoiding (i) and (ii) respectively.
We use $r_\mathrm{min}=0.9$ and $r_\mathrm{max}=1.1$ throughout.

In addition, an important aspect of the 3D dipole-dipole couplings in the experiments performed in this work is the homogeneous distribution of positive and negative couplings, i.e., in a sufficiently large system we find (iii) $\sum_{k<\ell} J_{k\ell} = 0$ as the spins are on-average uniformly distributed on the unit sphere in 3D.
This property is in general not fulfilled for the random $3$D graph as the number of spins is too small to lead to a uniform distribution. 
Instead, we enforce this condition by hand using the fact that the sign of the interactions,$(3\cos^2\theta_{k\ell} - 1)$, depends on the orientation with respect to the external $z$-field direction. 
Therefore, only after having drawn a random graph we (c) choose the orientation of the entire graph such that (iii) $\sum_{k<\ell} J_{k\ell} = 0$ is fulfilled.
To further mitigate finite size effects we average the magnetization dynamics over different realizations of random graphs~(usually $n_\mathrm{samples}{=}50$-$100$ samples).

Finally, in the experiment the initial state after hyperpolarization is $\rho_0 {\sim} 1+\mu I^z$ with finite polarization $\mu$. Here, instead we use the fully polarized pure state $\ket{\psi_0}=\ket{\uparrow \dots \uparrow}$ which has been shown to reproduce comparable dynamics at lower computational cost~\cite{DTC_Beatrez2022}.

\begin{figure}[t]
    \centering
    \includegraphics[width=\linewidth]{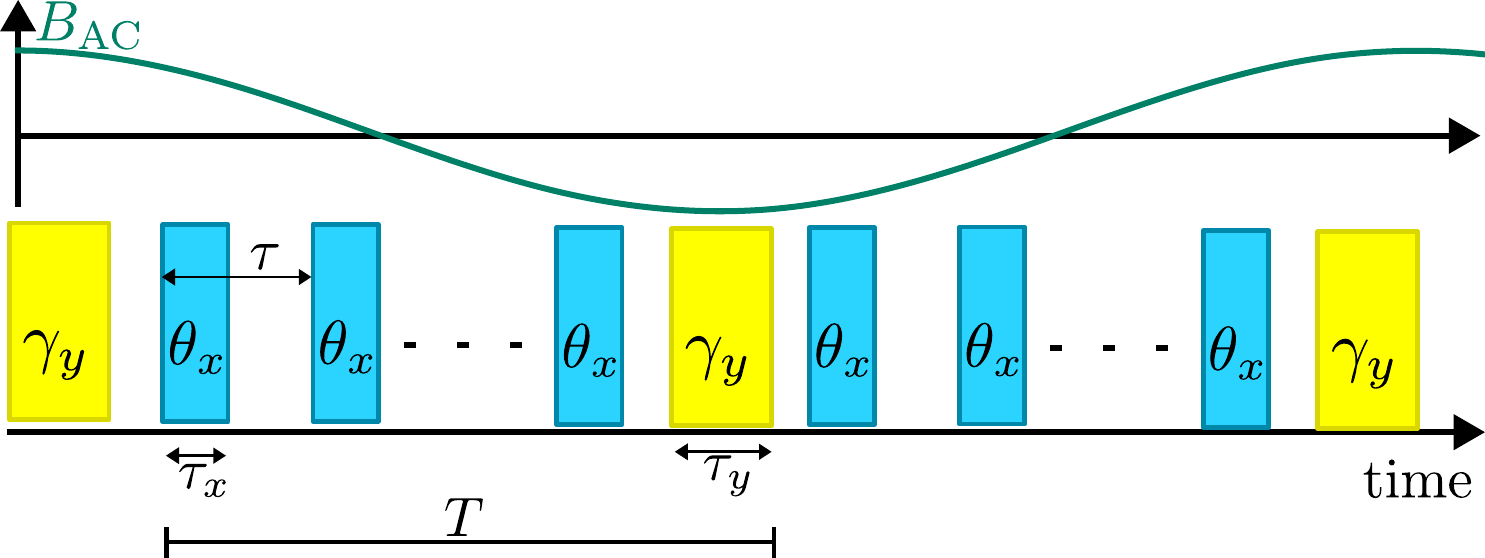}
    \caption{
        \textbf{Detailed DTC sequence.}
        Sketch of DTC sequence with individual pulse duration $\tau_x$ and $\tau_y$, as well as, spin-locking period $\tau$ and full period $T$; extended version of Fig~\ref{fig:fig1}B(i).
    }
    \label{supmat:fig:dtc_sequence}
\end{figure}

\paragraph*{Unitary Evolution.}
In the following, we describe the unitary evolution implemented in the numerical simulations.
Specifically, the experimental two-tone sequence, see Fig.~\ref{supmat:fig:dtc_sequence} and main text Fig.~\ref{fig:fig1}B, over a single DTC cycle can be recast into the following unitary evolution~\cite{sahin22_trajectory}
\begin{equation}
\label{supmat:eq:unitary_evolution}
    U_{\mathrm{DTC},\ell} = U_{y,\ell} U_{\mathrm{dd},\ell,N+1} \prod_{k=1}^N U_{x,\ell,k} U_{\mathrm{dd},\ell,k} \, ,
\end{equation}
\begin{subequations}
\label{supmat:eq:unitaries}
with the $\yhat$-pulse  
\begin{equation}
    U_{y,\ell}=\mathscr{T}\exp(-i\int^{(\ell + 1)T}_{(\ell+1) T - \tau_y} B_y I^y + \Bac(t) I^z  + \Hdd \mathrm{d}t)\, ,
\end{equation}
and $\xhat$-pulses
\begin{equation}
    U_{x,\ell,k}=\mathscr{T}\exp(-i\int^{\ell T + k \tau}_{\ell T + k \tau- \tau_x} B_x I^x + \Bac(t) I^z  + \Hdd \mathrm{d}t)
\end{equation}
and inter-pulse evolution
\begin{equation}
    U_{\mathrm{dd},\ell,k}=\exp\left[-i \left( \tau \Hdd + \int^{\ell T + k \tau - \tau_x}_{\ell T + (k-1) \tau} \mathrm{d}t \Bac(t) I^z \right) \right] \, ,
\end{equation}
where we used that $\Hdd$ conserves the $z$-magnetization, $\left[I^z,\,\Hdd \right]=0$, to solve the time-ordering $\mathscr{T}$ explicitly in the last equation.
\end{subequations}
Note that the AC field varies very slowly in comparison to the length of the $\xhat$ and $\yhat$-pulses, $\fac\ll 1/\tau_x,1/\tau_y$. Therefore, we assume a quasi-stationary AC field during those pulses, hence, simplifying the time-ordered integrals to
\begin{subequations}
\label{supmat:eq:unitaries_simple}
\begin{equation}
    U_{y,\ell}\approx \exp[-i \left( \gamma_y I^y + \vartheta^\mathrm{eff}_{y;\ell} I^z  + \Hdd \tau_y\right)]\, ,
\end{equation}
with $\vartheta^\mathrm{eff}_{y;\ell}=\int^{(\ell + 1)T}_{(\ell+1) T - \tau_y} \Bac(t) \mathrm{d}t$,
\begin{equation}
    U_{x,\ell,k}\approx \exp[-i \left( \theta_x I^y + \vartheta^\mathrm{eff}_{x;\ell,k} I^z  + \Hdd \tau_x\right)]\, ,
\end{equation}
with $\vartheta^\mathrm{eff}_{x;\ell,k}=\int^{\ell T + k \tau}_{\ell T + k \tau - \tau_x} \Bac(t) \mathrm{d}t$, and
\begin{equation}
    U_{\mathrm{dd},\ell,k}\approx \exp[-i \left( \vartheta^\mathrm{eff}_{\mathrm{dd};\ell,k} I^z  + \Hdd \tau\right)]\, ,
\end{equation}
with $\vartheta^\mathrm{eff}_{\mathrm{dd};\ell,k}=\int^{\ell T + k \tau - \tau_x}_{\ell T + (k-1) \tau} \mathrm{d}t \Bac(t) I^z$. 
\end{subequations}
The evolution with the approximate unitaries~\eqref{supmat:eq:unitaries_simple} can be efficiently implemented using exact diagonalization and is done via the Quspin python package.

Similarly, for the single-tone DTC one can use the same unitary evolution, Eqs.~\eqref{supmat:eq:unitary_evolution} and \eqref{supmat:eq:unitaries_simple}, by setting $\tau_x{=}0$.
Let us emphasize that, in contrast to previous work~\cite{DTC_Beatrez2022,sahin22_trajectory,harkins23_textures,moon24_rondeau}, we consider the full finite time $\xhat$ and $\yhat$ pulses. 
While the results in the absence of an AC field are qualitatively independent of the finite time of the pulses, in the presence of the AC field in the two-tone DTC they are vital to account for the observed behavior as described in Sec.~\ref{supmat:sec:two_tone}.

\section{Floquet engineering finite energy density}
\label{supmat:sec:engineering}

In this section, we will detail the average Hamiltonian analysis for single (\ref{supmat:sec:one_tone}) and two-tone (\ref{supmat:sec:two_tone}) DTC sequences.
We will demonstrate that the lifetime enhancement in both cases is due to the AC field effectively Floquet-engineering a coupling to the DTC order parameter.
This coupling introduces a finite energy density for DTC ordered states, thus energetically protecting those states from prethermalization to a featureless, infinite temperature state.
In subsection~\ref{supmat:sec:summary_engineering} we summarize the key ingredients required for the AC-induced lifetime extensions of DTC order.
Finally, in subsection~\ref{supmat:sec:dc_vs_ac} we compare our AC scheme to a previously introduced DC scheme~\cite{PrethermalWithoutTemperature_LuitzEtAl2020}.

\subsection{One-tone prethermal discrete time crystal}
\label{supmat:sec:one_tone}

Let us first focus on the conceptually simpler case of single-tone DTC: 
Here, the analysis is made easier since, the AC field $H_\mathrm{AC}=B_\mathrm{AC}(t) I^z$ and interactions $H_\mathrm{dd}$ commute~($\comm{H_\mathrm{AC}}{H_\mathrm{dd}}{=}0$), such that we can simply integrate the AC field between two consecutive $y$-pulses. 
Note that, even in the absence of the AC field for perfect $y$-pulses ($\gamma_y{=}\pi$) the symmetry-protected DTC is in principle infinitely long-lived, due to the perfect conservation of $I^z$. 
However, for finite~($\gamma_y \neq \pi$), but small~($\gamma_y \approx \pi$) deviations $\epsilon = \gamma_y-\pi$, this conservation law is broken leading to a fast decay of the polarization with heating rate $\Gamma_e$ determined through Fermi's Golden rule as 
\begin{equation}
\label{supmat:eq:FGR}
    \Gamma_e\propto (\epsilon/T)^2
\end{equation}
Such deviations are ubiquitous in the experiment due to spatially varying magnetic fields within the macroscopic sample~\cite{harkins23_textures,moon24_rondeau}.

We will now describe how an AC field can stabilize the PDTC order against symmetry-breaking terms, and, in fact, lead to an exponential enhancement in lifetime.
Integrating the AC field in between two $y$-pulses the dynamics is described by 
\begin{equation}
\label{supmat:eq:single_tone_unitary}
    U = \left(U_yU_{+z}U_\mathrm{dd}\right)\left(U_yU_{-z}U_\mathrm{dd}\right)\left(U_yU_{+z}U_\mathrm{dd}\right)\dots, 
\end{equation}
where $U_y=\exp(-i\gamma_y I^y)$, $U_{\pm z}=\exp(\pm i\overline{B} \tau I^z)$ with magnitude of integrated AC field $\overline{B}\propto \Bac$ and interactions $U_\mathrm{dd}=\exp(-i \tau H_\mathrm{dd})$. For simplicity, in Eq.~\eqref{supmat:eq:single_tone_unitary}, we have neglected the AC field during the $\yhat$ pulses, corresponding to the limit of infinitely fast pulses. 
A direct comparison with the numerical simulations shows a qualitatively good agreement and justifies this approximation, see Fig.~\ref{supmat:fig:proof_of_principle}A. However, neglecting the finite duration of the pulses is not generally valid as we will see in the two-tone driving case.

Note that, in the special case $\gamma_y=\pi$, we have, $U_y U_\mathrm{dd} = U_\mathrm{dd}U_y$ and $U_y U_{\pm z} = U_{\mp z}U_y$ such that after $2N$ $y$-pulses the dynamics are given by $U(2N)=\left(U_{+z}U_\mathrm{dd}\right)^{2N}$. 
For imperfect $y$-pulses, $\epsilon\neq 0$, the dynamics after even numbers of $y$-pulses is effectively described by $H^{\mathrm{1,AC}}_\mathrm{eff}=H_\mathrm{dd} + \overline{B} I^z + B_y I^y$. 
While $H^{\mathrm{1,AC}}_\mathrm{eff}$ no longer preserves the $z$-polarization it admits a finite energy density for the initial state $\mathcal{E}=\expval{H^{\mathrm{1,AC}}_\mathrm{eff}}_{\rho_0}/L = \overline{B} \mu$. Therefore, the system prethermalizes to a finite temperature state $\rho_\mathcal{T} \propto e^{-H^{\mathrm{1,AC}}_\mathrm{eff}}/\mathcal{T}$, with $\mathcal{T}$ such that $\Tr(H^{\mathrm{1,AC}}_\mathrm{eff} \rho_\mathcal{T}) = \mathcal{E}$. 

While Floquet heating leads to a slow increase of the effective temperature~($\mathcal{T}$)~\cite{Fleckenstein21_heating,Fleckenstein21L_heating} this process is exponentially suppressed in the driving period $T$,
\begin{equation}
\label{supmat:eq:exp}
    \Gamma_e^\mathrm{AC}\propto \exp(-1/JT) \, ,
\end{equation}
for (quasi-)short-range interacting systems~\cite{Abanin2015_Heating,Mori2016_Heating,pizzi2021classical}; this includes the sign-changing dipole-dipole interactions despite the interactions falling off as $1/r^3$~\cite{Ho18_longrangeheating}.
This is in stark contrast to the polynomial suppression due the Fermi's Golden Rule heating~\eqref{supmat:eq:FGR} in the absence of the AC field; thus, introducing the AC field leads to an exponential increase in the scaling of the lifetime with decreasing period $T$.
The AC induced lifetime enhancement is also supported by numerical simulations, taking into account the full dynamics, see Fig.~\ref{supmat:fig:proof_of_principle}A.

\subsection{Two-tone prethermal discrete time crystal}
\label{supmat:sec:two_tone}

We will now turn to the two-tone DTC. The two-tone DTC without an AC field is described in detail in Ref.~\cite{DTC_Beatrez2022}. Let us only summarize the key aspects here.

In the absence of an AC field, the two-tone DTC drive leads to the unitary evolution per DTC-cycle (period $T$)
\begin{equation}
\label{supmat:eq:two_tone_unitary}
    U_{\mathrm{2DTC}} = U_y U_\mathrm{dd} \prod_{\ell=0}^{N-1} U_x U_\mathrm{dd}\, ,
\end{equation}
with $U_y=\exp[-i\left(\gamma_y I^y + \tau_y \Hdd\right)]$ and $U_x=\exp[-i\left(\theta_x I^x + \tau_x \Hdd\right)]$ and $U_\mathrm{dd}=\exp(-i\tau\Hdd)$; note that, the product, $\prod_{\ell=0}^{N-1} O_\ell$, is time-ordered running from right to left, $\prod_{\ell=0}^{N-1} O_\ell=O_{N-1} O_{N-2} \dots O_{1} O_{0}$
Note that both $\theta_x$ and $\gamma_y$ are of order $O(1)=O(\tau^0,T^0)$, thus, preventing the application of the Baker-Campbell-Hausdorff~(BCH) formula.
Instead, to take care of the strong $\xhat$-pulses we consider a toggling frame expansion, i.e., using $U_x^{-1}U_x=1$ we can rewrite the spin-locking dynamics~(taking away the $\yhat$ pulse) of Eq.~\eqref{supmat:eq:two_tone_unitary} as
\begin{equation}
\label{supmat:eq:toggling}
    \begin{aligned}
     U_y^\dagger U_{\mathrm{2DTC}} 
        &= U_\mathrm{dd} \prod_{\ell=0}^{N-1} U_x U_\mathrm{dd} 
\\
        &= U_\mathrm{dd} U^{N}_x \pqty{\prod_{\ell=0}^{N-1} U^{-\ell}_x U_\mathrm{dd} U^{\ell}_x } 
\\
        &= U^{N+1}_x  \pqty{\prod_{\ell=0}^{N} U^{-\ell}_x U_\mathrm{dd} U^{\ell}_x}
\\
        &= U^{N+1}_x \bqty{\prod_{\ell=0}^{N} \exp(-i \tau \tilde{H}_\mathrm{dd, \ell}) } \,,
    \end{aligned}
\end{equation}
where we introduced $\tilde{H}_\mathrm{dd, \ell}=U^{-\ell}_x H_\mathrm{dd} U^{\ell}_x$ in the last line.
Since, $\norm{\tilde{H}_\mathrm{dd, \ell}}=\norm{\Hdd}$, with respect to the Frobenius norm, all instantaneous generators in the product in the final expression in Eq.~\eqref{supmat:eq:toggling} are of order $\norm{\Hdd}\tau$ allowing for application of the BCH formula, 
\begin{equation*}
    \prod_{\ell=0}^{N} \exp(-i \tau \tilde{H}_\mathrm{dd, \ell}) \approx \exp(-i \tau \sum_{\ell=0}^N \tilde{H}_\mathrm{dd, \ell})
\end{equation*}
up to $O(\tau)$ corrections~\cite{DTC_Beatrez2022}.
One can show that if $N+1{=}4n$~($n\in\mathbb{Z}$, and $\theta_x=\pi/2$ exactly, that the lowest order Floquet Hamiltonian
\begin{equation}
\label{supmat:eq:Hff}
    H_\mathrm{SL} \equiv  \frac{1}{N}\sum_{\ell=0}^N \tilde{H}_\mathrm{dd, \ell} 
     = \frac{1}{2} \sum_{k<\ell} J_{k\ell} \left( 3 I_k^x I_\ell^x - \boldsymbol{I}_k \cdot \boldsymbol{I}_\ell \right) 
\end{equation}
corresponds to the spin-lock Hamiltonian $H_\mathrm{SL}$ that preserves the $I^x$ magnetization, $\comm{H_\mathrm{SL}}{I^x}=0$.

For all other values of $N$ one can show that the emergent $U(1)$-symmetry is broken already in the lowest order $O(T^0)$. However, the explicit symmetry-breaking term is suppressed in $N$ as $O(1/N)$, such that symmetry-breaking effects are still suppressed out to longer times.
Therefore, the spin-locking sequence leads to an emergent $U(1)$-symmetry and the dynamics are well-described by replacing the spin-locking sequence $U_\mathrm{dd} \prod_{\ell=0}^{N-1} U_x U_\mathrm{dd}$ of duration $T_\mathrm{spin-lock}=(N+1)\tau-\tau_x$ by $U_\mathrm{spin-lock}=e^{-i T_\mathrm{spin-lock}H_\mathrm{SL}}$.
Then, the role of the $\yhat$-pulses is the same as in the single-tone DTC case in Sec.~\ref{sec:single_tone}.

The key difference to the single-tone DTC, however, is the fact that the AC-field and DTC-oscillation axes are orthogonal. Thus, the coupling to the DTC order parameter is not immediately evident.
In fact, naively one would expect the AC field to cause either an effective $z$-field that enhances symmetry breaking or be averaged out by the DTC sequence. 
Indeed, in the limit of instantaneous $y$-pulses this is exactly the case, as demonstrated below.

However, in the experiment a significant portion of the period is spent applying the $\yhat$-pulses---the duration $\tau_y$ of the $\yhat$-pulses exceeds the evolution time $\tau$ between pulses, $\tau_y>2\tau$---such that the AC-field also impacts these pulses.
Note that, the period of the AC field is much larger than the duration of the $\yhat$-pulses. 
Therefore, we will assume that the AC field is quasi-constant during the $\yhat$-pulses, such that the dynamics during the pulse are effectively described by 
\begin{equation}
\label{supmat:eq:two_tone_ydrive}
    U_y(n; t) = e^{-i t \left(\gamma_y I^y + (-1)^n \alpha I^z\right)/\tau_y}\, ,
\end{equation}
where $n$ indicates the $n$'th applied $\yhat$-pulse, $\tau_y$ is the length of the $y$-pulse, $0<t<\tau_y$ and $\alpha=\abs{\int \Bac(t) \mathrm{d}t}$ is the accumulated angle in $z$.
For simplicity, we have disregarded the action of the interactions during the $\yhat$ pulses which lead to minor corrections in the final result~(see simulations below).

\begin{figure}[t]
    \centering
    \includegraphics[width=\linewidth]{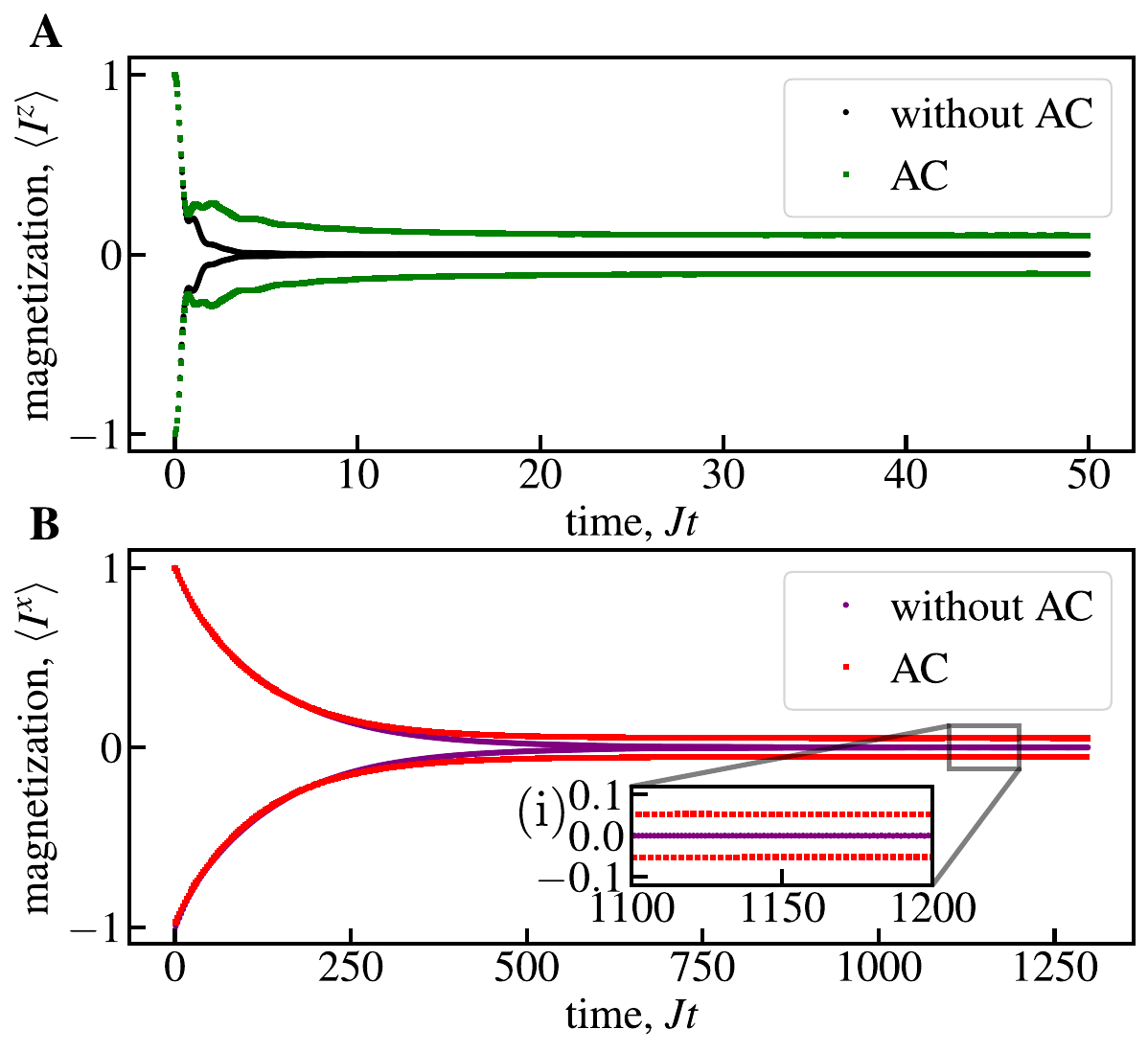}
    \caption{
    \textbf{Numerical simulation of AC induced lifetime enhancement}
    for single-tone~(\textbf{A}) and two-tone~(\textbf{B}) DTC, with $\Phi_\mathrm{AC}=0,\pi/2$, respectively, using the algorithm detailed in Sec.~\ref{supmat:sec:algorithm}. We use $N=16$, $\gamma_y=0.98\,\pi$, $\theta_x=\pi/2$, $J\tau=0.025$, $\tau_y{=}3\tau{=}2\tau_x$ and $\Bac=J/\pi$.
    \textbf{A} Magnetization $\expval{I^z}$ dynamics without~(black) and with~(green) AC field.
    \textbf{B} Magnetization $\expval{I^x}$ dynamics without~(purple) and with~(red) AC field;
        (i) zoom into late time regime.
    For both DTCs the additional AC field leads to an increase in signal and lifetime, in perfect agreement with experiments, see Figs.~\ref{fig:fig1} and \ref{fig:fig4}.
    }
    \label{supmat:fig:proof_of_principle}
\end{figure}

Going to a rotating frame with respect to the strong $I^y$ field and rotating back the dynamics are exactly described by 
\begin{equation}
\label{supmat:eq:two_tone_ydrive_rot}
    U_y(n; t) = e^{-i \gamma_y I^y} \mathscr{T}e^{-i (-1)^n \alpha \int_0^{\tau_y} \frac{\mathrm{d} t}{\tau_y} \left[\cos(\gamma_y t/\tau_{y}) I^z - \sin(\gamma_y t/\tau_{y}) I^x\right] } \, ,
\end{equation}
where $\mathscr{T}$ refers to time-ordering.
To leading order, $O(\frac{\alpha}{\gamma_y})$, we can approximate Eq.~\eqref{supmat:eq:two_tone_ydrive_rot} by 
\begin{equation}
\label{supmat:eq:two_tone_ydrive_eff}
    U_y(n; t) \approx e^{-i \gamma_y I^y} e^{i \frac{(-1)^n \alpha}{\gamma_y} I^x}\, ,
\end{equation}
where $\alpha\propto B_\mathrm{AC} \tau_y$.
Hence, the AC $z$-field on top of the finite time $y$-pulse leads to an effective AC-$x$ field. 
Therefore, the two-tone DTC is formally similar to the single-tone DTC. 
Indeed, in Fig.~\ref{supmat:fig:proof_of_principle}B we also provide numerical evidence for the AC-induced lifetime enhancement, taking into account the full dynamics. 

In contrast, to the single-tone DTC where the $U(1)$-symmetry, $[I^z,\Hdd]=0$, is exact to all orders, for the two-tone DTC the $U(1)$-symmetry is only quasi-conserved, i.e., higher order terms $O(JT)$ break the symmetry. 
Thus, the heating without AC field is expected to follow a power law in the period $\Gamma_e \propto (JT)^2$.
In contrast, via engineering a finite energy density via the AC field, this decay follows the usual exponentially suppressed Floquet heating decay rate $\Gamma_e^\mathrm{AC} \propto \exp(-1/JT)$.
This is the origin of the increase in lifetime, produced by the AC field, and observed in the experiment.
%Moreover, small deviations $\gamma_y=\pi+\epsilon$ may be averaged due to the interspersed spin-locking sequence.

\paragraph*{Limit of instantaneous pulses.}
To emphasize the importance of the finite duration pulses, we demonstrate that in the case of instantaneous pulses, the AC-field exactly cancels out at the resonance condition. 
Specifically, let us consider the fine-tuned case $\theta_x=\pi/2$, $\gamma_y=\pi$, with ideal angle $\Phi_\mathrm{AC}=\pi/2$ and on-resonance condition $\fac=\fres$. 
Moreover, let us take a closer look into the DTC evolution in the presence of an AC field over two cycles:
\begin{align*}
    U_\mathrm{2DTC}^2 
            &= U_y U_\mathrm{dd} U_{z,2N+2} \left( \prod_{\ell=1}^{N} U_x U_\mathrm{dd} U_{z,N+1+\ell} \right) 
        \\
            &\hspace{1cm} \cdot U_y U_\mathrm{dd} U_{z,N+1} \left( \prod_{\ell=1}^{N} U_x U_\mathrm{dd} U_{z,\ell} \right)
        \\
            &= U_y \left( \prod_{\ell=0}^{N} U^\ell_x U_\mathrm{dd} U_{z,N+2+\ell} U^{-\ell}_x \right)         
        \\
            &\hspace{1cm} \cdot U_y \left( \prod_{\ell=0}^{N} U^\ell_x U_\mathrm{dd} U_{z,\ell+1} U^{-\ell}_x \right)
        \\ 
            &= U_y \left( \prod_{\ell=0}^{N} \tilde{U}_{\mathrm{dd},\ell} \tilde{U}_{z,N+2+\ell} \right)
            U_y \left( \prod_{\ell=0}^{N} \tilde{U}_{\mathrm{dd},\ell} \tilde{U}_{z,\ell+1} \right)
        \\
            &\overset{O(T^0)}{\approx} U_y \left(\prod_{\ell=0}^{N} \tilde{U}_{\mathrm{dd},\ell}\right) \left(\prod_{\ell=0}^{N} \tilde{U}_{z,N+2+\ell}\right)
        \\
            &\hspace{1cm} \cdot U_y \left(\prod_{\ell=0}^{N} \tilde{U}_{\mathrm{dd},\ell}\right) \left(\prod_{\ell=0}^{N} \tilde{U}_{z,\ell+1}\right)
        \\
            &\overset{\eqref{supmat:eq:Hff}}{\approx} U_y e^{-i T_\mathrm{spin-lock} H_\mathrm{SL}} \left(\prod_{\ell=0}^{N} \tilde{U}_{z,N+2+\ell}\right)
        \\
            &\hspace{1cm} \cdot U_y e^{-i T_\mathrm{spin-lock} H_\mathrm{SL}} \left(\prod_{\ell=0}^{N} \tilde{U}_{z,\ell+1}\right)
        \\
            &\overset{O(T^0)}{\approx} e^{-i 2 T_\mathrm{spin-lock} H_\mathrm{SL}} U_y \left(\prod_{\ell=0}^{N} \tilde{U}_{z,N+2+\ell}\right) U_y \left(\prod_{\ell=0}^{N} \tilde{U}_{z,\ell+1}\right) \, ,
\end{align*}
where the definitions of $U_{z,\ell}$ are given below, and in the last line we used that $\comm{U_y}{H_\mathrm{SL}}=0$ for $\gamma_y=\pi$.

Hence, in summary, to lowest order $O(T^0)$ the interaction and single-particle fields decouple
\begin{equation}
\label{supmat:eq:twotone_twice}
    U_\mathrm{2DTC}^2 \approx e^{-i 2 T_\mathrm{spin-lock}} U_\mathrm{sp}^2 \,,
\end{equation}
with single-particle unitary
\begin{equation}
\label{supmat:eq:twotone_inst}
    U_\mathrm{sp}^2 = U_y U_{z,2N+2} \left( \prod_{\ell=1}^{N} U_x U_{z,N+1+\ell} \right) U_y U_{z,N+1} \left( \prod_{\ell=1}^{N} U_x U_{z,\ell} \right)\, ,
\end{equation}
where $U_y=\exp(-i \gamma_y I^y)$, $U_x=\exp(-i \theta_x I^x)$, and $U_{z,\ell}=\exp(-i B_\ell I^z)$ with integrated field $B_\ell=\int^{\ell \tau}_{(\ell-1) \tau} \Bac(t) \mathrm{d} t$.
Therefore, in the following we can focus on the single-particle contributions only, introducing the interactions only in the end.

Note that, (i) by symmetry of the cosine function we have $B_{N+1+\ell}=B_{N+1-\ell}$ and (ii) for $\gamma_y=\pi$ we have $U_y f\left(I^x, I^z\right) U_y = U_y^2 f\left(-I^x, -I^z\right)$ for any function $f$, i.e., in particular $U_y U_{x,z} U_y = U_y^2 U_{x,z}^\dagger$.
Thus, we can rewrite Eq.~\eqref{supmat:eq:twotone_inst} as 
\begin{align*}
    U^2_\mathrm{sp} &= U^2_y U^\dagger_{z,1} \left( \prod_{\ell=1}^{N} U^\dagger_x U^\dagger_{z,N+1-\ell} \right)  U_{z,N+1} \left( \prod_{\ell=1}^{N} U_x U_{z,\ell} \right)
    \\
        &= U^2_y U^\dagger_{z,1} \left( U^\dagger_x U^\dagger_{z,2} \dots U^\dagger_x U^\dagger_{z,N} U^\dagger_x \underline{U^\dagger_{z,N+1}}\right) \underline{U_{z,N+1}} 
        \\ & \hspace{1cm} \cdot \left( U_x U_{z,N} \dots U_x U_{z,2} U_x U_{z,1}  \right)
    \\
        &= U^2_y U^\dagger_{z,1} U^\dagger_x U^\dagger_{z,2} \dots U^\dagger_x U^\dagger_{z,N} \underline{ U^\dagger_x  U_x} U_{z,N} \dots U_x U_{z,2} U_x U_{z,1}
    \\
        &= \dots
    \\
        &= U^2_y = -1 \, ,
\end{align*}
where we repeatedly apply the unitarity of $U_x$ and $U_{z,\ell}$, i.e., $U_x^\dagger U_x=1=U_{z,\ell}^\dagger U_{z,\ell}$.
Therefore, by Eq.~\eqref{supmat:eq:twotone_twice}, the full two-cycle DTC evolution is given by
\begin{equation}
\label{supmat:eq:twotone_full_inst}
    U_\mathrm{2DTC}^2 \approx - e^{-i 2 T_\mathrm{spin-lock}} \,,
\end{equation}
to lowest order $O(T^0)$, which is equivalent to the two-cycle unitary in the absence of an AC field $\Bac=0$.

\begin{figure*}[t]
    \centering
    \includegraphics[width=\linewidth]{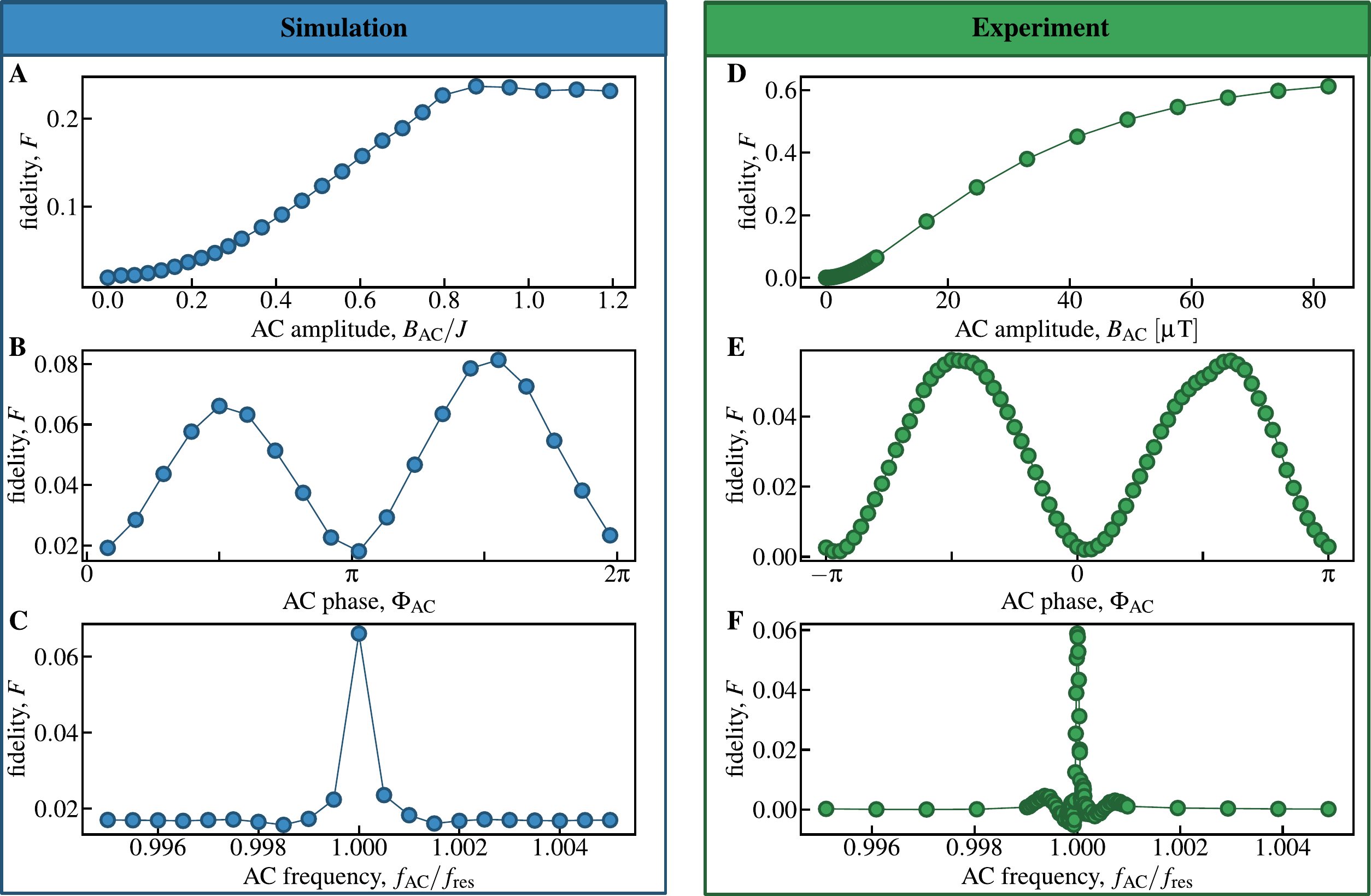}
    \caption{
    \textbf{Numerical simulation of properties of AC induced signal enhancement}
    for two-tone DTC. We depict the dependence of the fidelity metric on the properties of AC drive:
    (\textbf{A}), amplitude $\Bac$ sweep. The fidelity metric increases monotonically with the AC amplitude $\Bac$.
    (\textbf{B}), phase $\Phi_\mathrm{AC}$ sweep, $\Phi_\mathrm{AC}$ measured as shown in Fig.~\ref{fig:fig2}A. data shows a roughly $\sin^2$-like dependence on the phase with maximal signal achieved around $\Phi_\mathrm{AC}=\pi/2,3\pi/2$.
    (\textbf{C}), frequency $\fac$ sweep. The signal enhancement is a strongly resonant effect around the intrinsic DTC frequency $\fres$.
    (\textbf{D}-\textbf{F}), are the experimental analogs for (\textbf{A}-\textbf{C}); experimental data is same as shown in main text Fig.~\ref{fig:fig2} and Fig.~\ref{fig:fig3}.
    The simulated results are in qualitative agreement with the experimental results.
    Static parameters for simulation are as in Fig.~\ref{supmat:fig:proof_of_principle}, and experimental details are found in in main text Fig.~\ref{fig:fig2} and \ref{fig:fig3}.
    }
    \label{supmat:fig:thy_properties}
\end{figure*}

\subsection{Summary of AC-induced signal enhancement}
\label{supmat:sec:summary_engineering}

While the analysis above focused on dipolar interacting spins in a $3$D system driven by a specific Floquet-sequence, the results apply more broadly. 
For example, the key role of interactions is to drive thermalization in agreement with the eigenstate thermalization hypothesis~(ETH); however, this may also be achieved by considering a thermodynamic ensemble or by using open systems.

In a nutshell, the key ingredients are:
(i) (emergent) symmetry-protected period doubling response;
(ii) the system should (pre-)thermalize in agreement with ETH,
(iii) high-temperature initial state such that lifetime is limited by symmetry-breaking terms;
and 
(iv) ability to (effectively) couple the system to the DTC order parameter, e.g., via Floquet engineering.
Note that, since the DTC order parameter oscillates in time the coupling-inducing term will generally be time-dependent as well.
Then, by adding this time-varying coupling one can exponentially extend the lifetime of the PDTC order using the procedure presented above.

Let us further point out that the exponential suppression of Floquet heating in the drive frequency $1/T$ only applies to short-range and effectively short-range interacting systems~\cite{Abanin2015_Heating,Mori2016_Heating}. 
Note that in general the critical exponent for effective short-range interactions is $\alpha=d$, where the interactions scale as $J\propto 1/r^\alpha$ and $d$ is the spatial dimension. Hence, our dipole-dipole interacting system is critical long-range interacting and not short-range interacting.
However, it was shown that for sign-changing interactions the critical exponent is indeed $\alpha=d/2$~\cite{Ho18_longrangeheating}.
Thus, the sign-changing dipole-dipole interacting system behaves as a short-range system for the purpose of Floquet heating, i.e., displaying exponential suppression of Floquet heating in the drive frequency. 

Finally, let us emphasise that in order to observe the scaling with period $T$ one has to (i) reduce the pulse-durations $\tau_y$ and $\tau_x$ accordingly, while (ii) not only keeping the accumulated angles $\gamma_y$ and $\theta_x$ fixed but also decrease their deviations from the ideal values $(\gamma_y-\pi)$, $\theta_x\propto T$.
The required accuracy and high-power during the pulses makes observing the exponential scaling in the current experimental setup technically challenging.

\subsection{AC vs DC field comparison}
\label{supmat:sec:dc_vs_ac}
Let us emphasize that the idea of using finite energy density to stabilize $U(1)$-DTC order is not new to this work. In Ref.~\cite{PrethermalWithoutTemperature_LuitzEtAl2020}, the authors proposed to stabilize a sequence similar to the single-tone DTC above by adding a DC field, $B_\mathrm{DC}=h I^z$ with strength $h$, in the time-window where no $\yhat$ pulses are applied. 
An advantage of the DC field is that, depending on the strength of the field one can, (i) engineer a finite energy density~($h\tau\approx\pi/2$) similar to the procedure in this work or (ii) restore the broken $U(1)$-symmetry~($h\tau=\pi$) by averaging out errors in $\gamma_y$.
However, away from $h\tau=n\pi/2$ ($n\in\mathbb{Z}$) the $\gamma_y\approx\pi$ pulses `echo-out' the DC field, thus, requiring the persistent application of strong magnetic fields which can be experimentally challenging, due to the high-power required and the potential heating of the sample. 

In contrast, by applying an AC field the `echoing-out' is avoided, such that lifetime enhancement occurs for any finite value of the AC amplitude. This not only provides a practical advantage but forms the basis of the AC sensing application.
Moreover, the AC sensing scheme generalizes to the two-tone DTC; a similar extension for the DC protocol is not immediately clear, since, the additional $\theta_x=\pi/2$ spin-locking pulses would 'echo-out' the $\zhat$-DC field and persistently applying an $\xhat$-DC field would interfere with the signal readout.

%\mb{state why, otherwise prepare to answer Ref's q's on that ;)}.
%\mb{any advantages of AC over DC besides practical limitations in our specific system?}

% \begin{figure}
%     \centering
%     \includegraphics[width=\linewidth]{propertiesV1.pdf}
%     \caption{
%     \textbf{Numerical simulation of properties of AC induced signal enhancement}
%     for two-tone DTC. We depict the dependence of the fidelity metric on the properties of AC drive:
%     (\textbf{A}), amplitude $\Bac$ sweep. The fidelity metric increases monotonically with the AC amplitude $\Bac$.
%     (\textbf{B}), phase $\Phi_\mathrm{AC}$ sweep, $\Phi_\mathrm{AC}$ measured as shown in Fig.~\ref{fig:fig2}A. data shows a roughly $\sin^2$-like dependence on the phase with maximal signal achieved around $\Phi_\mathrm{AC}=\pi/2,3\pi/2$.
%     (\textbf{C}), frequency $\fac$ sweep. The signal enhancement is a strongly resonant effect around the intrinsic DTC frequency $\fres$.
%     The simulated results are in qualitative agreement with the experimental results, Fig.~\ref{fig:fig2} and Fig.~\ref{fig:fig3}.
%     %
%     Static parameters are as in Fig.~\ref{supmat:fig:proof_of_principle}.
%     }
%     \label{supmat:fig:thy_properties}
% \end{figure}

\section{Characteristics of AC enhanced signal}
\label{supmat:sec:thy_properties}

\begin{figure}
    \centering
    \includegraphics[width=\linewidth]{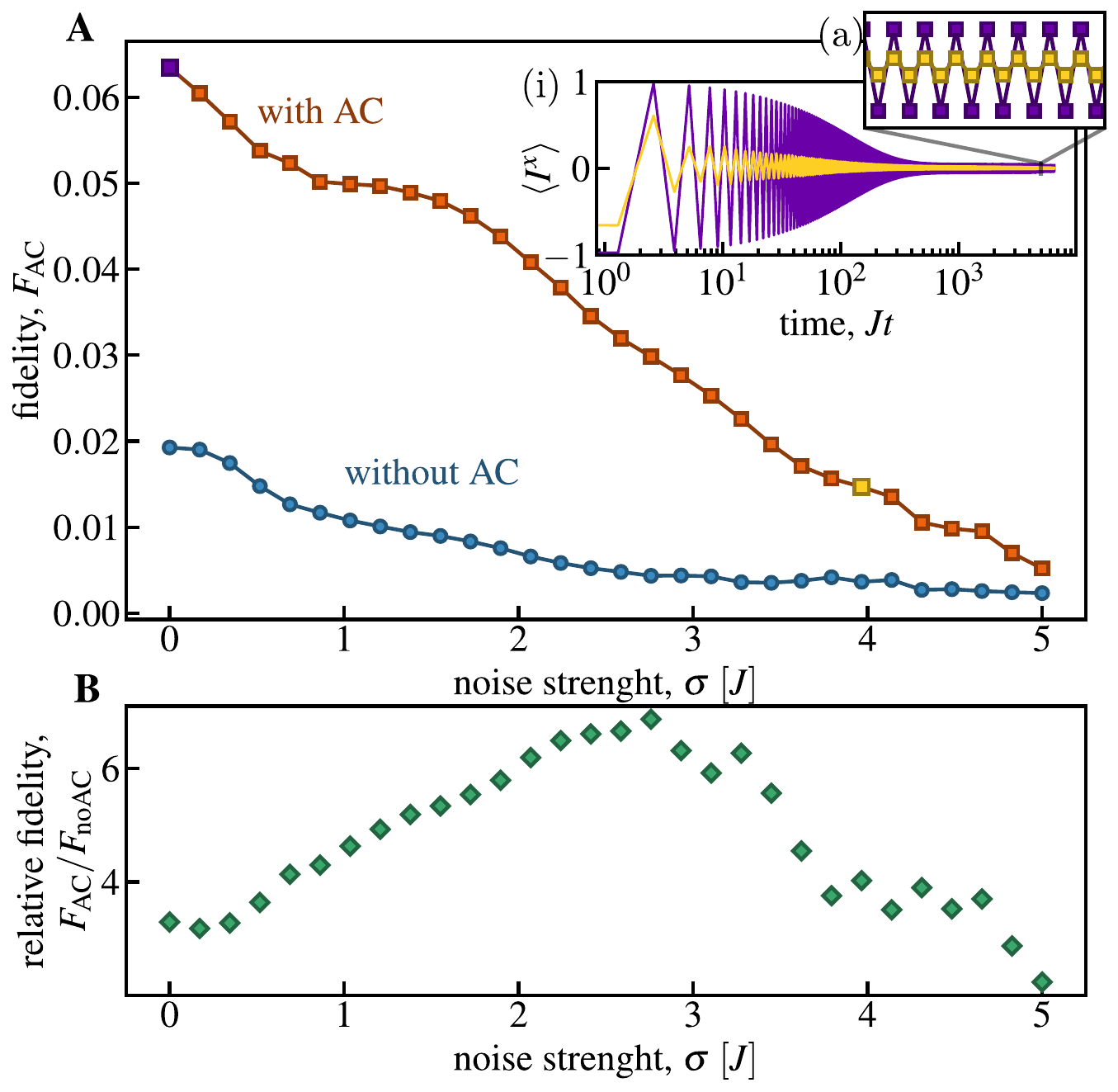}
    \caption{
    \textbf{Numerical simulation of noise resilience of AC induced signal enhancement}
    for two-tone DTC.
    \textbf{A}, fidelity metric $F$ for increasing on-site noise strength $\sigma$ for DTC without~(blue circles) and with~(orange squares) AC field, see Sec.~\ref{supmat:sec:noise} for definition of noise. 
    The noise strength reaches up to five times the strength of the median coupling.
    (i), DTC dynamics for specific points indicated in A, representing no noise~(purple) and large noise~(yellow), $\sigma\approx4\,J$. (a), zoom into late-time dynamics; even at large noise levels the period doubling dynamics remain observable, however, with decreased amplitude.
    \textbf{B}, Relative increase in fidelity comparing the case of additional and no AC field.
    The prethermal DTC order with and without AC field, is resilient to strong levels of on-site disorder; i.e., period doubling dynamics persist although with reduced amplitude.
    Notably, for moderate noise levels, the AC enhanced DTC is less affected by disorder resulting in an enhanced relative signal with increasing disorder.
    Other parameters are as in Figs.~\ref{supmat:fig:proof_of_principle}.
    }
    \label{supmat:fig:thy_noise}
\end{figure}

From the theoretical derivation in Sec.~\ref{supmat:sec:two_tone} we concluded that the AC field leads to an additional $I^x$-magnetization in the effective Hamiltonian $H_\mathrm{eff}=H_\mathrm{SL}+B_\mathrm{eff}I^x$ which causes prethermalization to a finite energy and magnetization state that is protected during the prethermal plateau.
To lowest order, the effective field is given by integrating the AC field during the y-pulse, $B_\mathrm{eff}=\int_{t_0}^{t_0+\tau_y} \Bac(t) \mathrm{d}t/\gamma_y$.
As we show below, this is sufficient to qualitatively explain the observed response of the DTC to different properties of the AC field $\Bac(t)=\Bac \sin(2\pi \fac t + \Phi_\mathrm{AC})$, i.e., the phase $\Phi_\mathrm{AC}$, amplitude $\Bac$ and frequency $\fac$ of the AC drive.
We support our theoretical findings with numerical simulations, see Fig.~\ref{supmat:fig:thy_properties}A-C.

To this end, let us explicitly derive the magnetization in the prethermal plateau. 
Assuming a high-temperature state, $J\mathcal{T}{\gg} 1$, we can write the prethermal state as $\rho_\mathcal{T}\approx 1- H_\mathrm{eff}{/}\mathcal{T}$. Then, the inverse temperature can be determined by quasi-energy conservation as $\expval{H_\mathrm{eff}}_{\rho_0}/L= \mu B_\mathrm{eff}\overset{!}{=} \expval{H_\mathrm{eff}}_{\rho_\mathcal{T}} = -\left[ B_\mathrm{eff}^2 + \norm{H_\mathrm{SL}}^2 \right]/\mathcal{T}$, hence, 
\begin{equation}
\label{supmat:eq:temperature}
    \mathcal{T}^{-1} {=} -\frac{\mu B_\mathrm{eff}}{B_\mathrm{eff}^2 + J^2_\mathrm{spin-lock}} \,,
\end{equation}
where $\mu$ is the magnetization of the initial state and where we defined $J^2_\mathrm{spin-lock}=\norm{H_\mathrm{SL}}^2$ which measures the effective interaction strength of the spin-locking Hamiltonian.
Consequently, the magnetization per spin in the prethermal plateau is given by
\begin{equation}
\label{supmat:eq:magnetization}
    \expval{I^x}_{\rho_\mathcal{T}}/L {\approx} - \mathcal{T}^{-1} B_\mathrm{eff} = \mu \frac{B^2_\mathrm{eff}}{B_\mathrm{eff}^2 + J^2_\mathrm{spin-lock}}\, .
\end{equation}
Note that, the Floquet heating dynamics after prethermalization correspond to an exponential decay of the magnetization magnitude in the prethermal plateau to zero, $\abs{\expval{I^x(t=nT)}}\sim e^{-\Gamma^\mathrm{AC}_e t}\abs{\expval{I^x}_{\rho_\mathcal{T}}}$.
This allows for a simple relationship between the fidelity metric and magnetization in the prethermal plateau in the resonant case:
there, the fidelity metric corresponds to time-integrating the absolute value of the signal, i.e., $F\propto \sum_{n=0}^N (-1)^n \expval{I^x(nT)} = \sum_n \abs{\expval{I^x(nT)}}\sim\expval{I^x}_{\rho_\mathcal{T}} \sum_n e^{-\Gamma^\mathrm{AC}_e n T}$. 
Thus, the fidelity metric is directly proportional to the magnetization in the prethermal plateau, i.e., $F=c \expval{I^x}_{\rho_\mathcal{T}}+F_0$ with some proportionality constant $c$ and a shift $F_0$ originating from the pre-thermalization dynamics.

%\mb{a bit of a confusing part below is that u cross-refer to Fig S10 and Figs 2/3 and it causes confusion; talk to Leo to make a single back-to-back comparison for SI if time permits}

\subsection{Amplitude}
Note that, the effective field is directly proportional to the AC field amplitude $B_\mathrm{eff}\propto \Bac$.
Further, notice that for small AC, and hence small effective fields $B_\mathrm{eff}$~($B_\mathrm{eff}{\ll} J_\mathrm{spin-lock}$), the magnetization~\eqref{supmat:eq:magnetization} increases quadratically with the field strength $\expval{I^x}_{\rho_\mathcal{T}}\propto B_\mathrm{eff}^2$.
This is in excellent agreement with the experimental~(Fig.~\ref{supmat:fig:thy_properties}D) and numerical results~(Fig.~\ref{supmat:fig:thy_properties}A).

With increasing field strength the ($B_\mathrm{eff}{\geq} J_\mathrm{spin-lock}$) the magnetization increase stalls and converges to a finite value, namely its initial value $\expval{I^x}_{\rho_\mathcal{T}} \overset{B_\mathrm{eff}\to \infty}{\longrightarrow} \mu L$.
This trend is in agreement with the observed experimental and theoretical results.
However, note that the slowing down observed in the numerical simulation, see Fig.~\ref{supmat:fig:thy_properties}A, is stronger than the experimentally observed, see Fig.~\ref{supmat:fig:thy_properties}D, and analytically predicted form; 
this is likely a finite system artefact caused by the field becoming comparable to the spectral width of the system.

\subsection{Phase}
Notice that if the AC field averages to zero during the $\yhat$-pulses the effective field vanishes $B_\mathrm{eff}=0$; thus, the dynamics agree with those in the absence of the AC field.
Further, assuming a quasi-constant AC field the effective field is given by $\abs{B_\mathrm{eff}} = \abs{\sin(\Phi_\mathrm{AC})} \Bac \tau_y/\gamma_y$ which takes its largest value around $\Phi_\mathrm{AC}=\pm \pi/2$ and vanishes around $\Phi_\mathrm{AC}=0,\pi$ in agreement with the strongest and weakest signal observed in the experiment, Fig.~\ref{fig:fig2}A.
Moreover, the functional dependence of the magnetization in the weak field regime, $\expval{I^x}_{\rho_\mathcal{T}}\propto B_\mathrm{eff}^2 \propto \sin^2(\Phi_\mathrm{AC})$, matches well in lowest order with experimental~(Fig.~\ref{supmat:fig:thy_properties}E), and numerical result~(Fig.~\ref{supmat:fig:thy_properties}B).

\subsection{Frequency}
The dependence on the frequency $\fac$ can be understood as follows. 
For simplicity, we focus on the regime $\Phi_\mathrm{AC}=\pi/2$.
In the far off-resonant regime the accumulated AC field during the $\yhat$-pulses oscillates wildly, thus averaging to zero over a few cycles.
In the near resonance regime, i.e., when the difference in frequency is small $\delta f {=} \abs{\fac-\fres}\ll \fac$, we may assume a separation of time-scales $\Bac(t) = \cos(2\pi \fac t ) = \cos(2\pi \fres t ) \cos(2\pi \delta f t ) + \sin(2\pi \fres t ) \sin(2\pi \delta f t ) $.
As we have seen above, the sine contribution can be neglected as it integrates to zero during the $\yhat$-pulses. 
Thus, the field is given by $\Bac(t) \sim \cos(2\pi \delta f t) \cos(2\pi \fres t)$, and hence the effective field attains a slowly varying component $B_\mathrm{eff}(t) = \cos(2\pi \delta f t) B_\mathrm{eff}$.
If the field varies slowly enough, the magnetization follows the change in external field adiabatically without changing the effective temperature, $\expval{I^x}_{\rho_\mathcal{T}}(t)\approx - \mathcal{T}^{-1} B_\mathrm{eff}(t)$, thus, leading to the observed beating, in agreement with the experiment~Fig.~\ref{supmat:fig:thy_properties}F and simulations~Fig.~\ref{supmat:fig:thy_properties}C.
As discussed in the main text, the narrow linewidth is thus a result of time-integrating the signal over the lifetime $T_2^\prime$.

\subsection{Noise resilience}
\label{supmat:sec:noise}

We use our numerical simulations to explore the robustness of the AC enriched PDTC towards errors in the pulse sequence and local fields on the spins. 
This is a common source of noise in experimental setups, due to spatial inhomogenities in the magnetic field and imperfections in tuning the $\yhat$ and $\xhat$ pulses.
Both the PDTC order and thermalization are expected to be resilient towards these errors, leading to enhanced sensing capabilities as imperfections in the sensor do not reduce sensitivity.
While the lack of local control in the experimental apparatus prevents a detailed exploration of this robustness, we can use our numerical simulations to investigate resilience of the sensing protocol.

Specifically, we consider constant-in-time but spatially-varying errors on the $\xhat$ and $\yhat$ pulses, as well as additional on-site fields in $\zhat$-direction, i.e., we replace $\theta_x I^x \to \sum_\ell (\theta_x + \tau_x \chi_\ell) I_\ell^x$, $\gamma_y I^y \to \sum_\ell (\gamma_y + \tau_y \eta_\ell) I_\ell^y$ and $\Hdd \to \Hdd + \sum_\ell \zeta_\ell I_\ell^z$ where $\chi_\ell, \, \eta_\ell,\, \zeta_\ell$ are uniformly distributed numbers in $\left[-\sigma/2, +\sigma/2\right]$.

We find a resilience of the response up to noise strengths exceeding the dipolar couplings strength, i.e., $\sigma>J$, see Fig.~\ref{supmat:fig:thy_noise}.
In fact, for moderate noise strength~($\sigma\leq3\,J$), the AC enriched DTC seems more robust than the DTC without an AC field, thus leading to relative increase in fidelity metric with increasing noise.
However, the overall decrease in signal will result in a smaller $T_2^\prime$ and, thus, less narrow linewidth.
In summary, the numerical simulations suggest that moderate levels of disorder in the system have marginal effects on the observed PDTC order.

\begin{figure}[t]
    \centering
    \includegraphics[width=\linewidth]{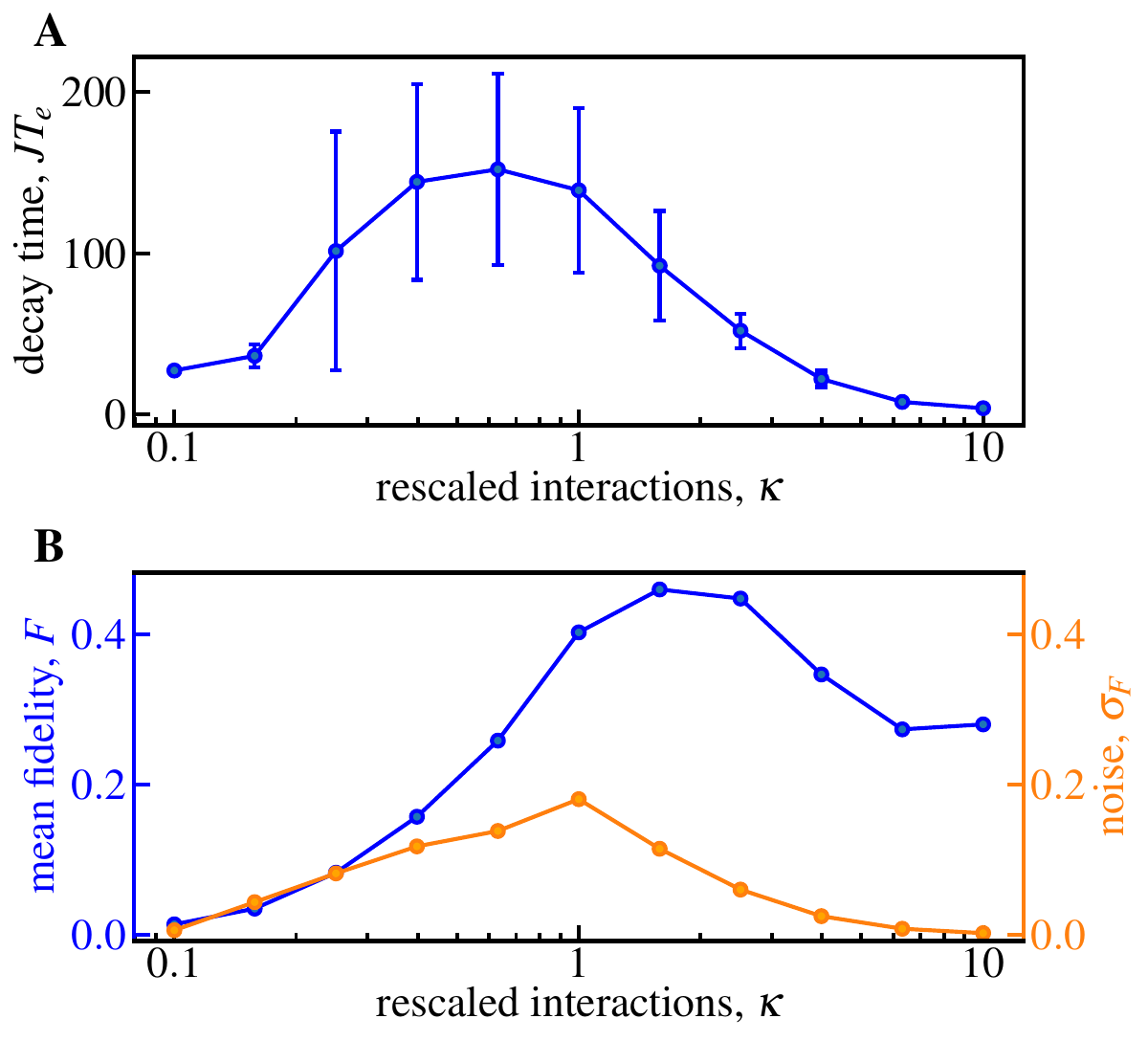}
    \caption{
    \textbf{Simulated scaling of lifetime and fidelity with spin density without AC field.}
    We simulate the quantum dynamics as described in Sec.~\ref{supmat:sec:algorithm} for $L=15$ spins for various rescalings of the median coupling $J\mapsto \kappa J$, corresponding to a change in spin-density $n\mapsto \kappa n$.
    \textbf{A}, $1/e$-lifetime, $T_e$, for different interaction strengths show non-monotonous behavior, peaking at finite $\kappa\neq0,\infty$~(here around $\kappa\approx0.6$); error-bars indicate standard deviation of lifetime over different seeds, estimated from the standard deviation of the signal $S$ via $\sigma_{T_e} = \dot{S} \sigma_S\mid_{t=T_e}$.
    \textbf{B}, fidelity metric $F$~(blue) and its standard deviation $\sigma_F$~(orange) computed using Eq.~\eqref{eq:fidelity_metric}; to obtain the result at fixed volume, from the fixed system size computation, we performed the rescaling $F = \kappa F_{L=15}$ and $\sigma_F = \sqrt{\kappa} \sigma_{F,\,L=15}$.
    Analogously to the decay time, the mean fidelity and its standard deviation show a non-monotonous behavior, peaking at a finite value of $\kappa\neq0,\infty$; notably, all three quantities peak at different values of $\kappa$.
    We choose $N=16$, $\gamma=0.98\pi$ and average over $10$ random graphs; other parameters as in Figs.~\ref{supmat:fig:proof_of_principle}.
    }
    \label{fig:scaling_spindensity}
\end{figure}

% \begin{figure}[t]
%     \centering
%     \includegraphics[width=\linewidth]{sensitivity_v1.pdf}
%     \caption{
%     \textbf{Simulated scaling of sensitivity with spin density.}
%     %
%     \textbf{A}, response $\partial F/\partial \Bac$ of the fidelity $F$ towards changes in the AC amplitude $\Bac$ for different $\kappa$; derivative is estimated via a second order accurate finite difference from $\Bac/J=0,0.1,0.2$. 
%     %
%     \textbf{B}, sensitivity
%     %
%     All data is normalized by the results at $\kappa=1$. Parameters are as in Fig.~\ref{fig:scaling_spindensity}.
%     }
%     \label{fig:senstivity_spindensity}
% \end{figure}

\begin{figure}[t]
    \centering
    \includegraphics[width=\linewidth]{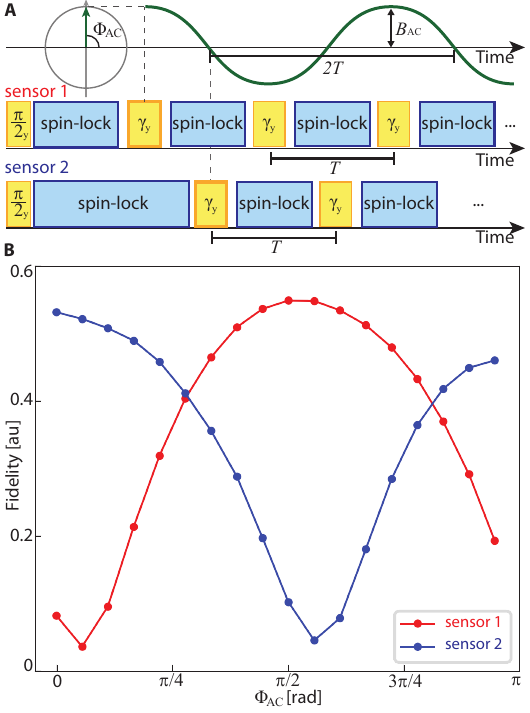}
    \caption{\textbf{Schematic for detecting field with unknown phase using two (or multiple) sensors.}, \textbf{A, } Two-sensor sensing schematic. An AC field of known frequency ($(2T)^{-1}$), but unknown phase ($\Phi_{\text{AC}}$) and amplitude ($B_{\text{AC}}$) is sensed by two sensors (sensor 1 and sensor 2). The two sensors have $\gamma_{y}$ pulses are offseted by $T/2$ to leverage the fidelity's characteristic to uniquely determine $B_{\text{AC}}$. \textbf{B, } Two sensors' response depending on the phase of the AC-field ($\Phi_{\text{AC}}$). Sensor 1 and sensor 2 are initialized such that the $\gamma_{y}$ pulses are offset by $T/2$. The response of sensor 1 (shown in red) and sensor 2 (shown in blue), depends on the phase of the AC-field ($\Phi_{\text{AC}}$). Here, $B_{\text{AC}} = 41.2 \mu T$, $T \approx 517.6 \mu s$, $\gamma_{y}$ pulses ($\approx 112.5\mu\text{s}$), $\theta_{x}$ pulses ($\approx 56.2\mu\text{s}$), and pulse separation $\tau$ ($\approx 36.0\mu\text{s}$).} 
    \label{fig:sensor_offset}
\end{figure}

\subsection{Dependence on spin density}
\label{supmat:sec:density}

%\paragraph*{Introduction}
A key characteristic of the \C-nuclear spin sensor is the abundance of \C, i.e., the density of nuclear spins, $n$.
While we use a diamond with natural abundance~(${\approx} 1\,\%$) of \C~in all our experiments, further enrichment is technically possible, providing another parameter to tune the properties of our sensing scheme.
Let us emphasize that estimating the impact of the nuclear spin density on the DTC sensing scheme is challenging due to the interplay of multiple factors, specifically:

\begin{enumerate}
    \item increasing the density of nuclear spins by a factor $\kappa$, $n \mapsto \kappa n$ while considering the same sample size increases the net signal by the same factor, $S\mapsto\kappa S$;
    \item at the same time, the interspin interaction increases by the same factor, $J\mapsto \kappa J$, which reduces the decoherence time, $T_2\mapsto T_2/\kappa$, and will also affect the Floquet heating rate. While the precise scaling dependence of the heating rate on the relative frequency, $\omega/J$, is not known, previous experiments are consistent with a power law $T_2^\prime \propto \pqty{\omega/J}^2$~\cite{Beatrez21_90s,DTC_Beatrez2022}, hence, $T_2^\prime \mapsto T_2^\prime / \kappa^2$, if the lifetime is limited by Floquet heating;
    \item conversely, if the lifetime is limited by noise, the relative noise strength $\sigma/J$ decreases, $\sigma/J \mapsto \kappa^{-1} \sigma/J$, extending the lifetime and increasing the fidelity, see also Sec.~\ref{supmat:sec:noise};
    \item additionally, the relative AC amplitude also decreases, $\Bac/J \mapsto \kappa^{-1} \Bac/J$, naively reducing the response of the DTC sensor to the same change in AC amplitude.
\end{enumerate}

Moreover, the spin density may impact the hyperpolarization mechanism and decoherence induced via coupling to $P1$ and NV$^-$-centers; however, precisely how increasing the spin density will affect either mechanism is unclear, since both are difficult to describe quantitatively from first principles.
Regardless, we expect the sensitivity to be minimal in both extremes of non-interacting and infinitely strongly interacting nuclear spins; in the former case, there is no interaction-induced stability of the DTC response away from $\gamma_y=\pi$, and in the latter case strong Floquet heating will quickly melt the DTC order.
However, away from these extreme cases the sensitivity increases and peaks at some finite interaction strength $J^\star$; the precise value of this sweet spot will depend on the precise details of the sensor.

\paragraph*{Numerical Analysis of the spin density.}
Since a thorough experimental analysis of the role of nuclear spin density on the sensing capabilities is technically challenging, instead, we focus on a numerical analysis.
Specifically, as our numerical analysis is constrained to small system sizes, we consider a fixed number of spins~($L=15$) and study the dependence of key characteristics of the experiment on the median coupling strength $J \mapsto \kappa J$, with scaling parameter $\kappa$. This corresponds to changing the density and volume of the system simultaneously, $n\mapsto \kappa J$ and $V \mapsto V/\kappa$.
To obtain the scaling for fixed volume we consider the extrapolation of the signal to $S\mapsto \kappa S$. To rescale the standard deviation of the signal, we note that the main source of noise in the simulation is the fluctuation over different samples of random graphs; therefore, we expect a self-averaging effect and hence a scaling of the standard deviation according to $\sigma_S \mapsto \sqrt{\kappa} \sigma_S$.
This is a major difference compared to the experimental apparatus, where we believe the noise to be dominated by the readout circuit; hence, in the experiment, scaling the signal $S$ scales the signal-to-noise ratio~(SNR) by the same factor.
Therefore, we generally expect better sensitivity scaling compared to the simulations below, at least up to a certain threshold where the SNR becomes limited by shot noise.

In Fig.~\ref{fig:scaling_spindensity}A and B, we report on the simulated dependence of the DTC lifetime and fidelity metric, respectively, as a function of the interaction rescaling $\kappa$ in the absence of any AC field; note that, $\kappa=1$ corresponds to the case studied in the previous sections.
In good agreement with the quantitative analysis above, we find that the lifetime decreases drastically for very weak~($\kappa\ll1$) and very strong interactions~($\kappa\gg1$), peaking at some intermediate value~(here $\kappa\approx0.6$) which strongly depends on the details of the sequence, sich as the noise strength, the pulses angle $\gamma$ and the spin-locking angle $\theta_x$, as well as the duration of the pulses.
We observe a similar behavior for the fidelity measure and the noise of the fidelity measure.
However, the maxima of these different quantities do not coincide.
As all three quantities, lifetime, signal, and noise, combined, determine the functionality of the sensor, finding the optimal operation point with respect to spin density is a complex task and will depend on the experimental details of the sensor.
% \\
% In Fig.~\ref{fig:senstivity_spindensity}A and B, we analyze the fidelity metric in the presence of the AC for varying spin densities and extract a simulated sensitivity, respectively.

% Use the widetext environment to span both columns
\lemma{
\textbf{Lemma}: Consider an AC-field with unknown amplitude $B^{(1)}$ and phase $\Phi^{(1)}$ is applied to the DTC sensor. We show that:
\textit{the fidelity values $F(B_{\text{AC}} = B^{(1)}, \Phi_{\text{AC}} = \Phi^{(1)})$ and $F(B_{\text{AC}}= B^{(1)}, \Phi_{\text{AC}} = \Phi^{(1)} +\pi/2)$ uniquely determine the amplitude $B^{(1)}$ and the phase $\Phi^{(1)}$ (up to [0, $\pi/2$]) of the AC field.
}

\smallskip
Let $B^{(1)}$ and $\Phi^{(1)}$ be the amplitude and the phase of the unknown AC magnetic field.

The fidelity of the discrete time crystal with an AC magnetic field is denoted as $F(B_{\text{AC}}, \Phi_{\text{AC}})$, where $B_{\text{AC}}$ is the amplitude and $\Phi_{\text{AC}}$ is the phase of the field.

$F(B_{\text{AC}}, \Phi_{\text{AC}})$ has two properties.

\begin{enumerate}
\item[{[A]}] $\forall\Phi_{\text{AC}} \in [0, \pi]$, $F(B_{\text{AC}}, \Phi_{\text{AC}})$ increases monotonically as $B_{\text{AC}}$ increase;

\item[{[B]}] $\forall B_{\text{AC}}$, $F(B_{\text{AC}}, \Phi_{\text{AC}})$ monotonically increases as $\Phi_{\text{AC}}$ increases for $\Phi_{\text{AC}} \in [0, \pi/2]$ and monotonically decreases as $\Phi_{\text{AC}}$ increases for $\Phi_{\text{AC}} \in [\pi/2, \pi]$.

\end{enumerate}

Properties [A] and [B] hold because the pulse overlap between the AC magnetic field and the $\gamma_{y}$ pulses increases as $B_{\text{AC}}$ increases, and the pulse overlap increases as $\Phi_{\text{AC}}$ increases for $\Phi_{\text{AC}} \in [0, \pi/2]$ and decreases for $\Phi_{\text{AC}} \in [\pi/2, \pi]$. We ignore the slight phase shift in the experimental data, but it can easily be incorporated into the proof by adjusting the bounds of $\Phi_{\text{AC}}$.

We assume the response is symmetric between $\Phi_{\text{AC}} \in [0, \pi]$ and $\Phi_{\text{AC}} \in [\pi, 2\pi]$, as shown in \ref{fig:fig2}A(ii) for $\gamma_{y} = 0.98$. If not, four sensors would be required instead of two to uniquely determine the phase.

\textit{Proof:}

We now show by proof of contradiction that $B^{(1)}$ and $\Phi^{(1)}$ (up to $[0, \pi/2]$) are uniquely determined if one measures $F(B^{(1)}, \Phi^{(1)})$ and $F(B^{(1)}, \Phi^{(1)} + \pi/2)$.

Let us assume that there exist $B^{(2)}$ and $\Phi^{(2)}$ such that 
\begin{align}
    F(B^{(1)},\Phi^{(1)}) &=  F(B^{(2)},\Phi^{(2)})
\label{eq:equality_1}
\intertext{\text{and}}
    F(B^{(1)},\Phi^{(1)} + \pi/2) &=  F(B^{(2)},\Phi^{(2)} + \pi/2) \, ,
\label{eq:equality_2}
\end{align}
where $B^{(1)} \neq B^{(2)}$ or $\Phi^{(1)} \neq \Phi^{(2)}$;
this would correspond to observing the same measurement for the two $\pi/2$-shifted sensor for different measurement signals.

Without loss of generality, we assume $0 \leq \Phi^{(1)}, \Phi^{(2)} \leq \pi/2$.

\smallskip

Further, let us focus on the case $\Bac^{(1)} < \Bac^{(2)}$; the case $\Bac^{(1)} > \Bac^{(2)}$ follows from similar arguments.
Since $F$ is monotonically increasing for both $\Bac$ [A] and $\Phi$ [B] for $\Phi\leq \pi/2$, the case $\Bac^{(1)} < \Bac^{(2)}$ and Eq.~\eqref{eq:equality_1} necessitate that $\Phi^{(1)}>\Phi^{(2)}$.
Hence, also $\Phi^{(1)}+\pi/2>\Phi^{(2)}+\pi/2$, such that the monotonic decrease of $F$ for $\Phi\geq \pi/2$ implies that
\begin{equation}
\label{eq:ineq_B}
    F(\Bac^{(1)},  \Phi^{(1)} + \pi/2) < F(\Bac^{(1)},  \Phi^{(2)} + \pi/2) \,,
\end{equation}
note that, we consider the same $\Bac$-value.
Likewise, since we consider $\Bac^{(1)} < \Bac^{(2)}$ and $F$ monotonically increasing in $\Bac$ we find
\begin{equation}
\label{eq:ineq_Phi}
    F(\Bac^{(1)},  \Phi^{(2)} + \pi/2) < F(\Bac^{(2)},  \Phi^{(2)} + \pi/2) \,,
\end{equation}
where the phase $\Phi=\Phi^{(2)}+\pi/2$ is fixed.
Combining Eqs.~\eqref{eq:ineq_B} and \eqref{eq:ineq_Phi}, we find
$$
     F(\Bac^{(1)},  \Phi^{(1)} + \pi/2) < F(\Bac^{(2)},  \Phi^{(2)} + \pi/2) \, ,
$$
which contradicts the assumption~\eqref{eq:equality_2}.
Hence, the conditions [A] and [B], and measuring $F(\Bac, \Phi)$ and $F(\Bac, \Phi+\pi/2)$, uniquely determine both the amplitude $\Bac$ and the phase $\Phi$ of the AC field, up to [0, $\pi/2$] the phase.
$\square$
\smallskip

Therefore, we have shown that using two $\pi/2$-shifted sensor is sufficient to uniquely determine the magnitude of an AC signal without requiring the knowledge of its phase.
}{proof:lemma}

\section{Sensing AC field with an unknown phase}
\label{supmat:two:sensor}

We have demonstrated the measurement of the amplitude of the AC field and the corresponding sensitivity of the DTC sensor when the phase of the AC field ($\Phi_{\text{AC}}$) is fixed at $\pi/2$. However, determining $B_{\text{AC}}$ for an AC field with an unknown phase presents a challenge, as the fidelity of the DTC sensor is influenced by both the amplitude and the phase of the AC field.

Here, we demonstrate how utilizing two (or more) sensors can accurately measure and determine the amplitude $B_{\text{AC}}$ of an AC field, even when its phase $\Phi_{\text{AC}}$ is unknown. Fig.~\ref{fig:sensor_offset}A illustrates the dual-sensor setup. By offsetting the $\gamma_{y}$ pulses by $T/2$ between the two sensors, where $T$ is the period of the DTC sequence, we can obtain two fidelity measurements: $F(B_{\text{AC}} = B^{(1)}, \Phi_{\text{AC}} = \Phi^{(1)})$ and $F(B_{\text{AC}}= B^{(1)}, \Phi_{\text{AC}} = \Phi_{\text{AC}}^{(1)} + \pi/2)$. Here, $F(B_{\text{AC}}, \Phi_{\text{AC}})$ represents the measured fidelity given the amplitude $B_{\text{AC}}$ and the phase $\Phi_{\text{AC}}$ of the applied AC field.

Fig.~\ref{fig:sensor_offset}B shows the response of the two sensors to an AC field with amplitude $B_{\text{AC}}$ across different phases $\Phi_{\text{AC}}$. By leveraging two key properties of the fidelity -- its monotonic increase with AC field amplitude $B_{\text{AC}}$ for all phases $\Phi_{\text{AC}}$, and its peak occurring at $\Phi_{\text{AC}} = \pi/2$ (disregarding a minor experimental phase offset) -- offsetting the two sensors by $T/2$ facilitates the unique determination of $B_{\text{AC}}$.

The following~\hyperlink{proof:main}{lemma} demonstrates that we can uniquely determine both the magnitude $B^{(1)}$, and the phase $\Phi^{(1)}$ of the AC field using the fidelity values $F(B_{\text{AC}} = B^{(1)}, \Phi_{\text{AC}} = \Phi^{(1)})$ and $F(B_{\text{AC}}= B^{(1)}, \Phi_{\text{AC}} = \Phi^{(1)} +\pi/2)$.

For practical usage, the response of the DTC sensor needs to be characterized for different values of $\Phi_{\text{AC}}$ and $B_{\text{AC}}$. This involves measuring the fidelity of the DTC across a range of phases and field strengths to approximate the function $F(B_{\text{AC}}, \Phi_{\text{AC}})$. If the sensor is operated at a bias AC field, the fidelity of the DTC at that bias AC field must be characterized for different $B_{\text{AC}}$ and $\Phi_{\text{AC}}$.

Given that the wavelength of the AC fields of interest is at least thousands of meters, the additional phase offset due to spatial separation of the sensors is negligible. For the detection of persistent signals, a single sensor could be initialized twice with two different phases.

The sensor's performance diminishes when not operated at the optimal phase, $\Phi_{\text{AC}} = \pi/2$. In a two-sensor scheme, sensitivity is expected to increase by approximately $\sqrt{2}$ in the worse case scenario, where the AC field has a phase of $\Phi_{\text{AC}} = \pi/4$ for sensor 1 and aliased to $\pi/4$ for sensor 2, due to fidelity measurement aliasing that restricts unique phase determination to $[0, \pi/2]$. We predict a sensitivity increase roughly by a factor of $\sqrt{2}$, because the overlap between $\gamma_{y}$ pulses and the AC field scales as $1/\sqrt{2}$ when the phase of the AC field is $\Phi_{\text{AC}} = \pi/4$, compared to the optimal phase $\Phi_{\text{AC}} = \pi/2$. Using more than two sensors can further enhance sensitivity in the worst-case scenario.

From the~\hyperlink{proof:main}{lemma}, it can be inferred that the phase $\Phi_{\text{AC}}$ (up to $[0, \pi/2]$) of an AC field can be detected using this two-sensor scheme, since $\Phi_{\text{AC}}$ is also uniquely determined with $B_{\text{AC}}$. Characterizing $\partial F/\partial \Phi_{\text{AC}}$ for various $B_{\text{AC}}$ to determine sensor sensitivity to phase is an interesting problem for future investigation.

In addition to the two-sensor scheme, an off-resonance sensing approach can potentially be used to mitigate the sensor's precise phase dependence. Extended coherence is observed for DTCs with a slightly off-resonant AC field applied, but with an additional beating effect (shown in \zfr{fig3}A(ii)). This beating signal could be leveraged to detect fields that are slightly off the DTC frequency, where the phase dependence would no longer be significant.

Furthermore, we can exploit the fact that maximum response is achieved when there is a maximum overlap between the AC field and the $\gamma_{y}$ pulses. We can initiate the DTC sequence slightly off-resonant to the AC field and monitor for maximum beating (when $\mathrm d\langle I_{x}\rangle/\mathrm dt = 0$). At this point, we adjust the sequence to be on-resonance, calibrating the phase at $\Phi_{\text{AC}} = \pi/2$, and proceed with the normal DTC sensing scheme. We leave the characterization of off-resonance sensing for future work.

\section{Effect of AC field during $\gamma_{y}$ pulses}
\label{supmat:only:gamma_y}

\begin{figure*}[t!]
    \centering
    \includegraphics[width=\textwidth]{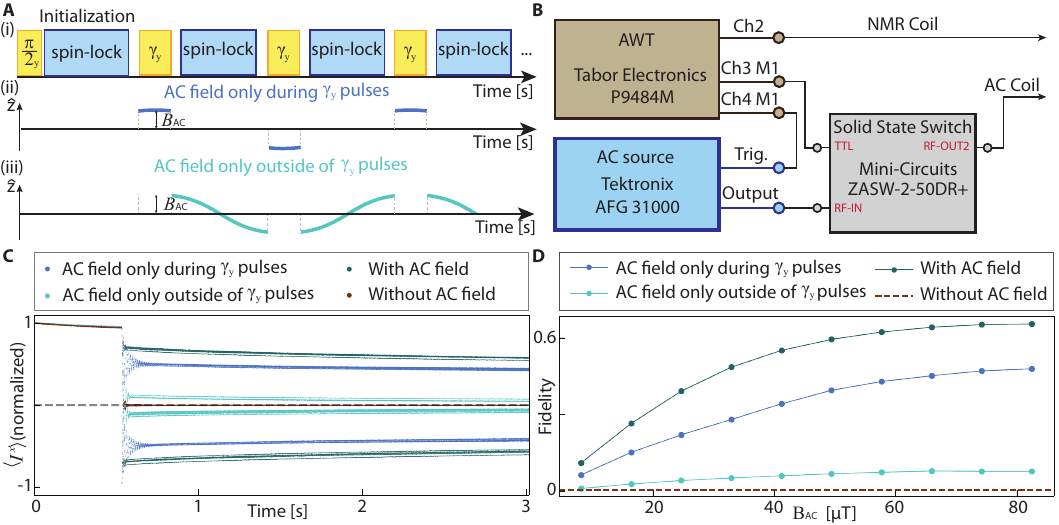}
    \caption{\textbf{AC-field only during $\gamma_{y}$ pulses and only outside of $\gamma_{y}$ pulses.} \textbf{A, Protocol:} (i) DTC sequence is applied after initializing \C\, nuclei in the $\mathbf{\hat{x}}$-axis by applying $(\pi/2)_{y}$ pulse and the spin-lock sequence (train of $\theta_{x}$ pulses). To examine the affect of the AC field during $\gamma_{y}$ pulses, AC-field is applied (ii) only during $\gamma_{y}$ pulses and (iii) only outside of $\gamma_{y}$ pulses. \textbf{B, Apparatus: } AWT (Arbitrary Waveform Transceiver) channel 2 generates pulses sent to the NMR coil after amplification. Marker 1 of channel 4 triggers the AC source, while marker 1 of channel 3 controls a solid-state switch, allowing the AC field to reach the AC coil only when the marker signal is high. \textbf{C, Signal:} $\langle I_{x}\rangle$ of the \C\, nuclei in the diamond, normalized to the amplitude of the initial signal. Dark blue: AC field applied only during $\gamma_{y}$ pulses; light blue: AC field applied only outside $\gamma_{y}$ pulses; dark green: AC field applied throughout the entire DTC sequence; brown: No AC field applied. $B_{\text{AC}} = \text{82.4}\mu\text{T}$, $\theta_{x}$ pulses ($\approx 52.8\mu\text{s}$), $\gamma_{y}$ pulses ($\approx 105.7\mu\text{s}$), and pulse separation $\tau$ ($\approx 36.0 \mu\text{s}$). \textbf{D, Fidelity vs $B_{\text{AC}}$} Same color scheme as C. The fidelity of the DTC is measured while varying $B_{\text{AC}}$.}
\label{fig:gated_AC-DTC}
\end{figure*}

In Sec.~\ref{supmat:sec:two_tone}, we theoretically show that the overlap between finite-length $\gamma_{y}$ pulses and the applied AC field is responsible for the observed extended coherence of \C~nuclear spins in their $\mathbf{\hat{x}}$-polarization. To confirm that this overlap is indeed the source of the extended coherence, we conduct experiments in which the AC field is applied exclusively during $\gamma_{y}$ pulses or exclusively outside of $\gamma_y$ pulses. 

Fig.~\ref{fig:gated_AC-DTC}A illustrates the schematic, while Fig.~\ref{fig:gated_AC-DTC}B details the experimental setup. A solid-state switch (Mini-Circuits ZASW-2-50DR+) is used to gate the continuous AC field applied from the source (Tektronix AFG 31000), allowing it to be applied only during or outside $\gamma_{y}$ pulses. We compare the fidelity of these two cases with the fidelity obtained when the AC field is applied throughout the entire DTC sequence and when no AC field is applied. Fig.~\ref{fig:gated_AC-DTC}C presents the raw data (time vs.~$\langle I_{x} \rangle$) of all four cases when $B_{\text{AC}} \approx 82.4\mu \text{T}$, while Fig.~\ref{fig:gated_AC-DTC}D shows how the fidelity changes across all four cases as $B_{\text{AC}}$ increases. Fig.~\ref{fig:gated_AC-DTC}C and D reveal that most of the increase in the fidelity originates from the overlap of the AC field with $\gamma_{y}$ pulses, as predicted by the theory, for all values of $B_{\text{AC}}$. Although there is a slight increase in fidelity when applying the AC field only outside of $\gamma_{y}$ pulses, it is minimal compared to the substantial increase observed when the AC field is applied only during $\gamma_{y}$ pulses.

Theoretically, an increase in fidelity is not expected when the AC magnetic field is applied solely outside of the $\gamma_{y}$ pulses. We attribute this slight increase to the $\theta_{x}$ pulses possessing a minor y-component in their pulse transients, with its effect accumulating over the train of $\theta_{x}$ pulses~\cite{mehring1972phase}.

\end{document}